\documentclass[11pt,a4paper]{article}
\usepackage{jcappub}
\usepackage[a4paper,top=1in,bottom=1in,left=15mm,right=15mm,footskip=0.25in]{geometry}

\usepackage{cleveref}[2012/02/15]
\crefformat{footnote}{#2\footnotemark[#1]#3}

\usepackage{doi}

\usepackage[flushleft]{threeparttable}

\usepackage{amsmath}
\usepackage{amsbsy}
\usepackage{mathtools}
\usepackage{float}
\usepackage{datetime}
\usepackage{textpos}
\usepackage{booktabs}
\usepackage{graphicx}
\usepackage{mathrsfs}
\usepackage{adjustbox}
\usepackage{epsfig}
\usepackage{color}
\usepackage{bm}
\usepackage{graphicx}
\usepackage{physics}
  
\usepackage[all]{hypcap}

\usepackage{natbib}

\providecommand{\adsurl}[1]{\href{#1}{ADS}}
 
\providecommand{\url}[1]{\href{#1}{#1}}
\usepackage{url}

\usepackage{accents}

\usepackage{amsmath}
\usepackage{graphicx}
\usepackage{amsmath}
\usepackage{float}

\usepackage{times}
\usepackage{mathptmx}

\usepackage{hhline}

\providecommand{\abs}[1]{\lvert#1\rvert}

\usepackage{subfigure}

\def\msun{{\,M_\odot}}

\DeclareMathAlphabet\mathbfcal{OMS}{cmsy}{b}{n}

\renewcommand{\d}{\textrm{d}}
\newcommand{\D}[1]{\pmb{D}_{#1}}
\newcommand{\obs}{\rm obs}

\def\alt{\raise0.3ex\hbox{$\;<$\kern-0.75em\raise-1.1ex\hbox{$\sim\;$}}}
\def\agt{\raise0.3ex\hbox{$\;>$\kern-0.75em\raise-1.1ex\hbox{$\sim\;$}}}

\newcommand{\thetab}{\pmb{\theta}}

\newcommand{\bw}{\begin{widetext}}
\newcommand{\ew}{\end{widetext}}
\def\d{{\rm d}}

\newcommand{\lsim}{\,\rlap{\raise 0.35ex\hbox{$<$}}{\lower 0.7ex\hbox{$\sim$}}\,}
\newcommand{\gsim}{\,\rlap{\raise 0.35ex\hbox{$>$}}{\lower 0.7ex\hbox{$\sim$}}\,}

\def\lesssim{\mathrel{\hbox{\rlap{\hbox{\lower3pt\hbox{$\sim$}}}\hbox{\raise2pt\hbox{$<$}}}}}
\def\gtrsim{\mathrel{\hbox{\rlap{\hbox{\lower3pt\hbox{$\sim$}}}\hbox{\raise2pt\hbox{$>$}}}}}

\def\xlinkspace#1 #2{%
 \ifx\relax#2%
 \xlinkdash#1-\relax
 \else
 \xlinkdash#1 -\relax
 \expandafter\xlinkspace\expandafter#2%
 \fi}

\def\xlinkdash#1-#2{%
 \ifx\relax#2%
 \tmp{#1}%
 \else
 \tmp{#1-}%
 \expandafter\xlinkdash\expandafter#2%
 \fi}

\title{Comparing the Galactic Bulge and Galactic Disk Millisecond Pulsars}

\author[a]{Harrison Ploeg}
\author[a]{Chris Gordon}
\author[b]{Roland Crocker}
\author[c]{Oscar Macias}

\affiliation[a]{School of Physical and Chemical Sciences, University of Canterbury, Christchurch, New Zealand}
\affiliation[b]{Research School of Astronomy and Astrophysics, Australian National University, Canberra, Australia}
\affiliation[c]{Kavli Institute for the Physics and Mathematics of the Universe, University of Tokyo, Kashiwa, Chiba 277-8583, Japan}
\affiliation[c]{GRAPPA Institute, University of Amsterdam, 1098 XH Amsterdam, The Netherlands}

\emailAdd{hzp10@uclive.ac.nz}
\emailAdd{chris.gordon@canterbury.ac.nz}
\emailAdd{roland.crocker@anu.edu.au}
\emailAdd{o.a.maciasramirez@uva.nl}

\abstract{
The Galactic Center Excess (GCE) is an extended gamma-ray source in the central region of the Galaxy found in Fermi Large Area Telescope (Fermi-LAT) data. One of the leading explanations for the GCE is an unresolved population of millisecond pulsars (MSPs) in the Galactic bulge. Due to differing star formation histories it is expected that the MSPs in the Galactic bulge are  older and therefore dimmer than those in the Galactic disk. Additionally, correlations between the spectral parameters of the MSPs and the spin-down rate of the corresponding neutron stars
have been observed. This implies that the bulge MSPs may be spectrally different from the disk MSPs. We perform detailed modelling of the MSPs from formation until observation. Although we  confirm the correlations, we do not find they are sufficiently large to significantly differentiate the spectra of the bulge MSPs and disk MSPs when the uncertainties are accounted for.
Our results demonstrate that the population of MSPs that can explain the gamma-ray signal from the resolved MSPs in the Galactic disk and the unresolved MSPs in the boxy bulge and nuclear bulge
can consistently be described as arising from a common evolutionary trajectory for some subset of astrophysical sources common to all these different environments.
We do not require that there is anything unusual about inner Galaxy MSPs to explain the GCE.
Additionally, we use a more accurate geometry for the distribution of bulge MSPs
and incorporate dispersion measure estimates of the MSPs' distances. We find that the elongated boxy bulge morphology means that some the bulge MSPs are closer to us and so easier to resolve.  We identify three resolved MSPs that have significant probabilities of belonging to the bulge population.

}


\begin{document}

\maketitle

\flushbottom

\section{Introduction}
\label{sec:intro}

 Initially, the GCE appeared to be  distributed spherically symmetrically 
around the Galactic Center with a radially-declining intensity.
This, and the fact that it exhibits a spectral peak at a few GeV, 
suggested the possibility that it was evidence of weakly interacting massive particles (WIMPs) self-annihilating with a Navarro-Frenk-White (NFW) density profile \citep{Goodenough:2009gk,Hooper_2011,Abazajian:2012pn,Gordon:2013vta}. Evidence now suggests that,
when examined in detail, the GCE is not in fact spherically symmetric but exhibits a spatial morphology correlated with stellar mass in the Galactic bulge \citep{Macias_2018,Bartels2017, Macias19, Abazajian2020,Coleman19}. 
This  disfavors a dark matter origin for the GCE and rather points to a scenario in which it is produced by a population of dim, unresolved, astrophysical point sources such as MSPs \citep{Abazajian:2010zy}. 
There is some debate about whether the resolved MSPs are consistent with the needed bulge population, see for example refs.~\cite{Hooper:2015jlu,Haggard_2017,Ploeg:2017vai,Bartels2018}. 

In previous investigations, the spectrum of the MSPs in the bulge have been assumed to be the same as those in the disk. However, the bulge MSPs are expected to be on average  older than the disk MSPs due to their different star formation histories \cite{Crocker:2016zzt}.
Therefore, we would expected them to have on average lower luminosities. In addition to this, a correlation between luminosities on the spectral parameters is seen in the data \cite{TheFermi-LAT:2013ssa,Kalapotharakos_2019}
and so one would expect the bulge MSPs to be spectrally different from the disk ones. This motivates the  more detailed modelling of the MSP populations that is performed in the current article.

We extend the model of the Galactic population of gamma-ray millisecond pulsars (MSPs) of Ploeg et al.~\cite{Ploeg:2017vai}. 
In the current article we use the dispersion measure estimates of the MSP distances and incorporate the corresponding uncertainties in the free electron densities.
Also, instead of empirically parameterising the MSP luminosity function, we start from empirical distributions describing MSP initial period, magnetic field strength, age, and gamma-ray spectra. We  assume there is a relationship between the luminosity of a pulsar and some of its other properties such as its period, period derivative or spectral energy cutoff. 
This supposition is motivated by work such as by Kalapotharakos et al.~\cite{Kalapotharakos_2019} in which, based on data on  resolved MSPs and young pulsars in Abdo et al.~\cite{TheFermi-LAT:2013ssa}, it was determined that  there is
a relationship between
MSP
luminosity, on the one hand, 
and MSP spectral energy cutoff $E_{\rm cut}$, magnetic field strength $B$, and the spin-down power $\dot{E}$,
on the other.
Specifically,
Kalapotharakos et al.~\cite{Kalapotharakos_2019}
found that gamma-ray emission from MSPs is via curvature radiation and scales like $L \propto E_{\rm cut}^{1.18 \pm 0.24} B^{0.17 \pm 0.05} \dot{E}^{0.41 \pm 0.08}$ where we use 68\% confidence intervals when quoting uncertainties unless otherwise specified.
Similarly, Gonthier et al.~\cite{Gonthier2018} assumed that the radio and gamma ray luminosities of MSPs
were dependent on period and period derivative 
and, under this assumption,
successfully
determined
the parameters of that relationship using a model of the distribution and properties of the Galactic population of MSPs, radio and gamma-ray detection thresholds, and a model of how the observed flux of an individual MSP depends on viewing angle and magnetic axis angle.

\section{Method}
To fit our model of the Galactic MSP population we used MSPs with confirmed gamma ray pulsations according to the Public List of LAT-Detected Gamma-Ray Pulsars\footnote{\label{footnote:pulsars}\url{https://confluence.slac.stanford.edu/display/GLAMCOG/Public+List+of+LAT-Detected+Gamma-Ray+Pulsars}}. From that list, we obtained names and periods of pulsars. We then used the gamma-ray data for the corresponding sources in the Fermi Large Area Telescope fourth source catalog data release 2 \citep[4FGL-DR2:][]{Ballet:2020hze}. Additional data was taken from the Australia Telescope National Facility (ATNF) pulsar catalog \citep{Manchester:2004bp} if available. As in Bartels et al.~\cite{Bartels2018} we used pulsars with periods less than $30$ ms that were not associated with globular clusters. For the Galactic Center Excess (GCE), we use the boxy bulge and nuclear bulge spectra\footnote{Available from \url{https://github.com/chrisgordon1/Galactic_bulge_spectra}.} from Macias et al.~\cite{Macias19}. We included both the systematic and statistical errors of these spectra which were added in quadrature to get the total error. The systematic error accounts for variation in the GCE spectra caused by using different maps of Inverse Compton emission.
\label{sec:method}
\subsection{Modeling the Galactic Millisecond Pulsar population}
\label{ssec:method_model_MSP_pop}
\subsubsection{Spatial distribution}
The spatial model of MSPs has three components: a disk distribution, a boxy bulge distribution, and a nuclear bulge distribution. The disk component models the population from which we expect the resolved MSPs to mainly come and has density:
\begin{equation}
\label{eq:rho_disk}
\rho_{\rm disk} (R, z) \propto \exp(-R^2/2\sigma_r^2) \exp(-\abs{z}/z_0)
\end{equation}
\noindent where $R^2 = x^2 + y^2$ is the radial coordinate in the Galactic disk and $z$ is the height above the Galactic Plane. We treat $\sigma_r$ and $z_0$ as free parameters to be fit to the data. The modeled GCE is produced by the boxy bulge and nuclear bulge components. The boxy bulge has density \citep{Freudenreich:1997bx,Macias19}:
\begin{equation}
\label{eq:rho_main_bulge}
\rho_{\rm boxy~bulge} (R_s) \propto \sech^2(R_s)
\times \begin{cases}
      1 & R \leq R_{\rm end} \\
      \exp(-(R - R_{\rm end})^2/h^2_{\rm end}) & R > R_{\rm end} \\
\end{cases}
\end{equation}
\noindent where $R_{\rm end} = 3.128$ kpc, $h_{\rm end} = 0.461$ kpc, and:
\begin{equation}
\label{eq:rho_main_bulge_r_perp}
R_\perp^{C_\perp} = \left(\frac{\abs{x'}}{1.696~\textrm{kpc}} \right)^{C_\perp} + \left(\frac{\abs{y'}}{0.6426~\textrm{kpc}} \right)^{C_\perp}
\end{equation}
\begin{equation}
\label{eq:rho_main_bulge_r_s}
R_s^{C_\parallel} = R_\perp^{C_\parallel} + \left(\frac{\abs{z'}}{0.4425~\textrm{kpc}} \right)^{C_\parallel}
\end{equation}
\noindent where $C_\parallel = 3.501$ and $C_\perp = 1.574$. The coordinates $x'$, $y'$ and $z'$ are Cartesian coordinates in the boxy bulge frame. Relative to the frame in which $x_\odot=-R_0$, $y_\odot=z_\odot=0$, this frame is rotated $13.79^{\circ}$ around the z-axis then $0.023^{\circ}$ around the new y-axis. We assume $R_0=8.3$ kpc as adopted from the YMW16 free electron density model \citep{Yao2017} which we use to convert between distance and dispersion measure 
The nuclear bulge MSP density is proportional to the sum of the mass densities of the nuclear stellar cluster (NSC) and nuclear stellar disk (NSD) \citep{Bartels2017}:
\begin{equation}
\label{eq:rho_nuclear_bulge}
\rho_{\rm nuclear~bulge} (r,z) \propto \rho_{\rm NSC}(r) + \rho_{\rm NSD}(r,z)
\end{equation}
\noindent where $r^2 = x^2 + y^2 + z^2$ and where the NSC has density:
\begin{equation}
\label{eq:rho_nuclear_stellar_cluster}
\rho_{\rm NSC} (r) = 
\begin{cases} 
      \frac{\rho_{0 \rm ,NSC}}{1 + \left( \frac{r}{r_0} \right)^2} & r \leq 6~\textrm{pc} \\
      \frac{\rho_{1 \rm ,NSC}}{1 + \left( \frac{r}{r_0} \right)^3} & 6~\textrm{pc} < r \leq 200~\textrm{pc} \\
      0 & r > 200~\textrm{pc}
   \end{cases}
\end{equation}
\noindent where $r_0 = 0.22~\textrm{pc}$, $\rho_{0 \rm ,NSC} = 3.3 \times 10^6~\textrm{M}_\odot~\textrm{pc}^{-3}$ and $\rho_{1 \rm ,NSC}$ is set so that $\rho_{\rm NSC} (r)$ is continuous at $r = 6~\textrm{pc}$. The NSD has density:
\begin{equation}
\label{eq:rho_nuclear_stellar_density}
\rho_{\rm NSD} (r,z) = 
\begin{cases} 
      \rho_{0 \rm ,NSD}\left(\frac{r}{1~\textrm{pc}} \right)^{-0.1} e^{-\frac{\abs{z}}{45~\textrm{pc}}} & r < 120~\textrm{pc} \\
      \rho_{1 \rm ,NSD}\left(\frac{r}{1~\textrm{pc}} \right)^{-3.5} e^{-\frac{\abs{z}}{45~\textrm{pc}}} & 120~\textrm{pc} \leq r < 220~\textrm{pc} \\
      \rho_{2 \rm ,NSD}\left(\frac{r}{1~\textrm{pc}} \right)^{-10} e^{-\frac{\abs{z}}{45~\textrm{pc}}} & r \geq 220~\textrm{pc}
   \end{cases}
\end{equation}
\noindent where $\rho_{0 \rm ,NSD} = 301~\textrm{M}_\odot~\textrm{pc}^{-3}$ and $\rho_{1 \rm ,NSD}$ and $\rho_{2 \rm ,NSD}$ are set so that $\rho_{\rm NSD} (r,z)$ is continous at both $r = 120~\textrm{pc}$ and $r = 220~\textrm{pc}$.

\subsubsection{Age distribution}
The MSP age distribution is dependent on two distributions: the star formation rate (SFR), and the delay time distribution (DTD). We assume for the DTD a five bin distribution between 0 Gyr and the age of the universe.

\begin{figure}
    \centering
    \includegraphics[width=0.75\linewidth]{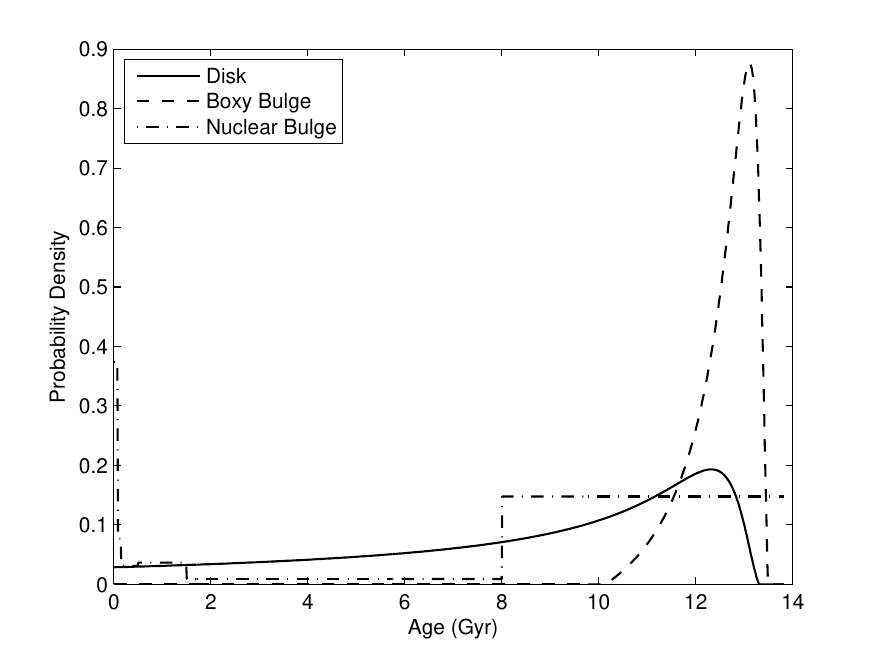}
    \caption{Star formation rates for the disk, boxy bulge, and nuclear bulge. Here, this is given as a normalized probability density of stellar mass having formed at a particular age. 
    }
    \label{fig:sfr}
\end{figure}

For the disk and boxy bulge components of the MSP population, the SFR is \citep{Crocker:2016zzt}:
\begin{equation}
\label{eq:SFR}
\textrm{SFR}(z) = \max(10^{A z^2 + B z + C} - D, 0)
\end{equation}
\noindent where for the disk MSPs $A = -4.06 \times 10^{-2}$, $B = 0.331$, $C = 0.338$ and $D = 0.771$. For the boxy bulge $A = -2.62 \times 10^{-2}$, $B = 0.384$, $C = -8.42 \times 10^{-2}$ and $D = 3.254$. The relationship between cosmological time $t$ and redshift  $z$ is \citep{Weinberg:2008zzc}:
\begin{equation}
\label{eq:z_to_t}
t(z) = \frac{9.778~\textrm{Gyr}}{h} \int^{1/(1+z)}_0 \frac{\dd x}{x \sqrt{\Omega_\Lambda + \Omega_K x^{-2} + \Omega_M x^{-3} + \Omega_R x^{-4}}}
\end{equation}
\noindent where we have used $h = 0.67$, $\Omega_\Lambda = 0.68$, $\Omega_M = 1 - \Omega_\Lambda$ and $\Omega_K = \Omega_R = 0$ \citep{Planck2014}. For the nuclear bulge, we use the MIST star formation rate from Nogueras-Lara et al.~\cite{Nogueras-Lara2020}. The three star formation rates are shown in Fig.~\ref{fig:sfr}.

From the SFR and DTD, we can find the probability density function of MSP ages:
\begin{equation}
    p({\rm age}) = \frac{\int_{0}^{t_0 - {\rm age}} {\rm SFR}(z(\tau)) {\rm DTD}(t_0 - {\rm age} - \tau) \dd\tau}{\int_{0}^{t_0}\int_{0}^{t'} {\rm SFR}(z(\tau)) {\rm DTD}(t' - \tau) \dd\tau \dd t'}
\end{equation}
\noindent where $t_0$ is the age of the universe, and $z(t)$ is the inverse of Eq.\ \ref{eq:z_to_t}.

As an alternative age distribution, we try a uniform distribution where MSPs form at a constant rate over the last 10 Gyr, similar to that assumed by Gonthier et al.~\cite{Gonthier2018}.

\subsubsection{Angular velocity}
In our model, the angular transverse velocities of pulsars in the directions of Galactic longitude and latitude, $\mu_l$ and $\mu_b$ respectively, are determined by assuming a pulsar travels in a circular orbit around the center of the Galaxy using a  parametric form of the   potential \citep{Carlberg1987,Kuijken1989}
with a random normally distributed peculiar velocity in every direction of scale $\sigma_v$ (i.e., a Maxwell distribution). For the Sun, we assume circular motion in the same potential with a peculiar velocity of (11.1, 12.24, 7.25) km s$^{-1}$ where the velocity components are in the direction of the Galactic Center, the direction of rotation and in the direction perpendicular to the plane, respectively \citep{Schonrich2010}. The relationship between velocity ($v$) and angular velocity ($\mu$) at a distance $d$ is:
\begin{equation}
\label{eq:proper_motion_to_linear_velocity}
    v \approx 4.74 \left(\frac{d}{\textrm{kpc}}\right) \left(\frac{\mu}{\textrm{ mas yr}^{-1}}\right) \textrm{ km s}^{-1}
\end{equation}

\subsubsection{Galactic Center Excess}
To simulate the GCE, we need to assign each bulge MSP a spectrum that describes its photon number flux ($N$) at energy $E$:%
\begin{equation}
\label{eq:plsec_spectrum}
    \frac{\dd N}{\dd E} = K E^{-\Gamma} \exp(-\left(E/E_{\rm cut}\right)^{2/3})
\end{equation}
\noindent where the spectral parameters are $K$, $\Gamma$, and $E_{\rm cut}$. The proportionality constant $K$ is determined via:
\begin{equation}
    F = \int_{0.1\textrm{ GeV}}^{100\textrm{ GeV}} E \frac{\dd N}{\dd E} \dd E
\end{equation}
{\textcolor{black}{where $F$ is the energy flux}}.
\noindent The simulated boxy bulge and nuclear bulge GCE spectra are then the sum of all MSP spectra in each of those two bulge populations of MSPs. This spectrum is equivalent to the one used to fit the resolved MSPs in the 4FGL-DR2 
{\textcolor{black}{catalog}}
\citep{Ballet:2020hze} and it will also be the one we use for our resolved MSPs. 

\subsubsection{Millisecond pulsar parameters}
Force-free electrodynamic solutions have given the following expression for the spin-down luminosity
 \cite{Spitkovsky2006}
\begin{equation}
    L_{\mathrm{sd}} \sim \frac{\mu^{2} \Omega^{4}}{c^{3}}\left(1+\sin ^{2} \alpha\right)
    \label{eq:spindownlum}
\end{equation}
where $\mu$ is the magnetic dipole moment, $\Omega$ is the rotational angular velocity, $c$ is the speed of light, and $\alpha$ is the angle between the rotation and magnetic field axes. The magnetic field strength ($B$)
at the magnetic pole of the star  is related to $\mu$ by \cite{Spitkovsky2006}
\begin{equation}
    \mu=\frac{B R^3}{2}
\label{eq:mu}
\end{equation}
where $R$ is the radius of the neutron star and we use $R = 12~\textrm{km}$.
The rotational kinetic energy  of the neutron star is given by the standard formula for a rotating body
\begin{equation}
E=\frac{1}{2} I \Omega^{2}
\end{equation}
where 
 $I$ is the neutron star's moment of inertia and we use
 $I = 1.7 \times 10^{45}~\textrm{g cm}^2$.
Therefore, the spin-down power  satisfies
\begin{equation}
    \dot{E} = 4 \pi^2 I \dot{P}_{\rm int} / P^3
    \label{eq:Edot}
\end{equation}
where $\dot{P}_{\rm int}$ is the time derivative of the intrinsic  period, which may be different to the observed period derivative ($\dot{P}$),
and $P=2\pi/\Omega$ is the period.
Equating the spin-down luminosity (Eq.~\ref{eq:spindownlum}) to the spin down power (Eq.~\ref{eq:Edot}) and using Eq.~\ref{eq:mu} yields
the following expression for the magnetic field strength of an MSP
\begin{equation}
\label{eq:magnetic_field_strength}
B^2 = \frac{c^3 I P \dot{P}_{\rm int}}{\pi^2 R^6 (1 + \sin^2(\alpha))}.
\end{equation}
%
%
The angle $\alpha$ is chosen randomly from the probability density:
\begin{equation}
    p(\alpha) = \frac{1}{2} \sin(\alpha)
\end{equation}
\noindent which corresponds to a uniformly random magnetic field axis relative to the rotation axis.

The intrinsic period time derivative is related to the observed period time  derivative $\dot{P}$ by \citep{TheFermi-LAT:2013ssa}:
\begin{equation}
\label{eq:period_derivative_observed_corrections}
    \dot{P} = \dot{P}_{\rm int} + \dot{P}_{\rm Shklovskii} + \dot{P}_{\rm Galactic}
\end{equation}
\noindent where the 
contribution to the observed period derivative
from the 
Shklovskii effect is given by 
\begin{equation}
\label{eq:Shklovskii}
    \dot{P}_{\rm Shklovskii} = 2.43 \times 10^{-21} \left(\frac{\mu}{\textrm{mas yr}^{-1}}\right)^2 \left(\frac{d}{\textrm{kpc}}\right) \left(\frac{P}{\textrm{s}}\right)
\end{equation}
\noindent and the contribution due to the relative acceleration in the Galactic potential is:
\begin{equation}
\label{eq:period_derivative_Galactic_correction}
    \dot{P}_{\rm Galactic} = \frac{1}{c} \boldsymbol{n}_{10} \cdot (\boldsymbol{a}_p - \boldsymbol{a}_\odot) P
\end{equation}
\noindent where $\boldsymbol{n}_{10}$ is the unit vector from the Sun to the pulsar, and $\boldsymbol{a}_p$ and $\boldsymbol{a}_\odot$ are the accelerations, due to the Galactic potential  \cite{Carlberg1987,Kuijken1989},  of the pulsar and Sun 
respectively.
Note that $P=\left(1+v_{\mathrm{R}} / c\right) P_{\mathrm{int}}$, where $v_{\mathrm{R}}$ is the radial velocity of the pulsar and $P_{\mathrm{int}}$ is the intrinsic period. 
Given $v_{\mathrm{R}} \ll c$, we approximate $P_{\mathrm{int}}=P$. Making the common assumption that magnetic field strength remains constant over time, Eq.~\ref{eq:magnetic_field_strength} results in:
\begin{equation}
\label{eq:current_period}
P = \sqrt{P_I^2 + \frac{2 \pi^2 R^6}{I c^3} (1 + \sin^2(\alpha)) B^2 t}
\end{equation}
\noindent where $P_I$ is the initial period of the MSP at birth and $t$ is the age.

We %
{\textcolor{black}{consider}}
multiple relationships between pulsar parameters and the $0.1$-$100$ GeV luminosity ($L$).
Our most general form is 
that     used by Kalapotharakos et al.~\cite{Kalapotharakos_2019}:
    \begin{equation}
    \label{eq:Model1}
        L = \eta E_{\rm cut}^{a_{\gamma}} B^{b_{\gamma}} \dot{E}^{d_{\gamma}}
    \end{equation}
    where $\eta$ is a proportionality factor.
 We also consider the model used by Gonthier et al.~\cite{Gonthier2018}:
    \begin{equation}
    \label{eq:Model6}
        L = \eta P^{\alpha_{\gamma}} \dot{P}^{\beta_{\gamma}}\,.
    \end{equation}
The simplest form we consider is that the luminosity is an entirely independent parameter as used by  Ploeg et al.~\cite{Ploeg:2017vai}:
    \begin{equation}
    \label{eq:Model9}
       L = \eta\,. 
    \end{equation}
\noindent

The likelihood probability density distributions of  $B$, $E_{\rm cut}$, and $\eta$ are assumed to be  log-normal
{\textcolor{black}{as this functional form gives a good fit to their histogrammed data of the resolved Fermi-LAT MSPs:}}
\begin{equation}
\label{eq:lognormal_def}
p(\log_{10}(x)|x_{\rm med},\sigma_x) = \frac{1}{\sqrt{2 \pi} \sigma_x} \exp\left(-\frac{\left(\log_{10}(x) - \log_{10}(x_{\rm med}) \right)^2}{2 \sigma_x^2}\right)
\end{equation}
\noindent where $x_{\rm med}$ is the median of $x$ and $\sigma_x$ is the standard deviation of $\log_{10}(x)$. 
We also assume $P_I$ has this form but our results are not sensitive to {\textcolor{black}{to this assumption.}}
In particular, we found that the cut off power law model used by Gonthier et al.~\cite{Gonthier2018}
{\textcolor{black}{gives similar results}}. 
Note that, 
even though $E_{\rm cut}$ is obtained from a fit of an individual MSP's spectral data,
for our purposes it is treated as a directly measured datum rather than a parameter to be estimated. 

Fits to the MSP gamma-ray data \citep{TheFermi-LAT:2013ssa} have uncovered correlations between the spectral parameters and $\dot{E}$. To allow  for this to potentially be an intrinsic property of the MSPs, we parameterise the median of the $E_{\rm cut}$ likelihood as:
\begin{equation}
    \log_{10}(E_{\rm cut, med} / \text{MeV}) = a_{E_{\rm cut}} \log_{10}(\dot{E}/(10^{34.5} \textrm{ erg s}^{-1})) + b_{E_{\rm cut}}
    \label{eq:Ecut}
\end{equation}
\noindent where $a_{E_{\rm cut}}$ and $b_{E_{\rm cut}}$ are allowed to vary in our model fits.
We also model the likelihood of spectral index $\Gamma$ using a normal distribution with mean $\mu_{\Gamma}$ and standard deviation $\sigma_{\Gamma}$. We assume that
\begin{equation}
    \mu_{\Gamma} = a_{\Gamma} \log_{10}(\dot{E}/(10^{34.5} \textrm{ erg s}^{-1})) + b_{\Gamma}
    \label{eq:Gamma}
\end{equation}
\noindent where $a_{\Gamma}$ and $b_{\Gamma}$ are parameters. We also have a correlation coefficient, $r_{\Gamma, E_{\rm cut}}$ between $\Gamma$ and $\log_{10}(E_{\rm cut})$, so the likelihood is:
\begin{equation}
    p(\Gamma, \log_{10}(E_{\rm cut})|\thetab) = \frac{1}{2 \pi \sigma_{\Gamma} \sigma_{E_{\rm cut}} \sqrt{1 - r_{\Gamma, E_{\rm cut}}^2}} \exp\left(-\frac{z_{\Gamma, E_{\rm cut}}}{2 (1 - r_{\Gamma, E_{\rm cut}}^2)}\right)
\end{equation}
\noindent where:
\begin{equation}
    z_{\Gamma, E_{\rm cut}} = \frac{\left(\Gamma - \mu_{\Gamma} \right)^2}{\sigma_{\Gamma}^2} + \frac{\left(\log_{10}(E_{\rm cut}) - \log_{10}(E_{\rm cut, med}) \right)^2}{\sigma_{E_{\rm cut}}^2} - \frac{2 r_{\Gamma, E_{\rm cut}} \left(\Gamma - \mu_{\Gamma} \right) \left(\log_{10}(E_{\rm cut}) - \log_{10}(E_{\rm cut, med}) \right)}{\sigma_{\Gamma} \sigma_{E_{\rm cut}}}\, .
\end{equation}
\noindent
Note that here and in the rest of this article we will use $\thetab$ to indicate the relevant model parameters. In this case they are $\thetab=\{a_{E_{\rm cut }},b_{E_{\rm cut }},a_\Gamma, B_\Gamma, \sigma_{E_{\rm cut}},\sigma_\Gamma,r_{\Gamma,E_{\rm cut}}\}$. See Table~\ref{tab:prior_ranges} for the corresponding  priors that we use.

As alternatives, we try two likelihoods of $P_I$ where there is a dependence on $B$. In the first,  a bivariate normal distribution relating $\log_{10}(B)$  and $\log_{10}(P_I)$ with a correlation parameter $r_{B, P_I}$ is assumed.
In the second, we adopt the relationship used by Gonthier et al.~\cite{Gonthier2018}:
\begin{equation}
\label{eq:gonthier_initial_period_dist}
    P_I = 0.18 \times 10^{C_{P_I} + 3 \delta / 7} B_8^{6/7} \textrm{ms}
\end{equation}
\noindent where there is a lower bound of $1.3$ ms, $\delta$ is drawn from a uniform distribution between $0$ and $2$. Also, $B_8 = (B/10^8 \textrm{G})$.
We fit $C_{P_I}$  as model parameter while it is set to $0$ by Gonthier et al.~\cite{Gonthier2018}.

\subsubsection{Millisecond pulsar detection}
The flux of an MSP is related to the luminosity in the usual way:
\begin{equation}
\label{eq:flux}
F = \frac{L}{4 \pi d^2}
\end{equation}
\noindent and the MSP detection threshold flux $F_{\rm th}$ is drawn from a log-normal distribution so that the probability of a detection is \citep{Hooper:2015jlu,Ploeg:2017vai,Bartels2018}:
\begin{equation}
\label{eq:detection_probability}
p(F_{\rm th} \leq F \vert l, b,\thetab) = \frac{1}{2} \left(1 + \erf \left(\frac{\log_{10}(F) - \left(\log_{10}(\mu_{\rm th}(l,b)) + K_{\rm th}\right)}{\sqrt{2} \sigma_{\rm th}}\right)\right)
\end{equation}
\noindent where a map in longitude ($l$) and latitude ($b$) of $\mu_{\rm th}(l,b)$ associated with the 4FGL-DR2 catalog can be found online\footnote{\url{https://fermi.gsfc.nasa.gov/ssc/data/access/lat/10yr\_catalog/}}.

\subsection{Fitting the model to data}
\label{ssec:method_fitting_data}
To fit the model parameters to the resolved MSP and GCE data, we use an adaptive Markov Chain Monte Carlo (MCMC) algorithm \citep{Haario01}. 
The resolved MSPs have  an unbinned Poisson likelihood \citep{Cash1979}:
\begin{equation}
\label{eq:likelihood_definition}
\mathcal{L}_{\rm res}
\propto \exp(-\lambda_{\rm res})  \prod_{i=1}^{N_{\rm res} }  \rho(\D{i})
\end{equation}
\noindent 
 where $N_{\rm res}$ is the number of resolved MSPs, $\lambda_{\rm res}$ is the expected number of resolved MSPs, 
and $\rho(\D{i})$ is the phase space density of the resolved MSPs which have 
the data 
\begin{equation}
    \D{i}=\{
l_i, b_i, d_i, P_i, \dot{P}_i, \mu_{l, i}, \mu_{b, i}, F_i, E_{{\rm cut}, i}, \Gamma_i
\} \,.
\end{equation}
That is, the resolved MSP
is located
at longitude $l_i$, latitude $b_i$,  distance $d_i$, period $P_i$, observed period derivative $\dot{P}_i$,  proper motion  in longitude  $\mu_{l, i}$, proper motion  in  latitude $\mu_{b, i}$, flux $F_i$, spectrum energy cut-off $E_{{\rm cut}, i}$, and  spectral index $\Gamma_i$.
The expected number of resolved MSPs can be obtained by integrating the phase space density over the  phase space volume as follows \cite{Cash1979}:
\begin{equation}
\label{eq:lambdarho}
    \lambda_{\rm res}=\int \rho(\D{i})\, \d{\D{i}}%
\end{equation}
This relation implies that 
\begin{equation}
\label{eq:rhop}
\rho(\D{i})=p(\D{i}\vert {\rm obs}, \thetab) \lambda_{\rm res}
\end{equation}
where ${\rm obs}$ indicates the MSP was observed, i.e.\ it was resolved. Also,
$p( \D{i}|{\rm obs},\theta) $ is 
the probability density that a resolved MSP has data $\D{i}$
given that  the model parameter values are $\theta$. It then follows from the above two equations that 
\begin{equation}
     \int p(\D{i}\vert {\rm obs}, \thetab) \, \d{\D{i}}=1
\end{equation}
as required.

There are many tens of thousands of unresolved MSPs in the Milky Way \cite{Ploeg:2017vai,Gonthier2018} and we have only resolved of order 100 in gamma rays. It follows that the probability of observing an individual MSP must be a very small. Therefore, from the law of rare events (see for example Section 1.1.1 of  \cite{Cameron1998}), 
the total number of resolved and unresolved MSPs ($N_{\rm tot}$) is well approximated by:
\begin{equation}
\label{eq:pobs1}
    N_{\rm tot} = \lambda_{\rm res} / p(\obs|\thetab)\, .
\end{equation}
We can find $p({\rm obs}|\thetab)$ by noting that a luminosity threshold distribution $p(L_{\rm th})$ is determined by the combination of the flux threshold and spatial distributions. Thus:
\begin{equation}
\label{eq:pobs}
    p({\rm obs}|\thetab) = \int \dd L_{\rm th} p(L_{\rm th}) p(L \geq L_{\rm th})
\end{equation}
\noindent where $p(L \geq L_{\rm th})$ is the probability that the luminosity is greater than or equal to the threshold.
Another useful relation that follows from standard probability theory is: 
\begin{equation}
\label{eq:PobsD}
    p( {\rm obs}, \D{i}\vert \thetab) = p(\D{i}\vert {\rm obs}, \thetab)p({\rm obs}|\thetab)\,.
\end{equation}
Combining the above equation with Eqs.~\ref{eq:pobs1} and \ref{eq:rhop} gives
 \begin{equation}
 \label{eq:rhoD}
   \rho(\D{i})=p({\rm obs}, \D{i}\vert \thetab) N_{\rm tot} \,.
\end{equation}

We have two distinct types of resolved MSPs: those with parallax measurements and those without. To accommodate this we have two separate probability density functions
\begin{equation}
p({\rm obs},\D{i}\vert \thetab)=  p({\rm obs},\D{i},{\rm parallax}_i\vert\thetab)+p({\rm obs},\D{i},\textrm{not parallax}_i|\thetab)
\end{equation}
The $\D{i}$ components are the same in both cases as we can estimate the distance for those MSPs that do not have parallax measurements by  their dispersion measures. 
The probability of a parallax measurement given distance $d$ is modelled as:
\begin{equation}
\label{eq:parallax_model}
    p(\textrm{parallax}_i \vert d_i,\thetab) = \min(1, C_{\rm parallax} \exp(-d_i / d_{\rm parallax})) \,.
\end{equation}
and $ p(\textrm{not parallax}_i \vert d_i,\thetab)=1- p(\textrm{parallax}_i \vert d_i,\thetab)$.
See Appendix~\ref{app:probability} for more details.

To take into account measurement uncertainty in our values of $\D{i}$ we marginalise over the true values $\hat{\pmb{D}}_i$
\begin{equation}
\label{eq:likelihood_integral_over_true_vals}
  p(\obs, \D{i}\vert \theta) = \int  p(\D{i}|\hat{\pmb{D}}_i)  p({\rm obs}, \hat{\pmb{D}}_i\vert\thetab)\, \d\hat{\pmb{D}}_i
\end{equation}
If MSP $i$ has the $j$th component of its data $\D{i}$ missing then we account for that by making $p(\D{i}|\hat{\pmb{D}}_i)$ uniform in the $j$th component.
See  Appendix~\ref{appendix:uncertainties} for more details.

We fit the boxy bulge and nuclear bulge GCE to spectra found by Macias et al.~\cite{Macias19}. We use a Gaussian likelihood for each bin in the GCE spectra:
\begin{equation}
\label{eq:likelihood_GCE}
    \mathcal{L}_{\rm GCE} \propto \prod_{i=1}^N \exp(-\left(\left(\frac{\dd N}{\dd E}\right)_{\textrm{sim,}i} - \left(\frac{\dd N}{\dd E}\right)_{\textrm{data,}i} \right)^2 / \left(2 \sigma^2_{\textrm{data,}i}\right))
\end{equation}
\noindent where $\left(\frac{\dd N}{\dd E}\right)_{\textrm{sim,}i}$ is the simulated GCE for bin $i$ and $\left(\frac{\dd N}{\dd E}\right)_{\textrm{data,}i}$ is the data with uncertainty $\sigma^2_{\textrm{data,}i}$. We fit only energy bins lower than $10$ GeV as the higher energy bins may contain significant secondary emission \cite{Macias19}.  The combined likelihood for our resolved MSPs and the GCE is then given by substituting Eqs.~\ref{eq:likelihood_GCE} and \ref{eq:likelihood_definition} into
\begin{equation}
    \mathcal{L}_{\rm total}=\mathcal{L}_{\rm res}\times \mathcal{L}_{\rm GCE} \,.
    \label{eq:Ltotal}
\end{equation}
This is then multiplied with the priors given in Table~\ref{tab:prior_ranges} to get the posterior which is then sampled using MCMC. To test our models we generate posterior predictive distributions \cite{Gelman2013}. See Appendix~\ref{appendix:sampling} for more details on the sampling methods we used.

\section{Results}
\label{sec:results}

In this section we present results for various 
assumed
luminosity functions, age distributions, and relationships between magnetic field strength and initial period. We rank these various models using the Watanabe Akaike Information Criterion (WAIC) \cite{Wantabe2010,Gelman2013} as described in Appendix \ref{appendix:WAIC}
and given in Eq.~\ref{eq:WAIC}. The WAIC provides a measure of the expected predictive accuracy of a model and it takes into account the number of model parameters and their posterior uncertainty. 
Under suitable regularity conditions, in the limit of a large amount of data, the difference in the WAIC between two models ($\Delta$WAIC) tends towards minus two times the log of their likelihood ratio \cite{Gelman2013}. So when adding an extra parameter to a model, the ``number of sigma'' in favour of adding  that parameter is given approximately by $\sqrt{\Delta{\rm WAIC}}$   \cite{Wilks1938}. This provides a rough benchmark in evaluating the significance of $\Delta$WAIC values.

We can gain a more detailed view of a model's fit
by comparing its posterior predictive distributions to data using single dimensional binned plots showing medians and $68\%$ and $95\%$ intervals, as well as corner plots constructed using the software by Foreman-Mackey~\cite{corner}. These posterior predictive distributions provide an effective method of evaluating the goodness of fit \citep{Gelman2013,Gelman2013a}. 
In particular, major failures of the model correspond to extreme posterior predictive p-values. These are defined as the proportion of posterior simulations which are more extreme than the data or some statistic of the data \citep{Gelman2013}.
Our corner plots show two dimensional distributions of real data and simulated data with $68\%$, $95\%$ and $99.7\%$ contours. In producing simulated data we model both missing data and uncertainties by picking a random real MSP and removing simulated data that is not available for the real MSP. For data with uncertainties attached, we take the relative error for the real MSP and 
add Gaussian noise to the simulated MSP which has a standard deviation with the same relative error as the real MSP.
We have a distance dependent model of the probability of a parallax measurement being available. If a simulated MSP has a parallax measurement, with probability given by Eq.\ \ref{eq:parallax_model}, we select a random real MSP out of those with available parallax measurements and use its relative uncertainty to simulate a parallax error.

The WAIC allows us to compare the various models used while accounting for the varying number of parameters they involve. In Table \ref{tab:WAIC} we show the WAIC averaged over the eight chains for each of a set of models of the Galactic MSP population relative to that of the best model. We also show the sample standard deviation over the WAIC for the eight MCMC chains run for each model. 
This variation occurs because we used Monte Carlo integration to compute the integrals in Section~\ref{sec:method}. The random numbers used to compute these integrals (such as over the resolved MSP data uncertainty distributions) were generated once per MCMC chain so that a calculation of the likelihood for a given set of parameters will always return the same result. However, between chains the likelihood may shift slightly as the set of randomly generated numbers used will be different.
As a result of this variation in the likelihood, we did not generate a single WAIC for all eight chains combined. We found that the posterior distributions of the model parameters were generally indistinguishable despite these variations, and the variation in WAIC for each model is typically small compared to $\Delta \textrm{WAIC}$ between models.

In Table \ref{tab:mcmc_results_parameters} we report medians and $68\%$ confidence intervals for the parameters of a subset of these models. The first three parameters in this table
are related to the number of MSPs in each of the three spatial components of the model: The parameter $\lambda_{\rm res}$ is the expected number of resolved MSPs; $\log_{10}(N_{\rm disk} / N_{\rm bulge})$ and $\log_{10}(N_{\rm nb} / N_{\rm bb})$  are parameters defining, respectively, the ratio of $N_{\rm disk}$ to $N_{\rm bulge}$ and $N_{\rm nb}$ to $N_{\rm bb}$. 
Here $N_{\rm disk}$ is the number of disk MSPs, $N_{\rm bulge}=N_{\rm nb}+N_{\rm bb}$ is the total number of  bulge MSPs, $N_{\rm nb}$ is the number of nuclear bulge MSPs, and $N_{\rm bb}$ is the number of boxy bulge MSPs. 

\begin{figure}
    \centering
    \includegraphics[width=0.99\linewidth]{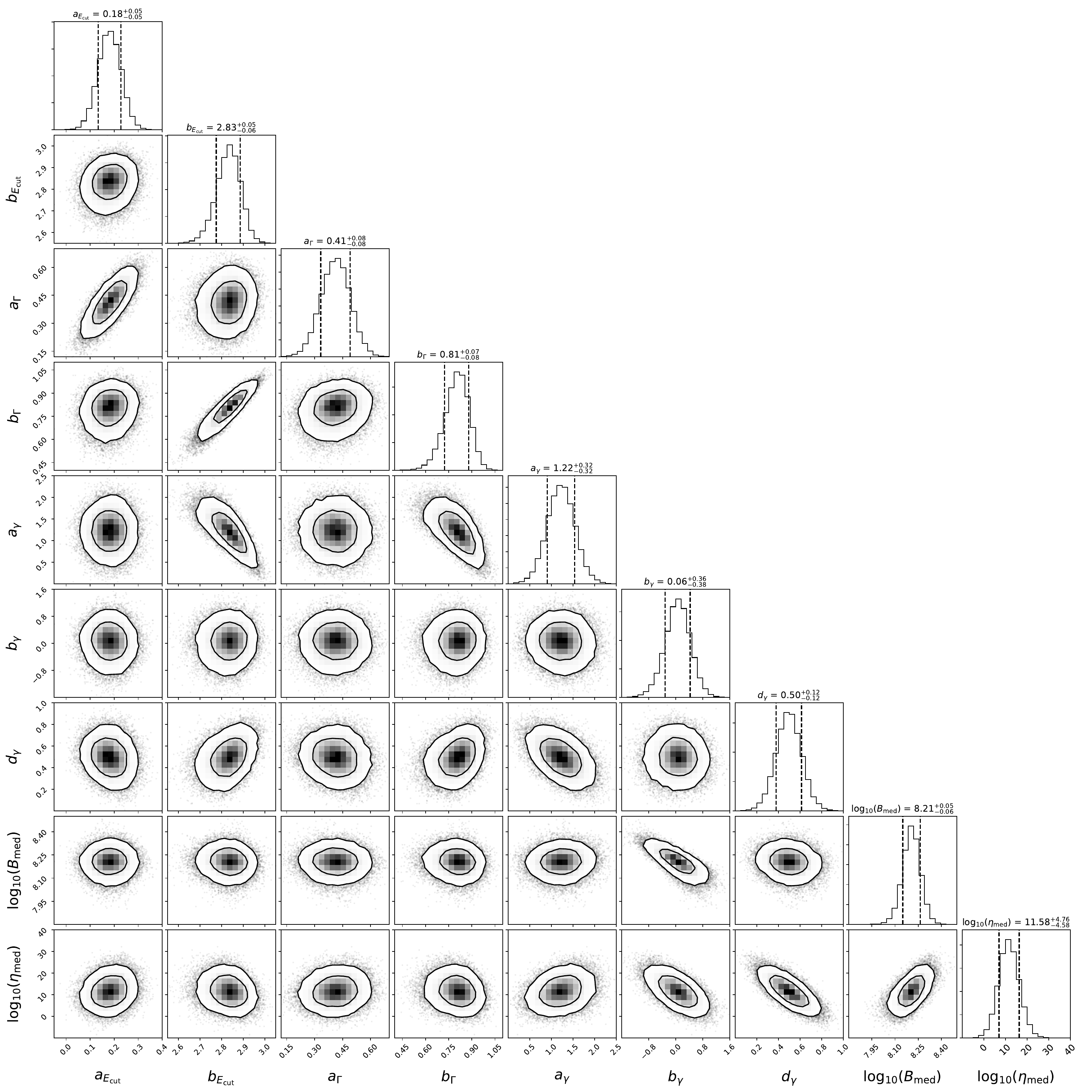}
    \caption{Corner plot showing a selection of parameters for {\textcolor{black}{Model A1}} ($L = \eta E_{\rm cut}^{a_{\gamma}} B^{b_{\gamma}} \dot{E}^{d_{\gamma}}$)  with $68\%$ and $95\%$ contours. These parameters relate to the luminosity, $E_{\rm cut}$, $\Gamma$ and $B$ distributions. The $B_{\rm med}$ parameter has units of Gauss.}
    \label{fig:E_cut_B_E_dot_MCMC_Params_1}
\end{figure}

\begin{figure}
    \centering
    \includegraphics[width=0.99\linewidth]{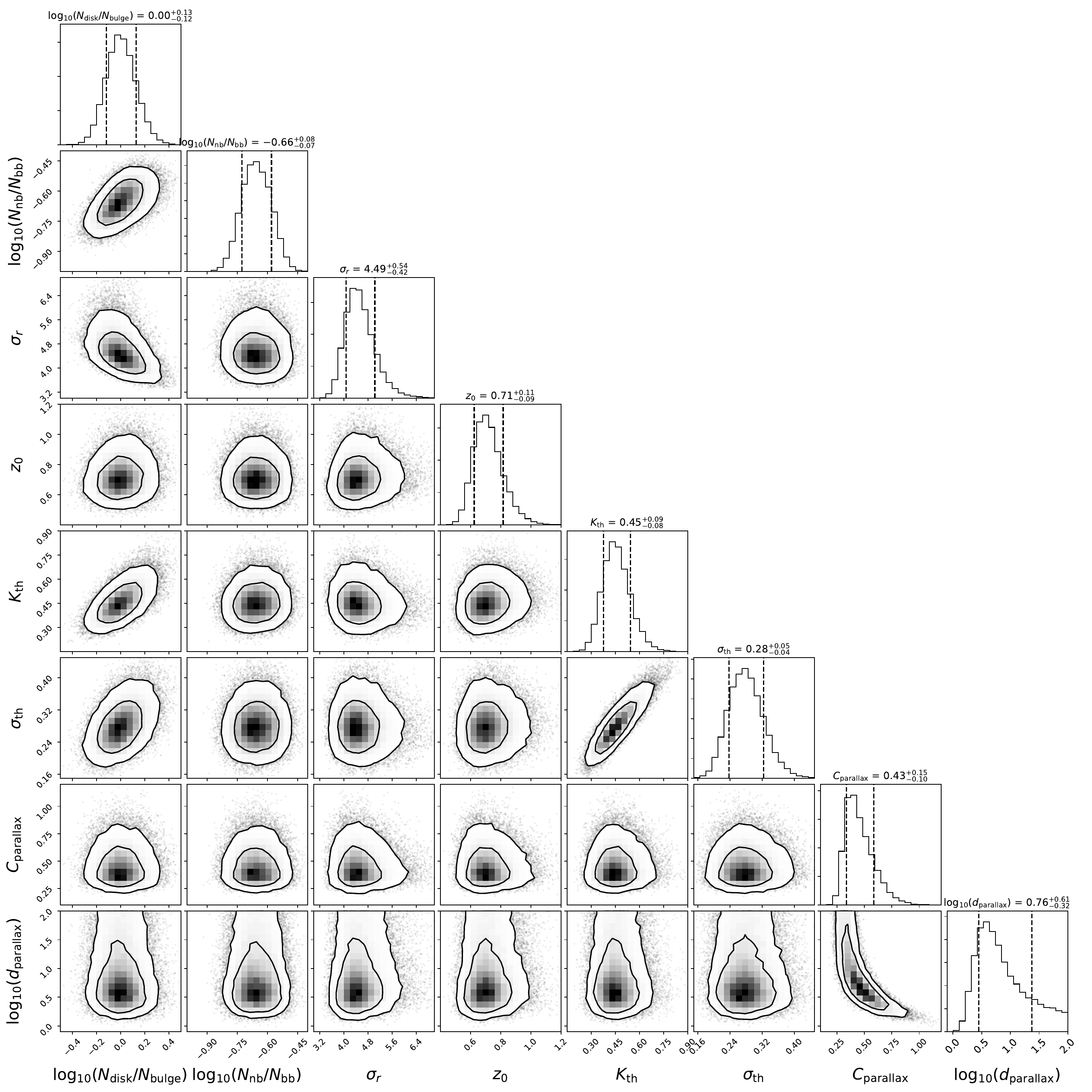}
    \caption{Corner plot showing a selection of parameters for Model A1 with $68\%$ and $95\%$ contours. These parameters relate to the number of MSPs in different components of the spatial distribution, the flux threshold, the model of the parallax measurement availability and the initial period distribution. 
    The parameters $\sigma_r$, $z_0$, and $d_{\rm parallax}$ have units of kpc.
    }
    \label{fig:E_cut_B_E_dot_MCMC_Params_2}
\end{figure}

\begin{figure}
    \centering
    \includegraphics[width=0.99\linewidth]{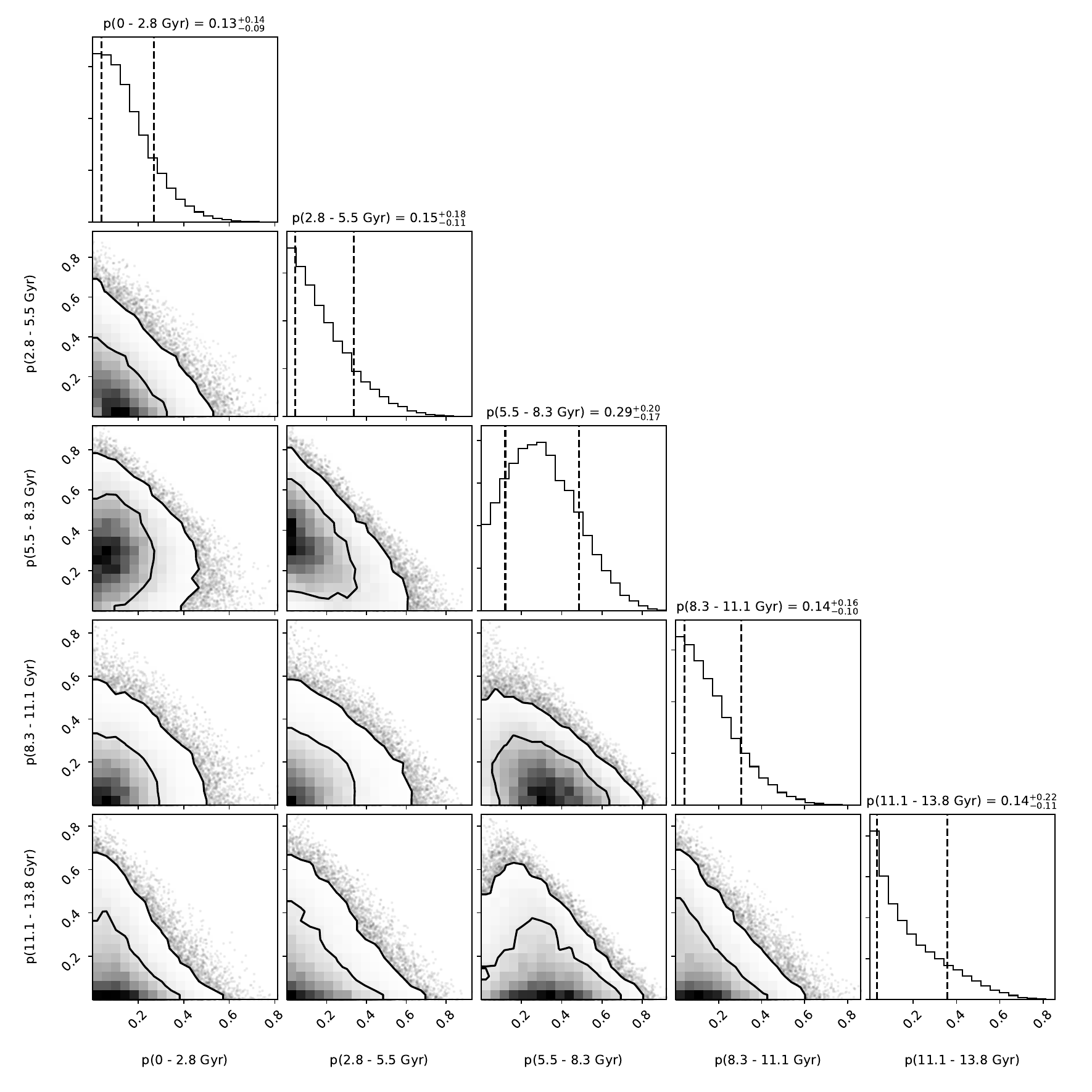}
    \caption{Model A1  probability of delay time within each DTD bin.}
    \label{fig:E_cut_B_E_dot_MCMC_DTD_bin_Params}
\end{figure}

\begin{figure}
\vspace{-2cm}
    \centering
    \subfigure{\centering\includegraphics[width=0.33\linewidth]{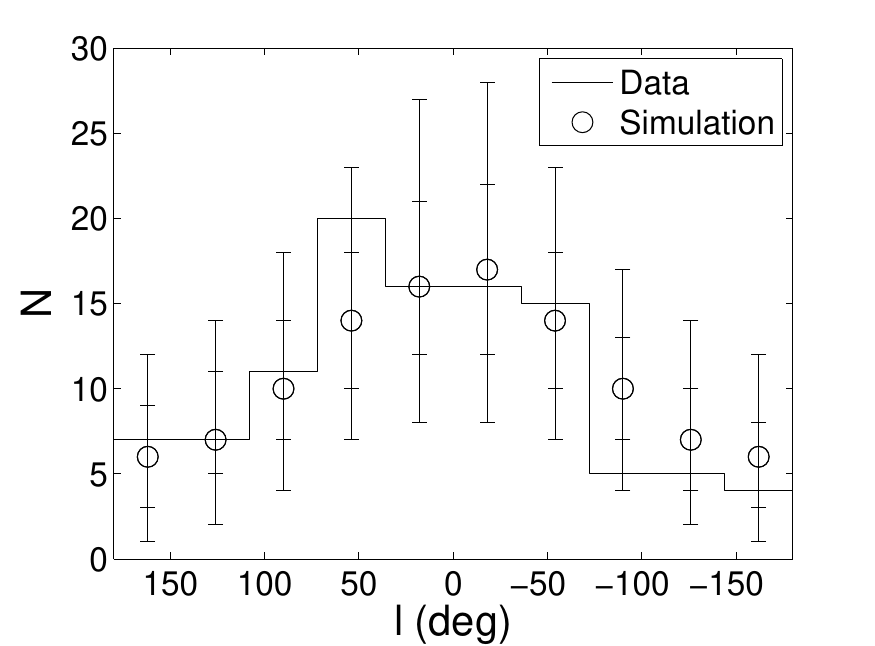}}
    \subfigure{\centering\includegraphics[width=0.33\linewidth]{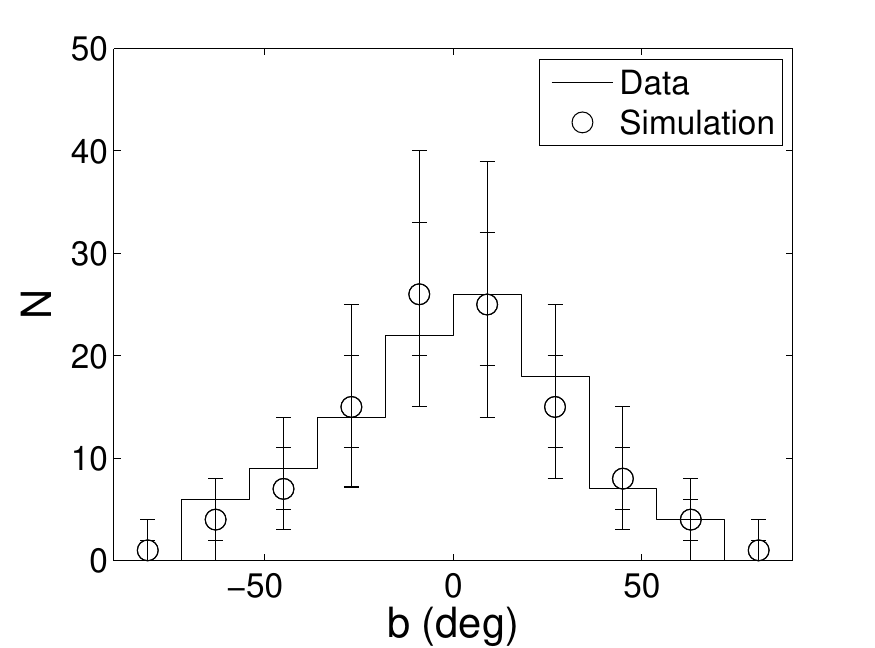}}
    \subfigure{\centering\includegraphics[width=0.33\linewidth]{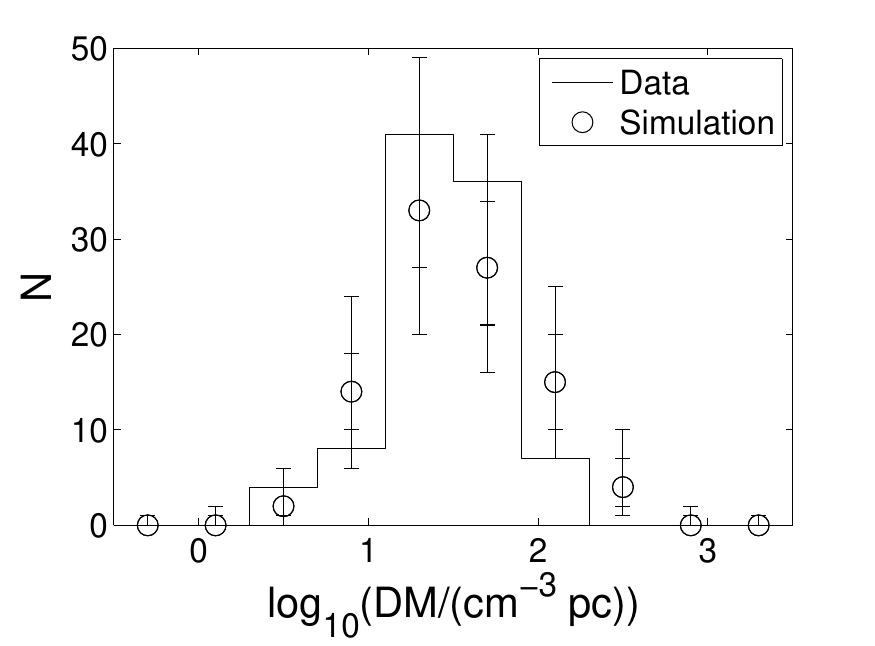}}
    \subfigure{\centering\includegraphics[width=0.33\linewidth]{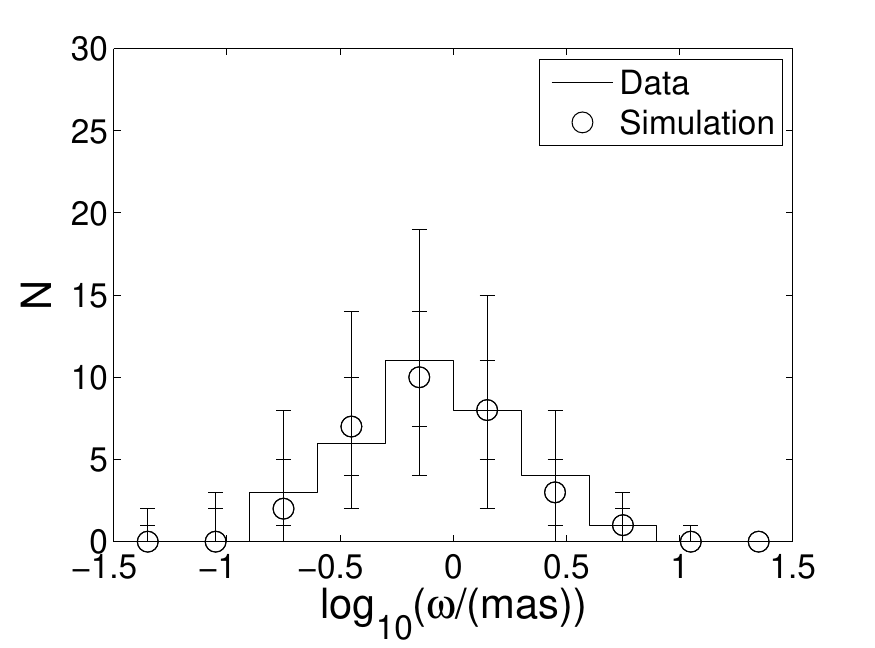}}
    \subfigure{\centering\includegraphics[width=0.33\linewidth]{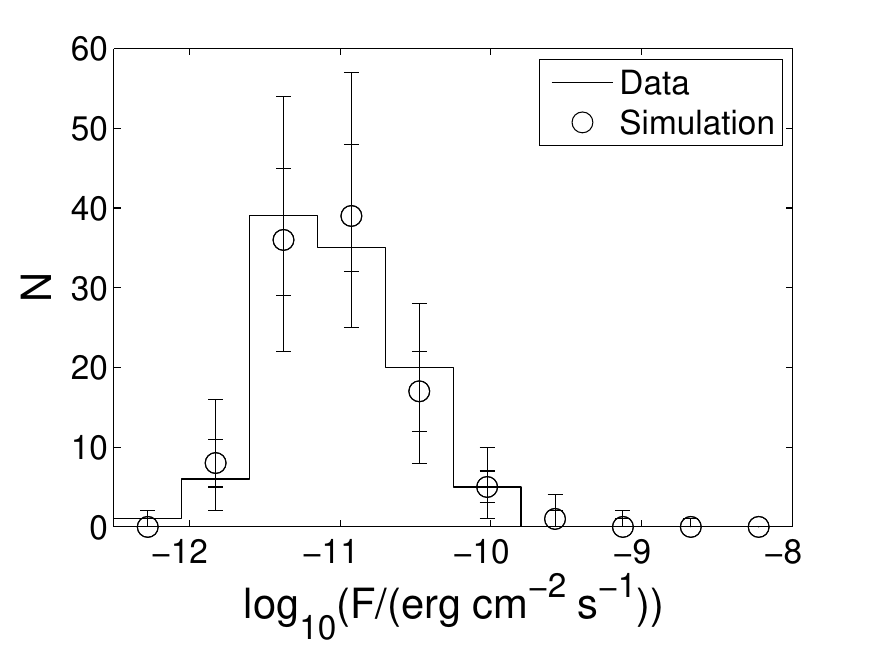}}
    \subfigure{\centering\includegraphics[width=0.33\linewidth]{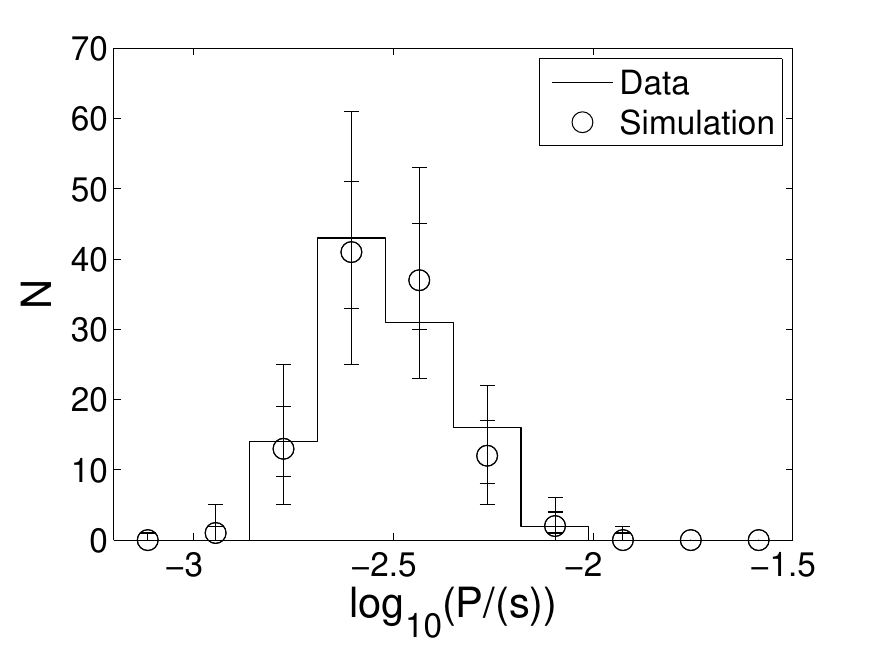}}
    \subfigure{\centering\includegraphics[width=0.33\linewidth]{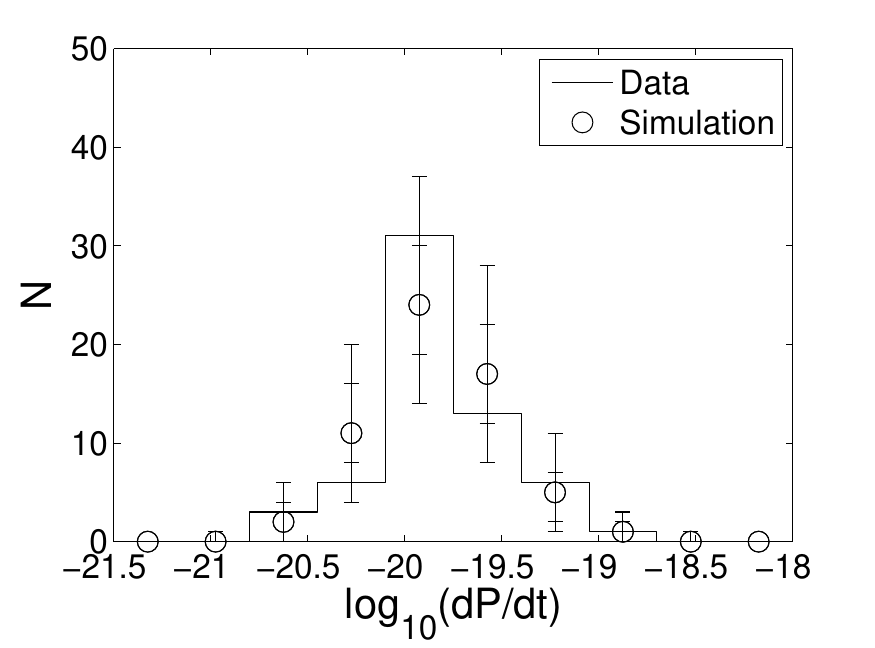}}
    \subfigure{\centering\includegraphics[width=0.33\linewidth]{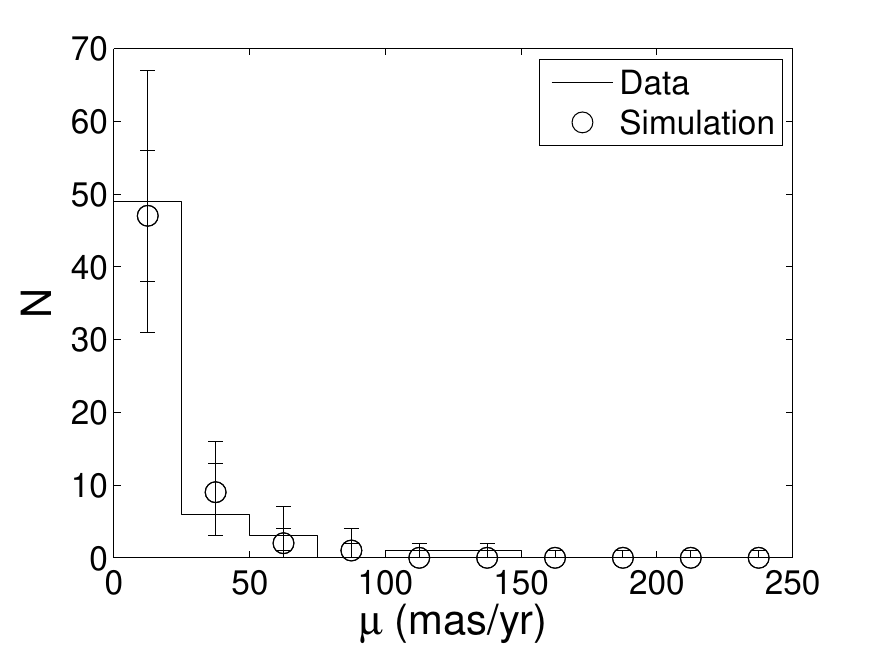}}
    \subfigure{\centering\includegraphics[width=0.33\linewidth]{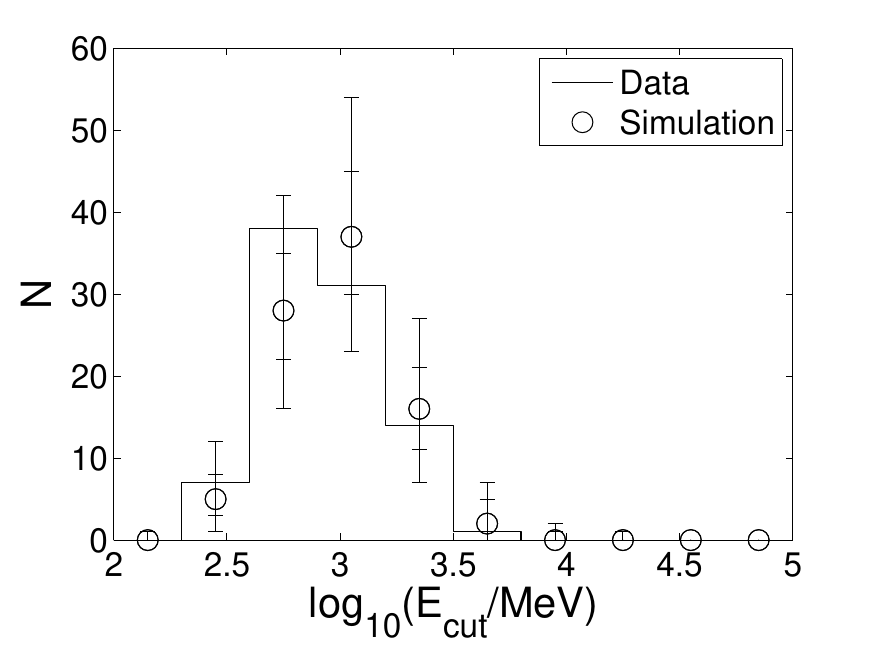}}
    \subfigure{\centering\includegraphics[width=0.33\linewidth]{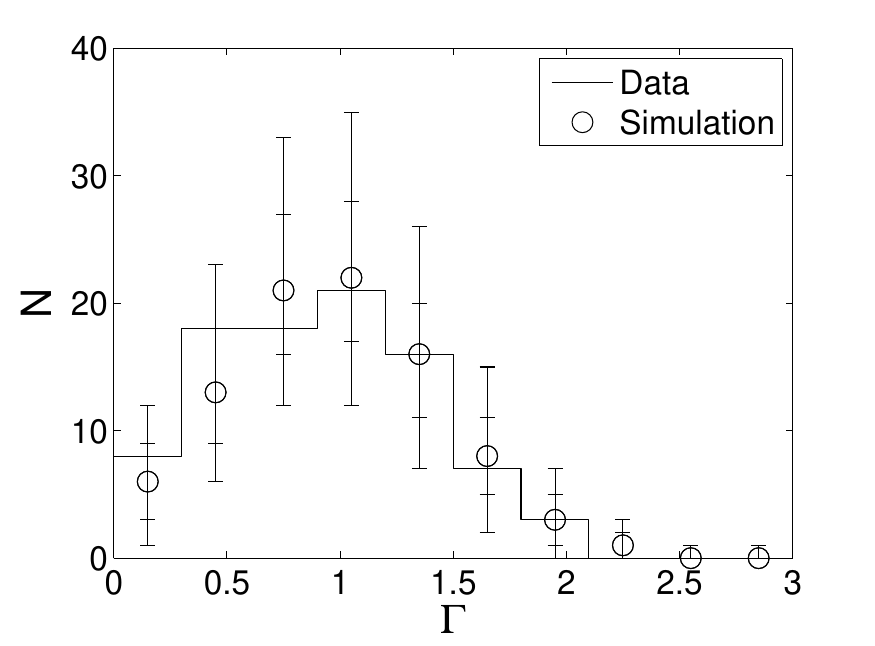}}
    \subfigure{\centering\includegraphics[width=0.33\linewidth]{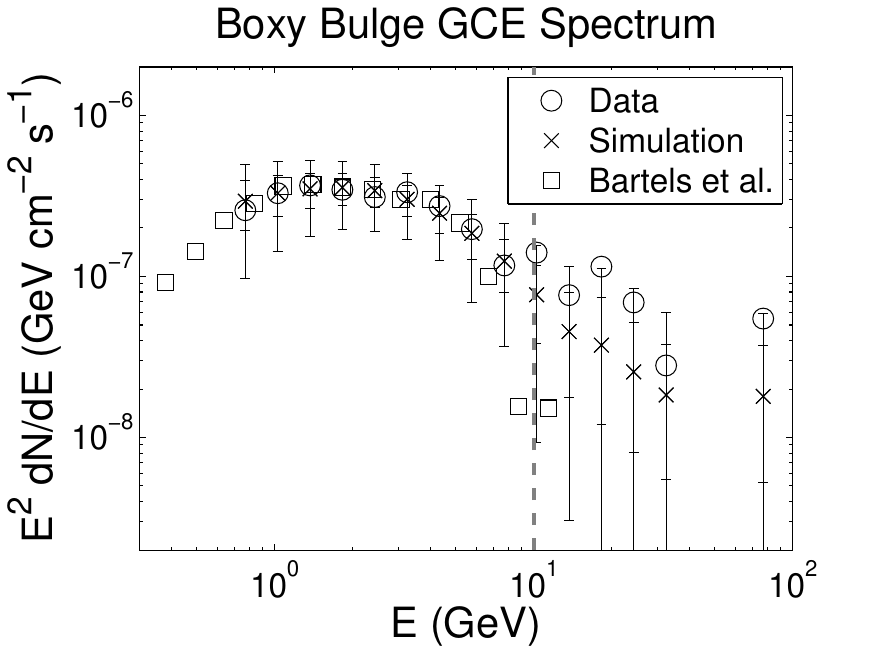}}
    \subfigure{\centering\includegraphics[width=0.33\linewidth]{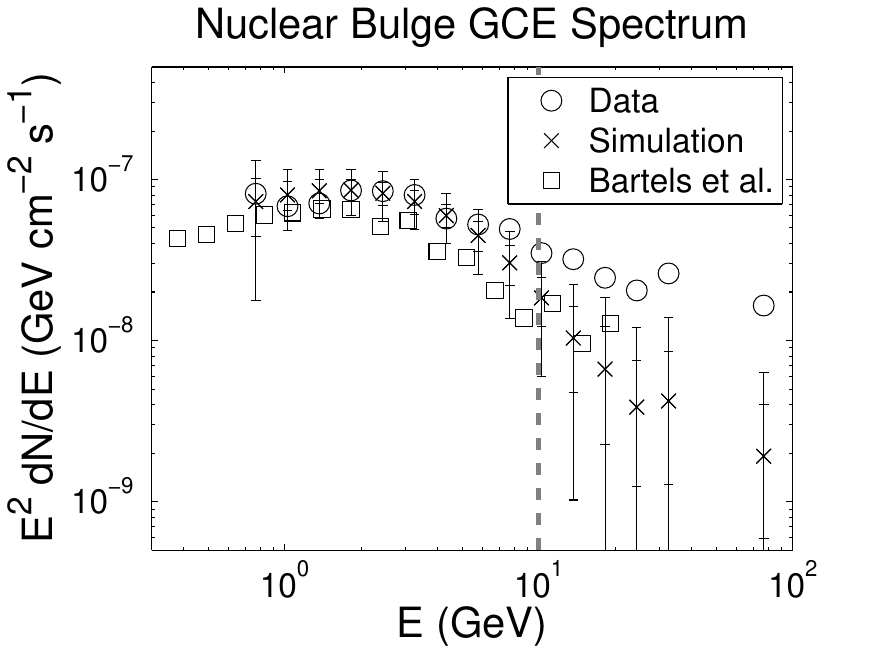}}
    \caption{Observed data compared to simulated observations for Model A1 of longitude $l$ (deg), latitude $b$ (deg), dispersion measure DM (cm$^{-3}$ pc), parallax $\omega$ (mas), flux $F$ (erg cm$^{-2}$ s$^{-1}$), period $P$ (s), period derivative $\dot{P}$, proper motion $\mu$ (mas yr$^{-1}$), spectral cutoff $E_{\rm cut}$ (MeV), spectral index $\Gamma$ and GCE spectra for the boxy and nuclear bulges. The GCE spectra are for the inner $40^{\circ} \times 40^{\circ}$ region and the vertical dashed line shows the energy up to which we fitted of $10$ GeV. The simulated data is shown as medians, $68\%$ and $95\%$ intervals in each bin. The intervals on the simulated GCE data include the errors on the observed data. For the GCE plots, we also show boxy bulge and nuclear bulge spectra from Bartels et al.~\cite{Bartels2017} in addition to those we fitted to of Macias et al.~\cite{Macias19}. }
    \label{fig:E_cut_B_E_dot_posterior_predictive_plots}
\end{figure}

\begin{figure}
    \centering
    \subfigure{\centering\includegraphics[width=0.49\linewidth]{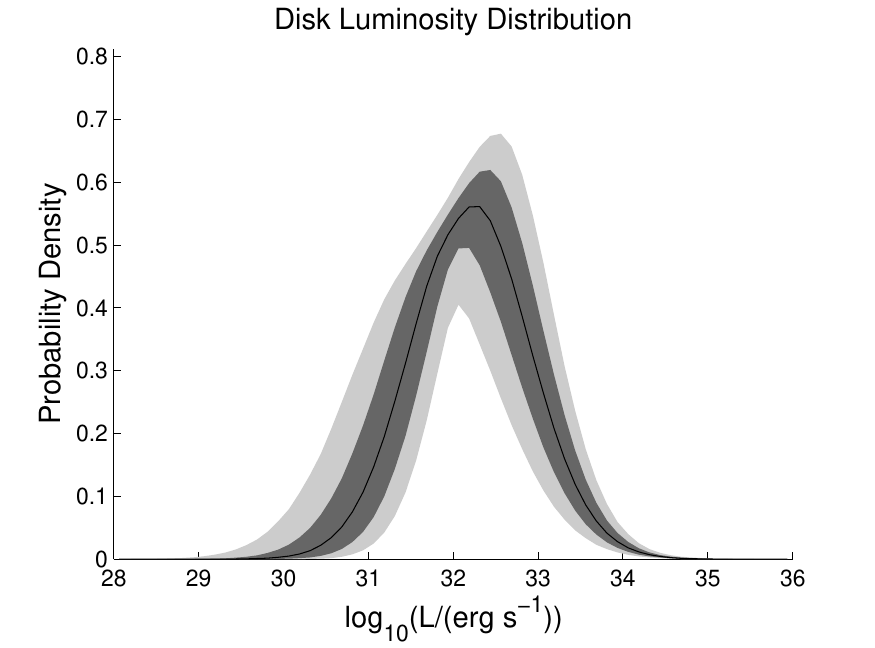}}
    \subfigure{\centering\includegraphics[width=0.49\linewidth]{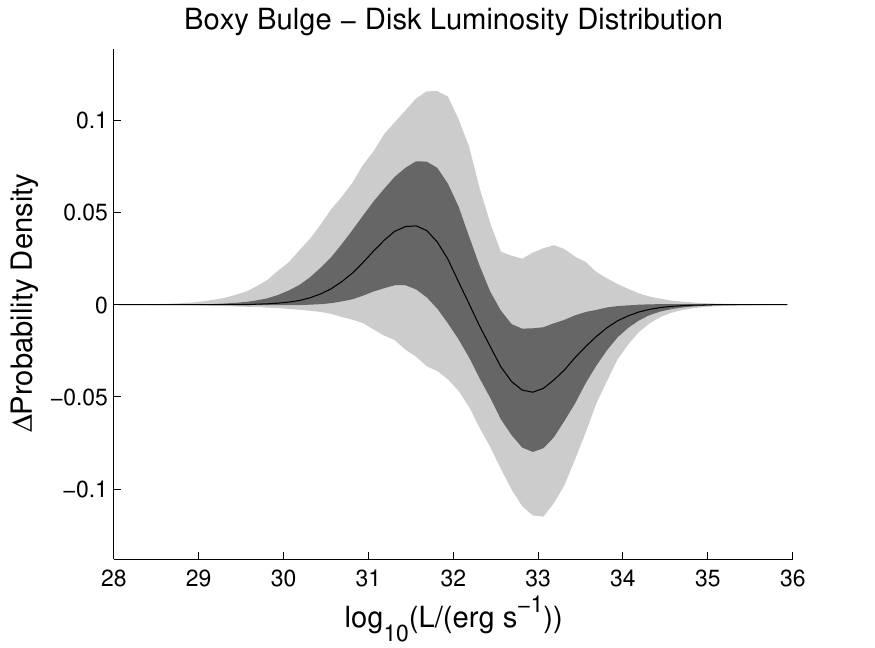}}
    \subfigure{\centering\includegraphics[width=0.49\linewidth]{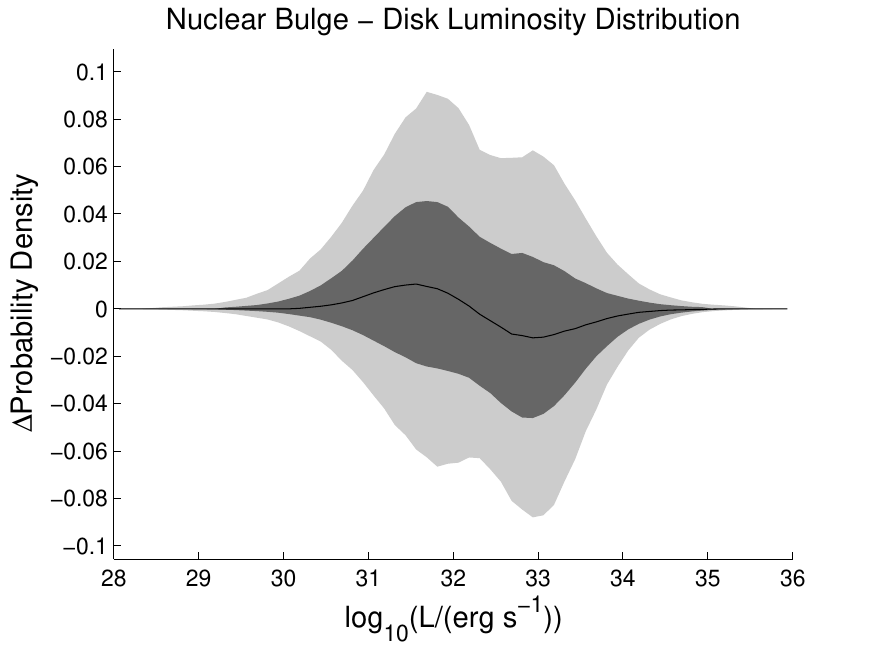}}
    \caption{MSP luminosity distribution in the disk and the change in probability density for the boxy bulge and nuclear bulge for Model A1. The black line shows the median probability density in each bin, dark grey the $68\%$ interval and light grey the $95\%$ interval. Note that the probability density is of $\log_{10}(L/({\rm erg\, s^{-1}}))$ rather than of $L/({\rm erg\, s^{-1}})$.
    }
    \label{fig:E_cut_B_E_dot_luminosity_distribution}
\end{figure}

\begin{figure}
    \centering
    \subfigure{\centering\includegraphics[width=0.49\linewidth]{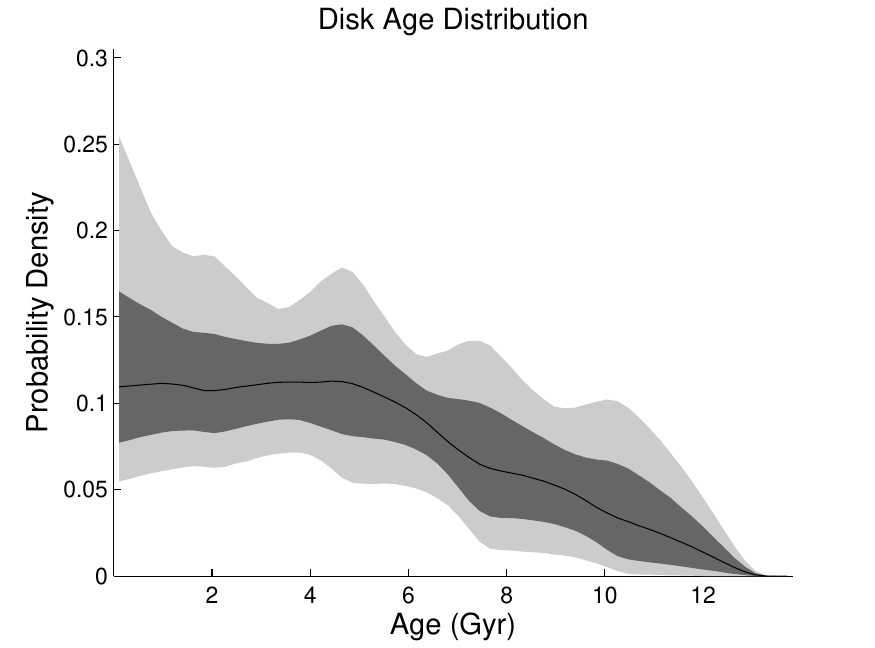}}
    \subfigure{\centering\includegraphics[width=0.49\linewidth]{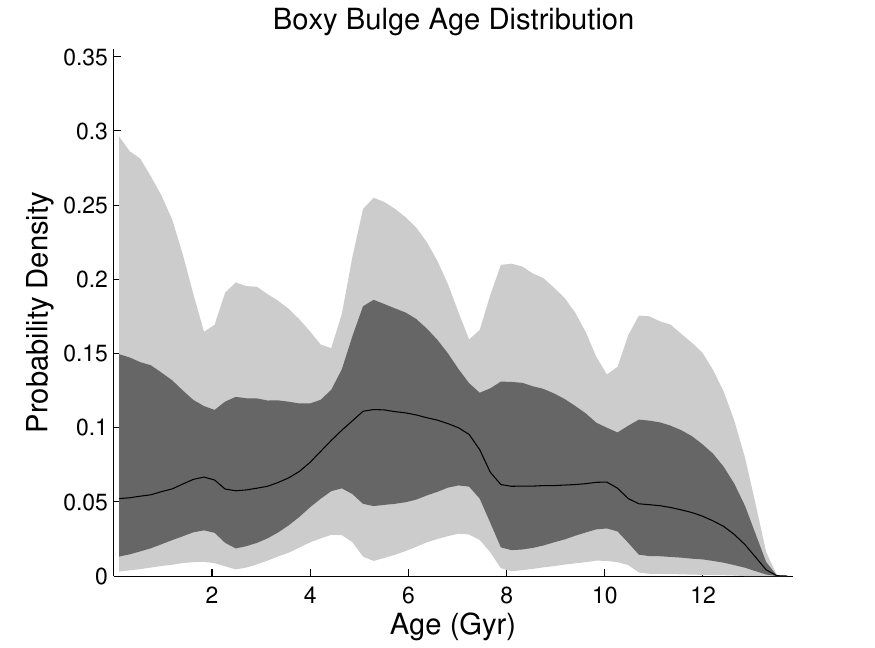}}
    \subfigure{\centering\includegraphics[width=0.49\linewidth]{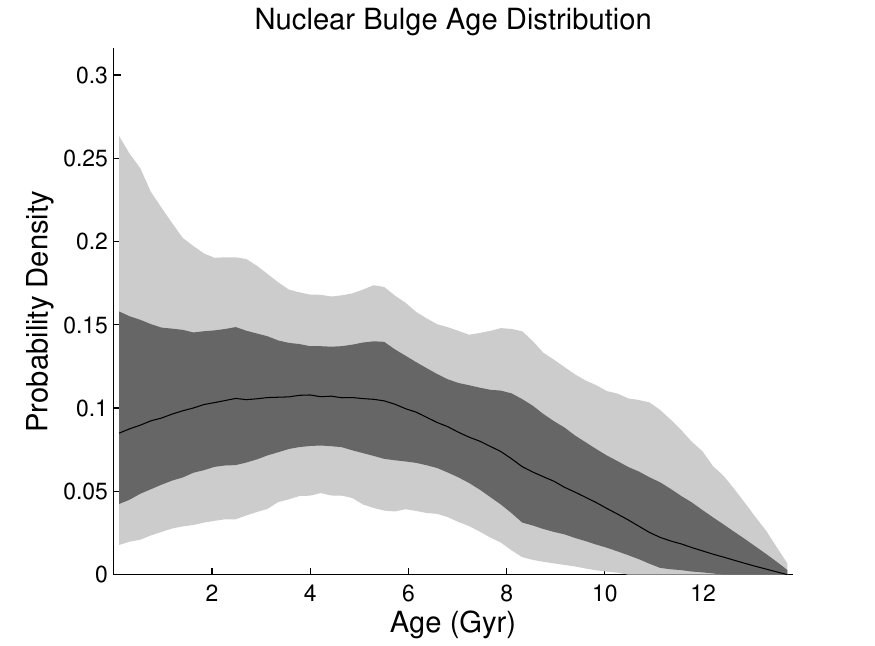}}
    \caption{MSP age distribution in the disk, boxy bulge and nuclear bulge for Model A1. The black line shows the median probability density in each bin, dark grey the $68\%$ interval and light grey the $95\%$ interval.}
    \label{fig:E_cut_B_E_dot_age_distribution}
\end{figure}

\begin{figure}
    \centering
    \includegraphics[width=0.75\linewidth]{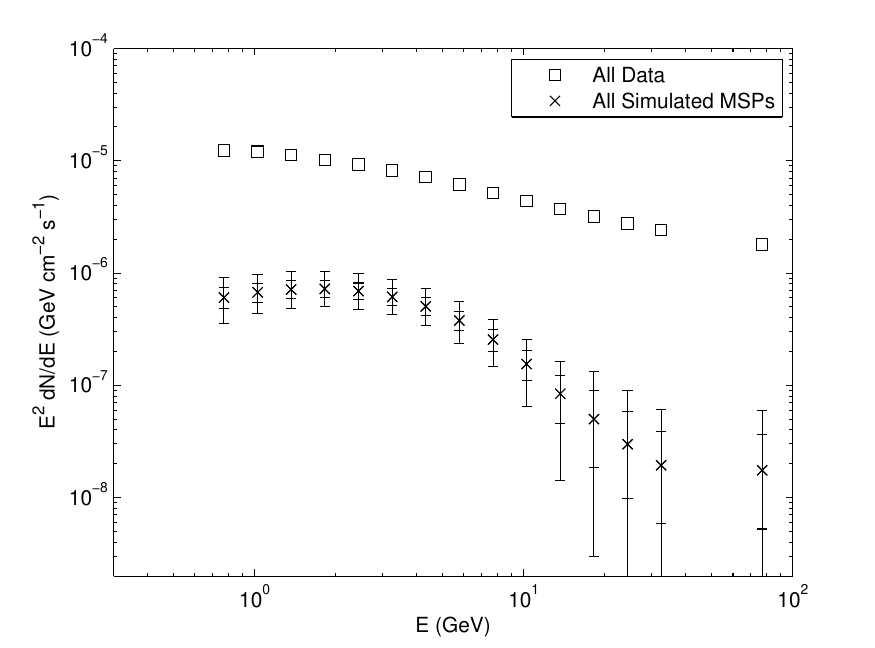}
    \caption{ Total observed gamma ray emission compared to all simulated MSP emission from inner $40^{\circ} \times 40^{\circ}$ region for Model A1. Data is from Abazajian et al.~\cite{Abazajian2020}. }
    \label{fig:E_cut_B_E_dot_all_msp_emission}
\end{figure}

\begin{figure}
    \centering
    \subfigure{\centering\includegraphics[width=0.49\linewidth]{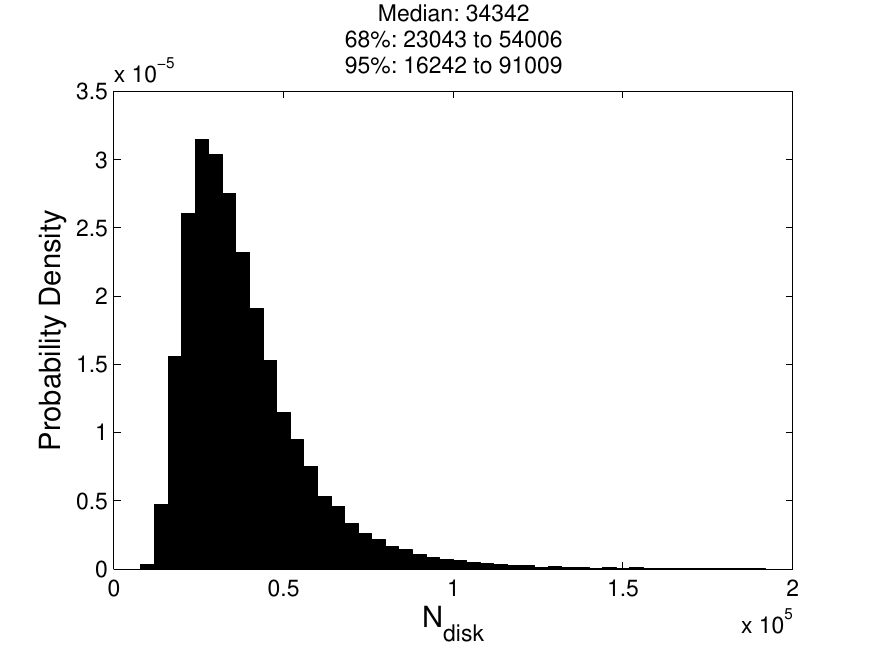}}
    \subfigure{\centering\includegraphics[width=0.49\linewidth]{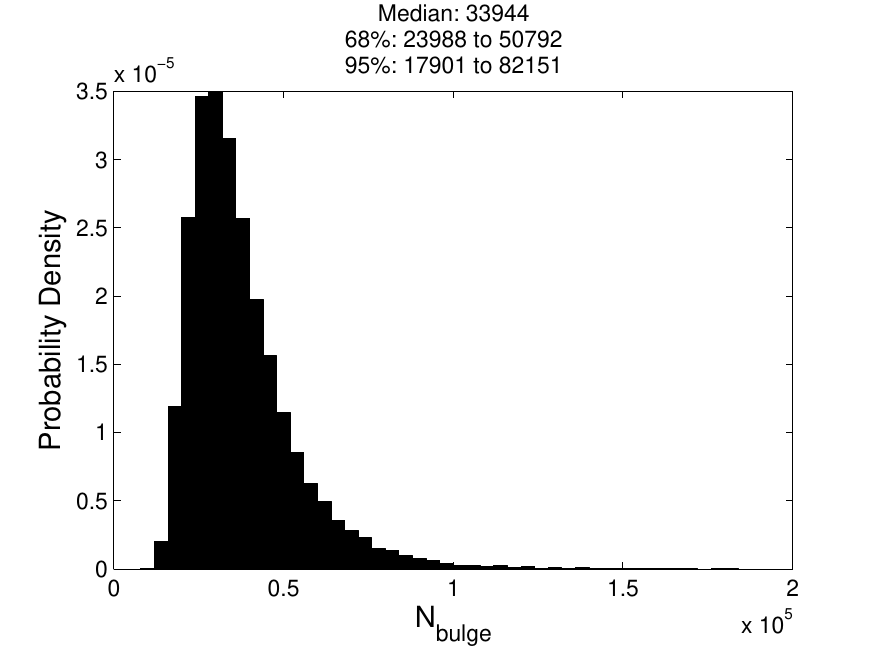}}
    \caption{Distribution of the number of disk and bulge MSPs for Model A1.}
    \label{fig:E_cut_B_E_dot_N_MSPs}
\end{figure}

\begin{figure}
    \centering
    \subfigure{\centering\includegraphics[width=0.49\linewidth]{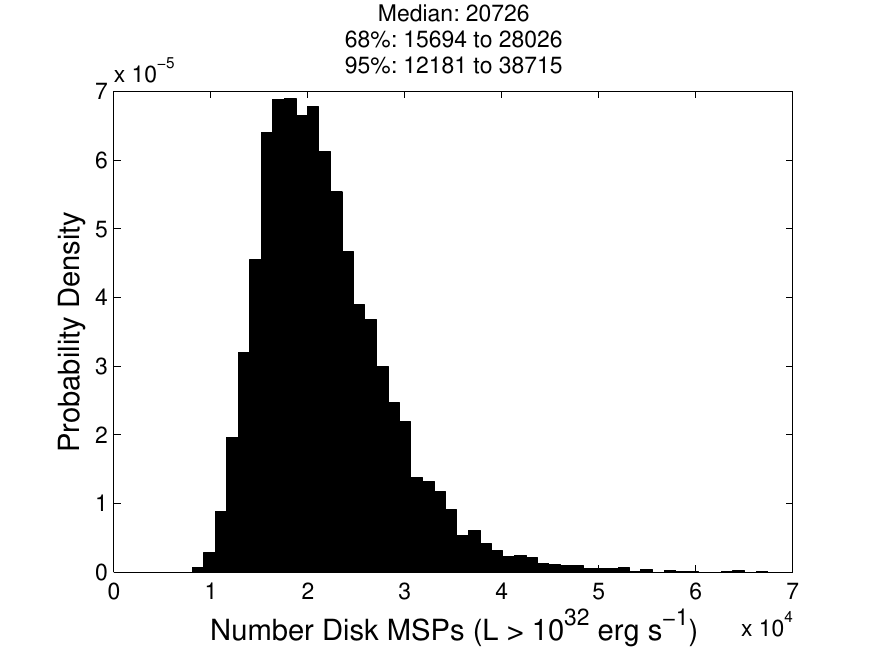}}
    \subfigure{\centering\includegraphics[width=0.49\linewidth]{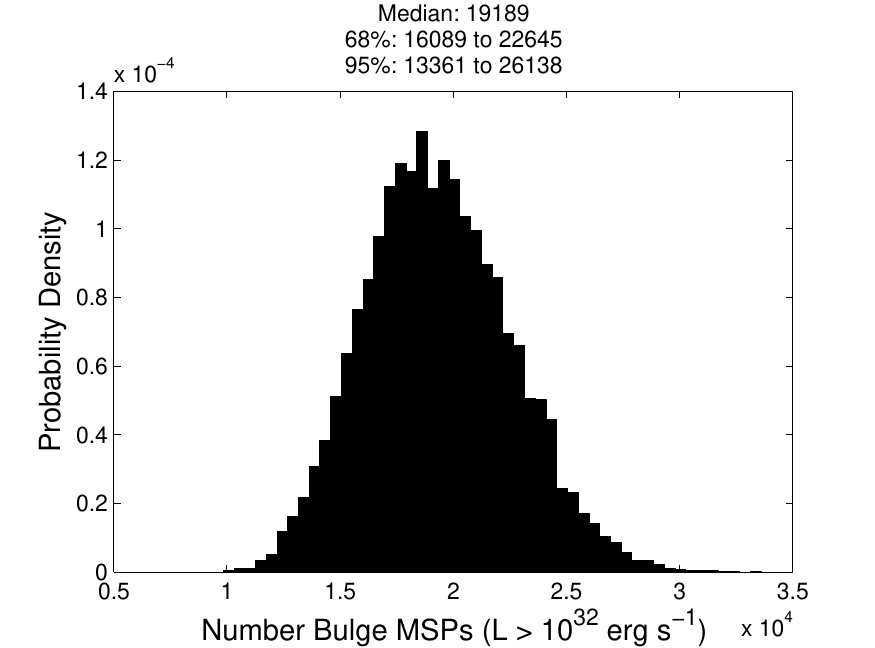}}
    \caption{Distribution of the number of disk and bulge MSPs with $L > 10^{32} \textrm{ erg s}^{-1}$ for 
    Model A1.}
    \label{fig:E_cut_B_E_dot_N_MSPs_greater_than_log_L_32}
\end{figure}

\begin{figure}
    \centering
    \subfigure{\centering\includegraphics[width=0.49\linewidth]{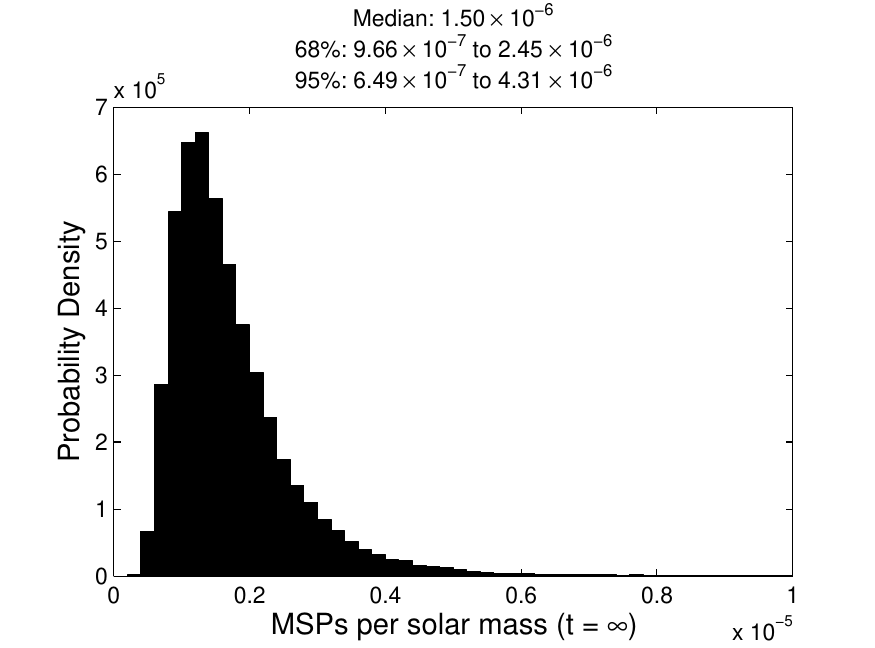}}
    \subfigure{\centering\includegraphics[width=0.49\linewidth]{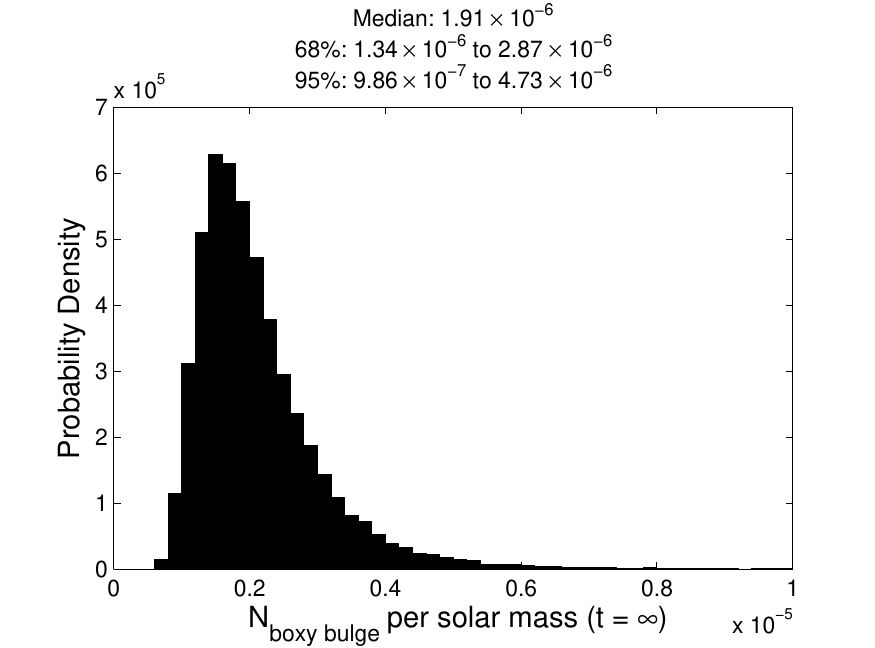}}
    \subfigure{\centering\includegraphics[width=0.49\linewidth]{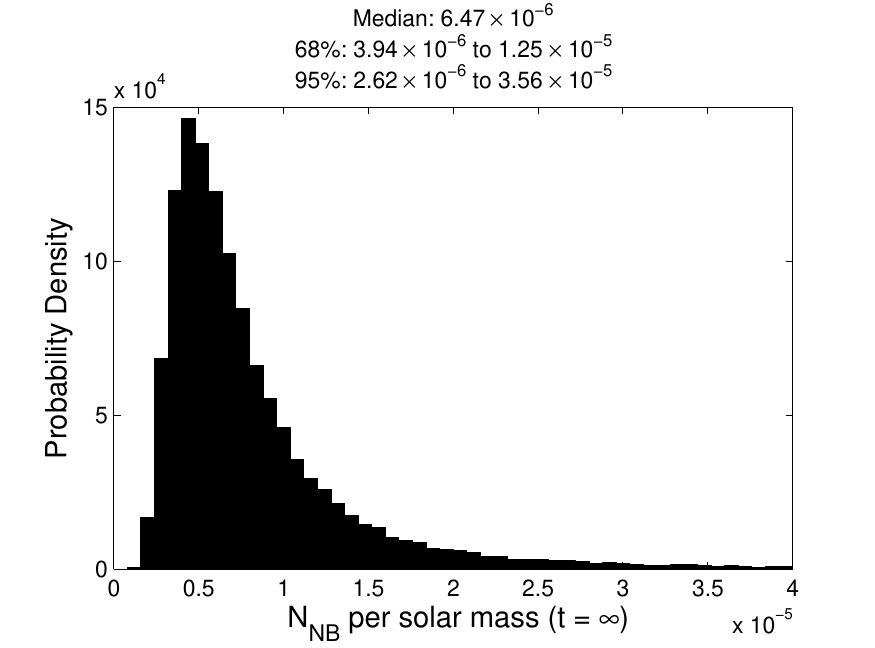}}
    \subfigure{\centering\includegraphics[width=0.49\linewidth]{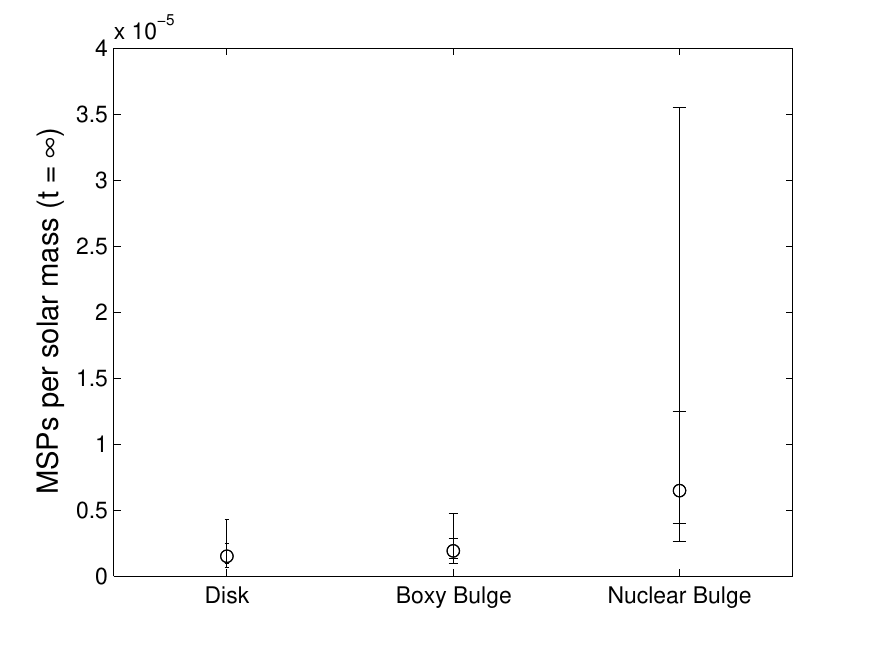}}
    \caption{Distribution of the number of disk, boxy bulge and nuclear bulge MSPs produced  per solar mass at $t = \infty$ assuming no further star formation beyond today for Model A1. In the bottom right plot, the medians, $68\%$ and $95\%$ intervals are shown side by side. }
    \label{fig:E_cut_B_E_dot_N_MSPs_per_solar_mass}
\end{figure}

\begin{figure}
    \centering
    \subfigure{\centering\includegraphics[width=0.49\linewidth]{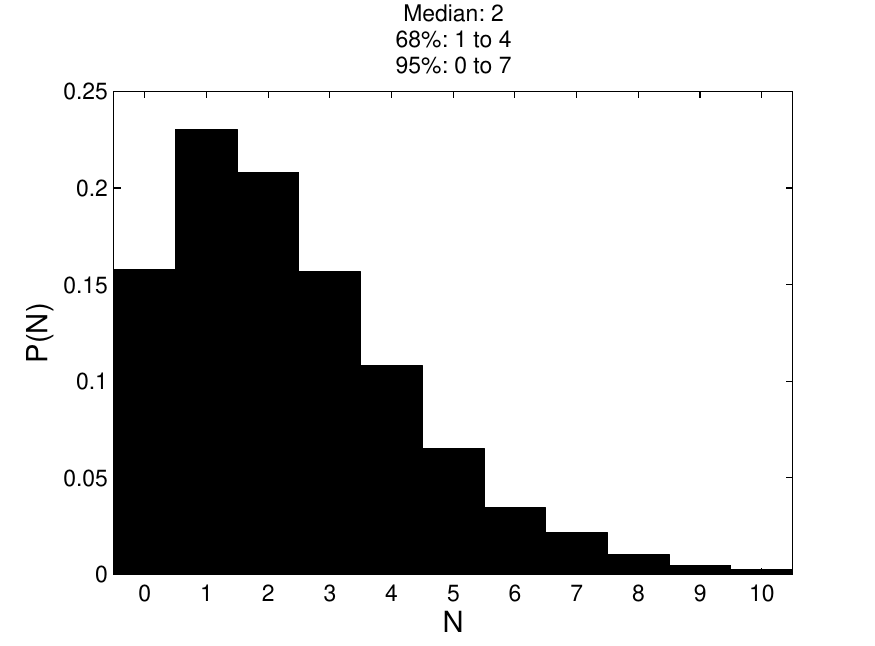}}
    \subfigure{\centering\includegraphics[width=0.49\linewidth]{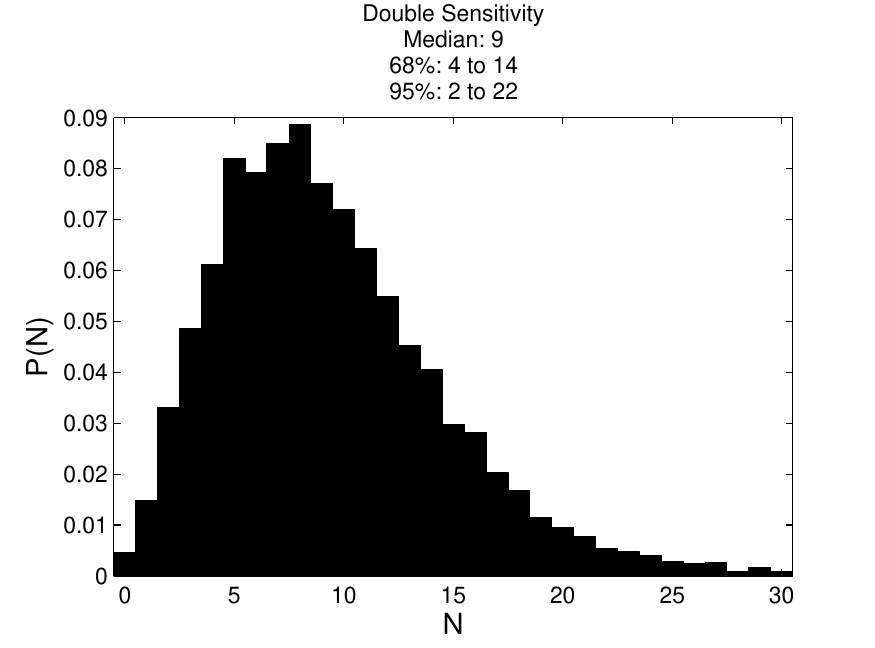}}
    \subfigure{\centering\includegraphics[width=0.49\linewidth]{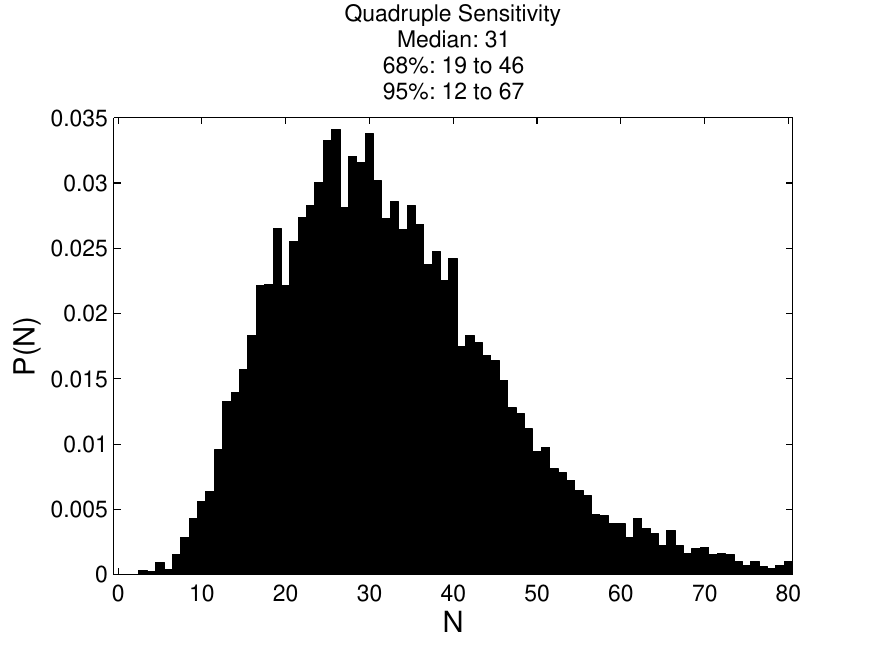}}
    \caption{Probability of observing $N$ bulge MSPs with the current sensitivity as well as for two and four times the current sensitivity for Model A1. The double and quadruple sensitivity distributions were produced by subtracting $0.3$ and $0.6$ respectively from $K_{\rm th}$.}
    \label{fig:E_cut_B_E_dot_N_observed_bulge_MSPs}
\end{figure}

\begin{figure}
    \centering
    \includegraphics[width=0.75\linewidth]{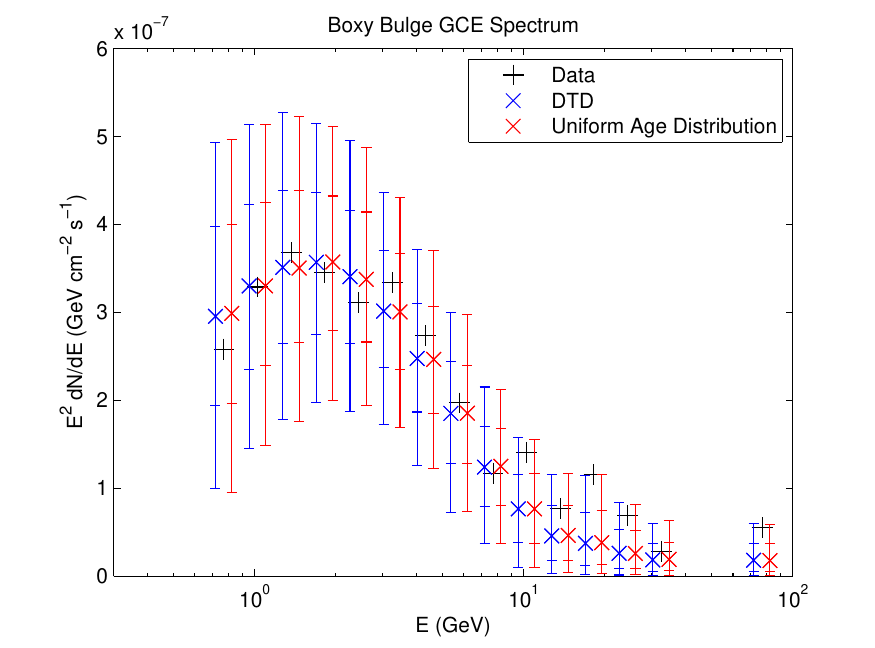}
    \caption{ Comparison of posterior predictive boxy bulge GCE spectra for the $L = \eta E_{\rm cut}^{a_{\gamma}} B^{b_{\gamma}} \dot{E}^{d_{\gamma}}$ model in the case where we have a DTD 
    (Model A1) and the case where a uniform MSP age distribution was used (Model A2). The DTD case has been shifted slightly to the left and the uniform case to the right in order to make comparison easier. Only bins with $E<10$~GeV were fitted. }
    \label{fig:E_cut_B_E_dot_dtd_vs_uniform_boxy_bulge_spectra}
\end{figure}

In Figs.~\ref{fig:E_cut_B_E_dot_MCMC_Params_1}, \ref{fig:E_cut_B_E_dot_MCMC_Params_2}, and \ref{fig:E_cut_B_E_dot_MCMC_DTD_bin_Params}, we display corner plots showing some of these parameters for the best model, which was A1 which had  $L = \eta E_{\rm cut}^{a_{\gamma}} B^{b_{\gamma}} \dot{E}^{d_{\gamma}}$.
In Fig.~\ref{fig:E_cut_B_E_dot_posterior_predictive_plots} the resolved MSP and GCE data is compared to simulated data. 
{\textcolor{black}{We exclude data that led to an apparently negative $\dot{P}_{\rm int}$ from the binned $\dot{P}$ data.}} This exclusion affects all bins with data except the highest two. The luminosity distribution is shown in Fig.~\ref{fig:E_cut_B_E_dot_luminosity_distribution}, and  the MSP age distributions are shown in Fig.~\ref{fig:E_cut_B_E_dot_age_distribution}. In Fig.~\ref{fig:E_cut_B_E_dot_all_msp_emission} we compare the total gamma ray emission from MSPs in the region of interest to the observed total. In Fig.~\ref{fig:E_cut_B_E_dot_N_MSPs} we show the number of MSPs in the disk and in the bulge.
The number with luminosity greater than $10^{32} \textrm{ erg s}^{-1}$ is shown in Fig.~\ref{fig:E_cut_B_E_dot_N_MSPs_greater_than_log_L_32}, and in Fig.~\ref{fig:E_cut_B_E_dot_N_MSPs_per_solar_mass} we show the number of MSPs produced per solar mass at $t = \infty$ assuming no further star formation {\textcolor{black}{after today}}. The masses used were $(3.7 \pm 0.5) \times 10^{10} \msun$ for the disk, $(1.6 \pm 0.2) \times 10^{10} \msun$ for the boxy bulge, and $(1.4 \pm 0.6) \times 10^{9} \msun$ for the nuclear bulge \citep{Bland-Hawthorn2016, Crocker:2016zzt}. In Fig.~\ref{fig:E_cut_B_E_dot_N_observed_bulge_MSPs} we show the modeled probability of resolving $N$ bulge MSPs, as well as the number for double and quadruple the current sensitivity. In Fig.~\ref{fig:E_cut_B_E_dot_dtd_vs_uniform_boxy_bulge_spectra} we compare the posterior predictive distributions for the boxy bulge spectra in the cases where we fitted the DTD and where the MSPs were uniformly distributed in age. %

\begin{figure}
    \centering
    \includegraphics[width=0.99\linewidth]{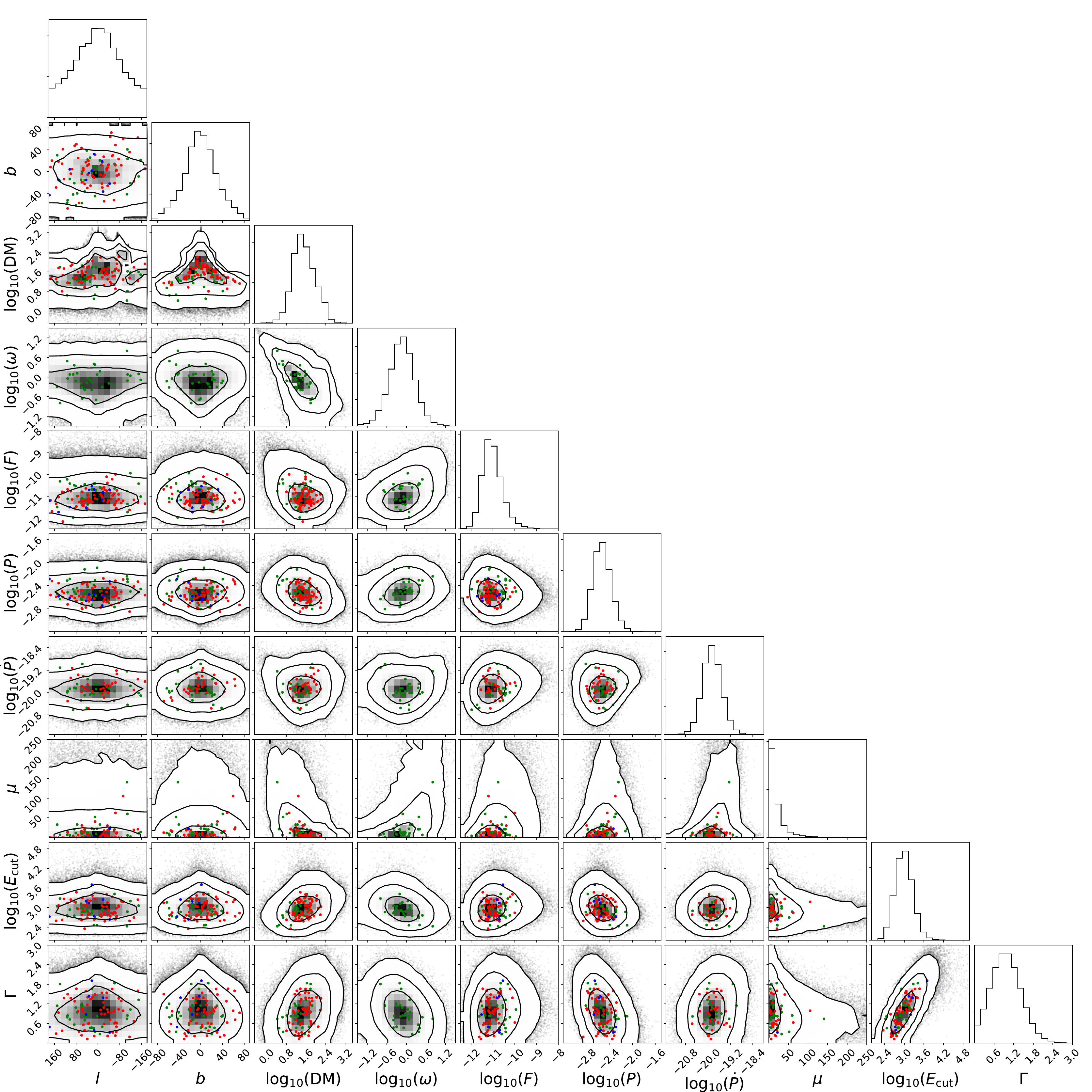}
    \caption{Model A1 ($L = \eta E_{\rm cut}^{a_{\gamma}} B^{b_{\gamma}} \dot{E}^{d_{\gamma}}$) posterior predictive corner plots showing longitude $l$ (deg), latitude $b$ (deg), dispersion measure DM (cm$^{-3}$ pc), parallax $\omega$ (mas), flux $F$ (erg cm$^{-2}$ s$^{-1}$), period $P$ (s), period derivative $\dot{P}$, proper motion $\mu$ (mas yr$^{-1}$), spectral cutoff $E_{\rm cut}$ (MeV) and spectral index $\Gamma$ simulated and real data. Red points are MSPs with dispersion measure distances, green is parallax, and blue is no distance measurement. The contours show regions containing $68\%$, $95\%$ and $99.7\%$ of the simulated data.}
    \label{fig:E_cut_B_E_dot_MSP_Params}
\end{figure}

\begin{figure}
    \centering
    \includegraphics[width=0.99\linewidth]{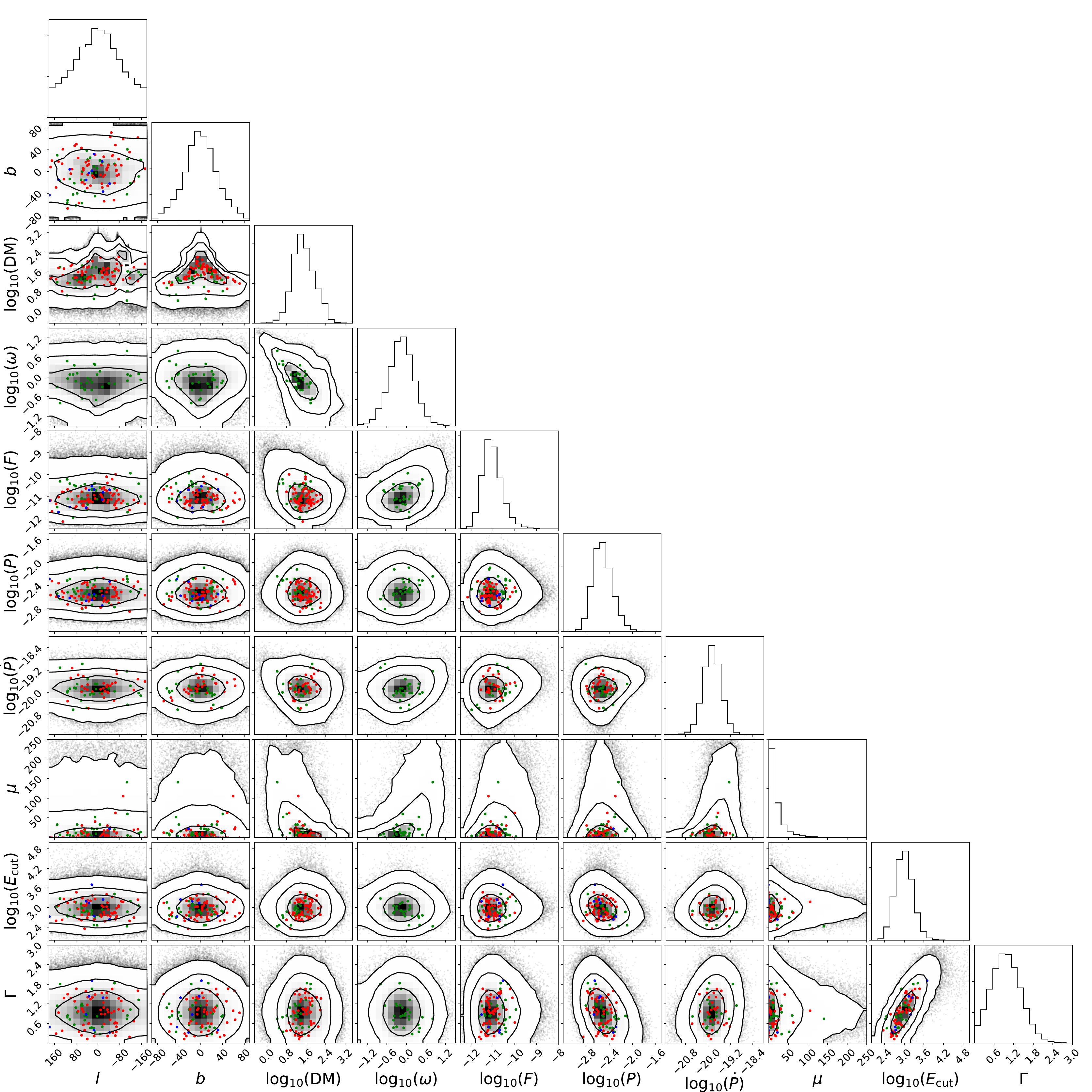}
    \caption{
    The same as Fig.~\ref{fig:E_cut_B_E_dot_MSP_Params} except for 
    Model A9  which has $L=\eta$.}
    \label{fig:efficiency_only_MSP_Params}
\end{figure}

\begin{figure}
    \centering
    \includegraphics[width=0.99\linewidth]{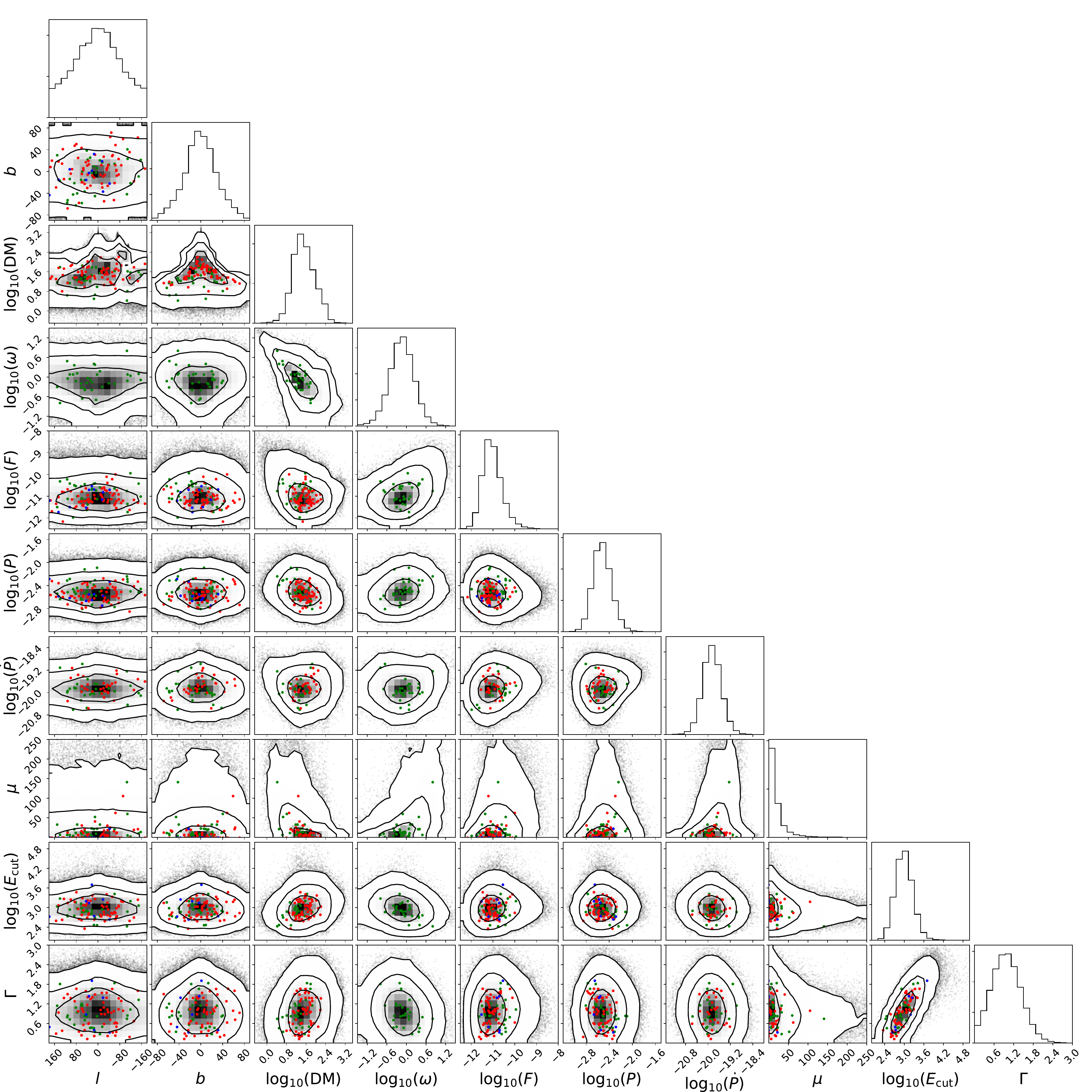}
    \caption{The same as Fig.~\ref{fig:E_cut_B_E_dot_MSP_Params} except for 
    Model A7  which has the same luminosity function as Model A1 but has $E_{\rm cut}$ independent of $\dot{E}$ ($a_{E_{\rm cut}} = a_{\Gamma} = 0$).}
    \label{fig:E_cut_B_E_dot_NoSpectrumDependenceOnEdot_MSP_Params}
\end{figure}

Finally, we also display corner plots showing the distributions of simulated, resolved MSPs for different models of the MSP luminosity function: Fig.~\ref{fig:E_cut_B_E_dot_MSP_Params} shows the $L = \eta E_{\rm cut}^{a_{\gamma}} B^{b_{\gamma}} \dot{E}^{d_{\gamma}}$ case, Fig.~\ref{fig:efficiency_only_MSP_Params} shows the $L = \eta$ case, and Fig.~\ref{fig:E_cut_B_E_dot_NoSpectrumDependenceOnEdot_MSP_Params} shows the case for the $L = \eta E_{\rm cut}^{a_{\gamma}} B^{b_{\gamma}} \dot{E}^{d_{\gamma}}$ with $E_{\rm cut}$ and $\Gamma$ independent of $\dot{E}$.

\begin{table}
\begin{center}
    \begin{tabular}{c|c|c}
         Parameter & Prior Minimum & Prior Maximum \\ \hline \hline
         
         $\lambda_{\rm res}$ & $0$ & $1000$ \\ \hline
         $\log_{10}(N_{\rm disk} / N_{\rm bulge})$ & $-100$ & $100$ \\ \hline
         $\log_{10}(N_{\rm nb} / N_{\rm bb})$ & $-100$ & $100$ \\ \hline
         $\sigma_r$ (kpc) & $0$ & $15$ \\ \hline
         $z_0$ (kpc) & $0$ & $1.5$ \\ \hline
         $K_{\rm th}$ & $-20$ & $20$ \\ \hline
         $\sigma_{\rm th}$ & $0$ & $5$ \\ \hline
         $C_{\rm parallax}$ & $0$ & $5$ \\ \hline
         $\log_{10}(d_{\rm parallax}/{\rm kpc})$ & $-1$ & $2$ \\ \hline
         $a_{E_{\rm cut}}$ & $-5$ & $5$ \\ \hline
         $b_{E_{\rm cut}}$ & $2$ & $5$ \\ \hline
         $\sigma_{E_{\rm cut}}$ & $0$ & $3$ \\ \hline
         $a_{\Gamma}$ & $-5$ & $5$ \\ \hline
         $b_{\Gamma}$ & $0$ & $5$ \\ \hline
         $\sigma_{\Gamma}$ & $0$ & $5$ \\ \hline
         $r_{\Gamma, E_{\rm cut}}$ & $-1$ & $1$ \\ \hline
         $a_{\gamma}$ & $-10$ & $10$ \\ \hline
         $b_{\gamma}$ & $-10$ & $10$ \\ \hline
         $d_{\gamma}$ & $-10$ & $10$ \\ \hline
         $\alpha_{\gamma}$ & $-10$ & $10$ \\ \hline
         $\beta_{\gamma}$ & $-10$ & $10$ \\ \hline
         $\textrm{DTD } p(0 \textrm{ - } 2.8 \textrm{ Gyr})$ & $0$ & $1$ \\ \hline
         $\textrm{DTD } p(2.8 \textrm{ - } 5.5 \textrm{ Gyr})$ & $0$ & $1$ \\ \hline
         $\textrm{DTD } p(5.5 \textrm{ - } 8.3 \textrm{ Gyr})$ & $0$ & $1$ \\ \hline
         $\textrm{DTD } p(8.3 \textrm{ - } 11.1 \textrm{ Gyr})$ & $0$ & $1$ \\ \hline
         $\textrm{DTD } p(11.1 \textrm{ - } 13.8 \textrm{ Gyr})$ & $0$ & $1$ \\ \hline
         $\log_{10}(P_{I, \textrm{ med}}/\textrm{s})$ & $-4$ & $10$ \\ \hline
         $\sigma_{P_{I}}$ & $0$ & $10$ \\ \hline
         $C_{P_I}$ & $-2$ & $5$ \\ \hline
         $\log_{10}(B_{\rm med}/\textrm{G})$ & $-1$ & $20$ \\ \hline
         $\sigma_{B}$ & $0$ & $10$ \\ \hline
         $r_{B, P_I}$ & $-1$ & $1$ \\ \hline
         $\log_{10}(\eta_{\rm med})$ & $-50$ & $50$ \\ \hline
         $\sigma_{\eta}$ & $0$ & $10$ \\ \hline
         $\sigma_{v}$ (km s$^{-1}$) & $0$ & $10000$ \\ \hline
    \end{tabular}
    \label{tab:prior_ranges}
    \caption{Prior ranges for model parameters. The priors are uniform within these ranges except for the DTD bin probabilities which have  a Dirichlet prior \citep{Betancourt2013}. I.e.\ the prior for the DTD bin probabilities is uniform on a four dimensional hyperplane in the five dimensional DTD bin space. The hyperplane consist of all points for which the five DTD bin probabilities add up to one. }
    \end{center}
\end{table}

\begin{table}
    \begin{tabular}{c|c|c|c|c}
         Model Label & Description & $\Delta \textrm{WAIC}$ & WAIC & $p_\text{WAIC}$ \\
          &  &  & Std. Dev. &  \\ \hline \hline
         
      A1&   $L = \eta E_{\rm cut}^{a_{\gamma}} B^{b_{\gamma}} \dot{E}^{d_{\gamma}}$ & $0$ & $0.8$ & $26.9$ \\ \hline
      A2&   $L = \eta E_{\rm cut}^{a_{\gamma}} B^{b_{\gamma}} \dot{E}^{d_{\gamma}}$, Uniform Age Distribution & $0.4$ & $1.0$ & $26.1$ \\ \hline
    A3&     $L = \eta E_{\rm cut}^{a_{\gamma}} B^{b_{\gamma}} \dot{E}^{d_{\gamma}}$, Covariance $B$ and $P_I$ & $1.5$ & $1.0$ & $27.5$ \\ \hline
        A4& $L = \eta E_{\rm cut}^{a_{\gamma}} B^{b_{\gamma}} \dot{E}^{d_{\gamma}}$, Eq.\ \ref{eq:gonthier_initial_period_dist} Initial Period Distribution & $7.3$ & $1.5$ & $25.9$ \\ \hline
        A5& $L = \eta \dot{E}^{\alpha_{\gamma}}$  & $9.5$ & $1.2$ & $24.4$ \\ \hline
        A6& $L = \eta P^{\alpha_{\gamma}} \dot{P}^{\beta_{\gamma}}$ & $11.0$ & $0.7$ & $25.3$ \\ \hline
        A7& $L = \eta E_{\rm cut}^{a_{\gamma}} B^{b_{\gamma}} \dot{E}^{d_{\gamma}}$, $a_{E_{\rm cut}} = a_{\Gamma} = 0$ & $19.4$ & $0.5$ & $25.1$ \\ \hline
        A8& $L = \eta E_{\rm cut}^{a_{\gamma}} B^{b_{\gamma}} \dot{E}^{d_{\gamma}}$, $a_{E_{\rm cut}} = a_{\Gamma} = 0$, Uniform Age Distribution & $21.1$ & $0.7$ & $24.0$ \\ \hline
    A9&     $L = \eta$ & $42.2$ & $0.7$ & $24.1$ \\ \hline
         \hline
      B1&   $L = \eta E_{\rm cut}^{a_{\gamma}} B^{b_{\gamma}} \dot{E}^{d_{\gamma}}$, Uniform Age Distribution, No GCE & $0$ & $1$ & $24$ \\ \hline
    B2&     $L = \eta E_{\rm cut}^{a_{\gamma}} B^{b_{\gamma}} \dot{E}^{d_{\gamma}}$, No GCE & $1.2$ & $0.8$ & $25.3$ \\ \hline
      B3&   $L = \eta E_{\rm cut}^{a_{\gamma}} B^{b_{\gamma}} \dot{E}^{d_{\gamma}}$, Covariance $B$ and $P_I$, No GCE & $2.3$ & $0.8$ & $26.1$ \\ \hline
    B4&     $L = \eta E_{\rm cut}^{a_{\gamma}} B^{b_{\gamma}} \dot{E}^{d_{\gamma}}$, Eq.\ \ref{eq:gonthier_initial_period_dist} Initial Period Distribution, No GCE & $9$ & $4$ & $24$ \\ \hline
      B5&  $L = \eta \dot{E}^{\alpha_{\gamma}}$, No GCE & $13$ & $1$ & $23$\\ \hline
    B6&     $L = \eta P^{\alpha_{\gamma}} \dot{P}^{\beta_{\gamma}}$, No GCE & $15.2$ & $0.9$ & $23.8$ \\ \hline
      B7&  $L = \eta E_{\rm cut}^{a_{\gamma}} B^{b_{\gamma}} \dot{E}^{d_{\gamma}}$, $a_{E_{\rm cut}} = a_{\Gamma} = 0$, No GCE & $21.3$ & $0.9$ & $23.6$ \\ \hline
    B8&     $L = \eta E_{\rm cut}^{a_{\gamma}} B^{b_{\gamma}} \dot{E}^{d_{\gamma}}$, $a_{E_{\rm cut}} = a_{\Gamma} = 0$, Uniform Age Distribution, No GCE & $22.9$ & $0.9$ & $22.5$ \\ \hline
      B9&  $L = \eta$, No GCE & $47$ & $2$ & $23$ \\ \hline
    \end{tabular}
    \label{tab:WAIC}
    \caption{Average WAIC for each model relative to the model with best (lowest) WAIC. The average is taken over eight MCMC chains run for each model. We separate cases where we do not fit a GCE (B1-B9) from those where we do (A1-A9). The WAIC standard deviation column shows the sample standard deviation of WAIC for the eight MCMC chains run for each model. The $p_{\rm WAIC}$ column gives a measure of the effective number of parameters for the corresponding model \citep{Wantabe2010, Gelman2013}. }
\end{table}

\begin{table}
\begin{center}
    \begin{tabular}{c|c|c|c|c|c}
    Parameter & Model A1 & Model B2 & Model A5 & Model A7 & Model A9 \\
 \hline \hline
         $\lambda_{\rm res}$ & $108\substack{+11 \\ -10}$ & $107\substack{+11 \\ -10}$ & $108\substack{+10 \\ -10}$ & $108\substack{+11 \\ -10}$ & $108\substack{+11 \\ -10}$ \\ \hline
         $\log_{10}(N_{\rm disk} / N_{\rm bulge})$ & $0.00\substack{+0.13 \\ -0.12}$ & $-$ & $0.00\substack{+0.13 \\ -0.12}$ & $0.04\substack{+0.12 \\ -0.11}$ & $0.09\substack{+0.11 \\ -0.10}$ \\ \hline
         $\log_{10}(N_{\rm nb} / N_{\rm bb})$ & $-0.66\substack{+0.08 \\ -0.07}$ & $-$ & $-0.66\substack{+0.08 \\ -0.07}$ & $-0.61\substack{+0.06 \\ -0.07}$ & $-0.58\substack{+0.04 \\ -0.04}$ \\ \hline
         $\sigma_r$ (kpc) & $4.5\substack{+0.5 \\ -0.4}$ & $4.5\substack{+0.5 \\ -0.4}$ & $4.4\substack{+0.5 \\ -0.4}$ & $4.4\substack{+0.5 \\ -0.4}$ & $4.5\substack{+0.6 \\ -0.4}$ \\ \hline
         $z_0$ (kpc) & $0.71\substack{+0.11 \\ -0.09}$ & $0.70\substack{+0.10 \\ -0.09}$ & $0.71\substack{+0.11 \\ -0.09}$ & $0.71\substack{+0.11 \\ -0.09}$ & $0.72\substack{+0.11 \\ -0.09}$ \\ \hline
         $K_{\rm th}$ & $0.45\substack{+0.09 \\ -0.08}$ & $0.43\substack{+0.09 \\ -0.08}$ & $0.45\substack{+0.09 \\ -0.08}$ & $0.46\substack{+0.09 \\ -0.08}$ & $0.46\substack{+0.09 \\ -0.08}$ \\ \hline
         $\sigma_{\rm th}$ & $0.28\substack{+0.05 \\ -0.04}$ & $0.27\substack{+0.05 \\ -0.04}$ & $0.28\substack{+0.05 \\ -0.04}$ & $0.28\substack{+0.05 \\ -0.04}$ & $0.28\substack{+0.05 \\ -0.04}$ \\ \hline
         $C_{\rm parallax}$ & $0.43\substack{+0.15 \\ -0.10}$ & $0.45\substack{+0.16 \\ -0.11}$ & $0.43\substack{+0.15 \\ -0.10}$ & $0.43\substack{+0.15 \\ -0.10}$ & $0.40\substack{+0.13 \\ -0.08}$ \\ \hline
         $\log_{10}(d_{\rm parallax}/{\rm kpc})$ & $0.8\substack{+0.6 \\ -0.3}$ & $0.7\substack{+0.6 \\ -0.3}$ & $0.8\substack{+0.6 \\ -0.3}$ & $0.8\substack{+0.6 \\ -0.3}$ & $0.9\substack{+0.6 \\ -0.4}$ \\ \hline
         $a_{E_{\rm cut}}$ & $0.18\substack{+0.05 \\ -0.05}$ & $0.21\substack{+0.05 \\ -0.05}$ & $0.19\substack{+0.05 \\ -0.05}$ & $-$ & $0.17\substack{+0.05 \\ -0.05}$ \\ \hline
         $b_{E_{\rm cut}}$ & $2.83\substack{+0.05 \\ -0.06}$ & $2.79\substack{+0.06 \\ -0.07}$ & $2.99\substack{+0.03 \\ -0.02}$ & $2.79\substack{+0.06 \\ -0.06}$ & $2.99\substack{+0.03 \\ -0.03}$ \\ \hline
         $\sigma_{E_{\rm cut}}$ & $0.23\substack{+0.02 \\ -0.02}$ & $0.23\substack{+0.02 \\ -0.02}$ & $0.22\substack{+0.02 \\ -0.02}$ & $0.25\substack{+0.02 \\ -0.02}$ & $0.23\substack{+0.02 \\ -0.02}$ \\ \hline
         $a_{\Gamma}$ & $0.41\substack{+0.08 \\ -0.08}$ & $0.43\substack{+0.08 \\ -0.08}$ & $0.43\substack{+0.07 \\ -0.08}$ & $-$ & $0.39\substack{+0.08 \\ -0.08}$ \\ \hline
         $b_{\Gamma}$ & $0.81\substack{+0.07 \\ -0.08}$ & $0.77\substack{+0.08 \\ -0.10}$ & $1.00\substack{+0.04 \\ -0.04}$ & $0.70\substack{+0.09 \\ -0.10}$ & $0.99\substack{+0.04 \\ -0.04}$ \\ \hline
         $\sigma_{\Gamma}$ & $0.36\substack{+0.04 \\ -0.03}$ & $0.36\substack{+0.04 \\ -0.03}$ & $0.34\substack{+0.03 \\ -0.03}$ & $0.42\substack{+0.04 \\ -0.03}$ & $0.35\substack{+0.03 \\ -0.03}$ \\ \hline
         $r_{\Gamma, E_{\rm cut}}$ & $0.75\substack{+0.05 \\ -0.06}$ & $0.77\substack{+0.05 \\ -0.06}$ & $0.73\substack{+0.06 \\ -0.07}$ & $0.79\substack{+0.04 \\ -0.05}$ & $0.75\substack{+0.05 \\ -0.06}$ \\ \hline
         $a_{\gamma}$ & $1.2\substack{+0.3 \\ -0.3}$ & $1.4\substack{+0.3 \\ -0.3}$ & $-$ & $1.3\substack{+0.3 \\ -0.3}$ & $-$ \\ \hline
         $b_{\gamma}$ & $0.1\substack{+0.4 \\ -0.4}$ & $0.1\substack{+0.4 \\ -0.4}$ & $-$ & $0.1\substack{+0.4 \\ -0.4}$ & $-$ \\ \hline
         $d_{\gamma}$ & $0.50\substack{+0.12 \\ -0.12}$ & $0.48\substack{+0.12 \\ -0.12}$ & $-$ & $0.56\substack{+0.11 \\ -0.10}$ & $-$ \\ \hline
         $\alpha_{\gamma}$ & $-$ & $-$ & $0.74\substack{+0.11 \\ -0.10}$ & $-$ & $-$ \\ \hline
         $\textrm{DTD } p(0 \textrm{ - } 2.8 \textrm{ Gyr})$ & $0.13\substack{+0.14 \\ -0.09}$ & $0.13\substack{+0.15 \\ -0.09}$ & $0.13\substack{+0.14 \\ -0.09}$ & $0.10\substack{+0.11 \\ -0.07}$ & $0.02\substack{+0.02 \\ -0.01}$ \\ \hline
         $\textrm{DTD } p(2.8 \textrm{ - } 5.5 \textrm{ Gyr})$ & $0.15\substack{+0.18 \\ -0.11}$ & $0.15\substack{+0.19 \\ -0.11}$ & $0.15\substack{+0.18 \\ -0.11}$ & $0.12\substack{+0.16 \\ -0.09}$ & $0.02\substack{+0.03 \\ -0.02}$ \\ \hline
         $\textrm{DTD } p(5.5 \textrm{ - } 8.3 \textrm{ Gyr})$ & $0.29\substack{+0.20 \\ -0.17}$ & $0.26\substack{+0.18 \\ -0.16}$ & $0.29\substack{+0.20 \\ -0.17}$ & $0.25\substack{+0.18 \\ -0.14}$ & $0.09\substack{+0.06 \\ -0.05}$ \\ \hline
         $\textrm{DTD } p(8.3 \textrm{ - } 11.1 \textrm{ Gyr})$ & $0.14\substack{+0.16 \\ -0.10}$ & $0.11\substack{+0.14 \\ -0.08}$ & $0.15\substack{+0.17 \\ -0.10}$ & $0.14\substack{+0.17 \\ -0.10}$ & $0.06\substack{+0.11 \\ -0.05}$ \\ \hline
         $\textrm{DTD } p(11.1 \textrm{ - } 13.8 \textrm{ Gyr})$ & $0.14\substack{+0.22 \\ -0.11}$ & $0.22\substack{+0.20 \\ -0.15}$ & $0.13\substack{+0.21 \\ -0.10}$ & $0.27\substack{+0.24 \\ -0.19}$ & $0.78\substack{+0.08 \\ -0.14}$ \\ \hline
         $\log_{10}(P_{i, \textrm{ med}}/\textrm{s})$ & $-2.61\substack{+0.05 \\ -0.04}$ & $-2.60\substack{+0.05 \\ -0.04}$ & $-2.61\substack{+0.05 \\ -0.04}$ & $-2.63\substack{+0.04 \\ -0.03}$ & $-2.69\substack{+0.03 \\ -0.03}$ \\ \hline
         $\sigma_{P_{i}}$ & $0.13\substack{+0.02 \\ -0.02}$ & $0.13\substack{+0.02 \\ -0.02}$ & $0.13\substack{+0.02 \\ -0.02}$ & $0.12\substack{+0.02 \\ -0.02}$ & $0.13\substack{+0.02 \\ -0.02}$ \\ \hline
         $\log_{10}(B_{\rm med}/\textrm{G})$ & $8.21\substack{+0.05 \\ -0.06}$ & $8.21\substack{+0.06 \\ -0.06}$ & $8.21\substack{+0.03 \\ -0.04}$ & $8.21\substack{+0.05 \\ -0.05}$ & $8.25\substack{+0.02 \\ -0.02}$ \\ \hline
         $\sigma_{B}$ & $0.21\substack{+0.03 \\ -0.02}$ & $0.21\substack{+0.03 \\ -0.02}$ & $0.22\substack{+0.03 \\ -0.03}$ & $0.21\substack{+0.03 \\ -0.03}$ & $0.19\substack{+0.02 \\ -0.02}$ \\ \hline
         $\log_{10}(\eta_{\rm med})$ & $12\substack{+5 \\ -5}$ & $12\substack{+5 \\ -5}$ & $7\substack{+4 \\ -4}$ & $9\substack{+4 \\ -4}$ & $32.17\substack{+0.23 \\ -0.31}$ \\ \hline
         $\sigma_{\eta}$ & $0.52\substack{+0.06 \\ -0.05}$ & $0.53\substack{+0.06 \\ -0.05}$ & $0.58\substack{+0.06 \\ -0.05}$ & $0.51\substack{+0.06 \\ -0.05}$ & $0.72\substack{+0.08 \\ -0.06}$ \\ \hline
         $\sigma_{v}$ (km s$^{-1}$) & $77\substack{+6 \\ -6}$ & $76\substack{+6 \\ -5}$ & $77\substack{+6 \\ -6}$ & $77\substack{+6 \\ -6}$ & $78\substack{+6 \\ -6}$ \\ \hline
    \end{tabular}
    \label{tab:mcmc_results_parameters}
    \caption{ Medians and $68\%$ confidence intervals for a selection of different models of the Galactic MSP population. See Table~\ref{tab:WAIC} for a description of the model labels.
    }
    \end{center}
\end{table}

\section{Discussion}
\label{sec:discussion}

\begin{figure}
    \centering
    \subfigure{\centering\includegraphics[width=0.49\linewidth]{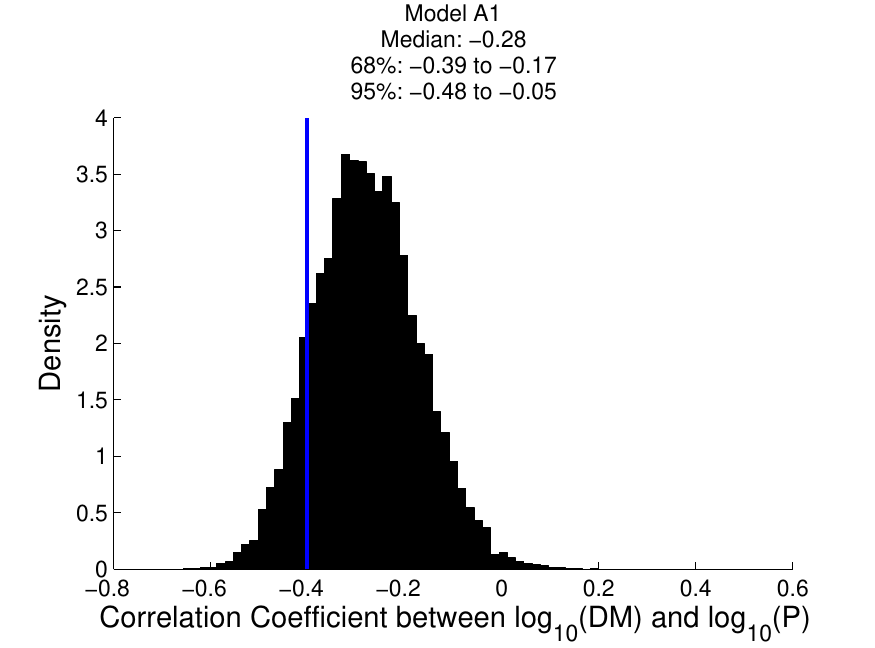}}
    \subfigure{\centering\includegraphics[width=0.49\linewidth]{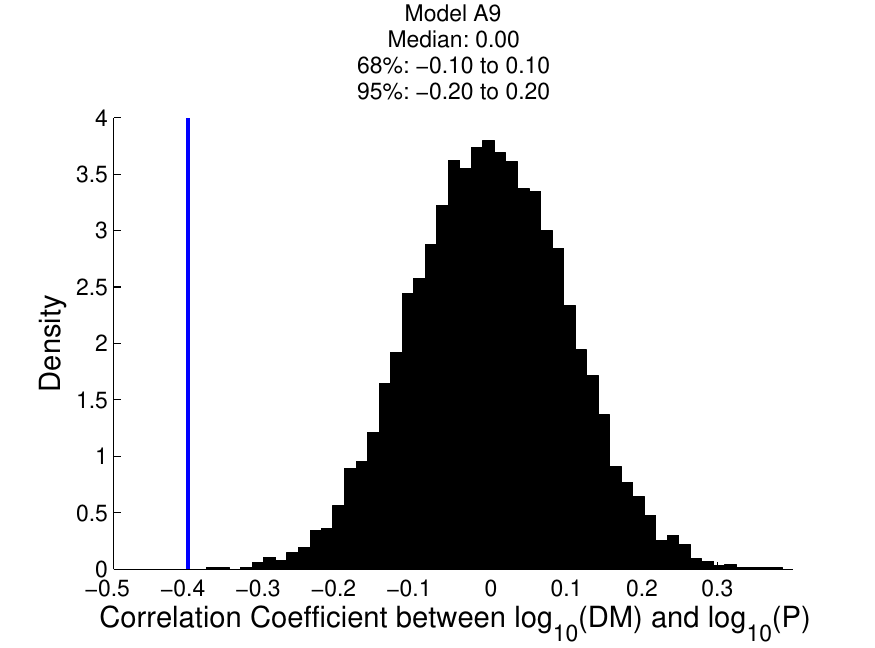}}
    \caption{Comparison of correlation coefficients between $\log_{10}(P)$ and $\log_{10}({\rm DM})$ for resolved MSP simulated and real data for  Model A1 
    ($L = \eta E_{\rm cut}^{a_{\gamma}} B^{b_{\gamma}} \dot{E}^{d_{\gamma}}$) 
    on the left and Model A9 ($L=\eta$) on the right. The blue line shows the correlation coefficient for the resolved MSP data at -0.40. The posterior predictive p-values are $0.13$ for A1, and $0$ for A9. }
    \label{fig:E_cut_B_E_dot_vs_efficiency_only_log_DM_log_P_correlation_coefficients}
\end{figure}

\begin{table}
\begin{center}
    \begin{tabular}{|c||c|c|c|c|c|c|c|}
         \cline{1-2}
        $\log_{10}(\omega)$ & $0.417$ \\ \cline{1-3}
        $\log_{10}(F)$ & $0.627$ & $0.159$ \\ \cline{1-4}
        $\log_{10}(P)$  & $0.126$ & $0.948$ & $0.752$ \\ \cline{1-5}
        $\log_{10}(\dot{P})$ & $0.911$ & $0.245$ & $0.451$ & $0.159$ \\ \cline{1-6}
        $\mu$ & $0.034$   & $0.398$ & $0.294$ & $0.938$ & $0.338$ \\ \cline{1-7}
        $\log_{10}(E_{\rm cut})$ & $0.789$ & $0.039$ & $0.437$ & $0.527$ & $0.382$ & $0.477$ \\ \hline
        $\Gamma$   & $0.550$ & $0.163$ & $0.773$ & $0.583$ & $0.352$ & $0.707$ & $0.612$ \\ \hline \hline
       & $\log_{10}(\textrm{DM})$ & $\log_{10}(\omega)$ & $\log_{10}(F)$ & $\log_{10}(P)$ & $\log_{10}(\dot{P})$ & $\mu$ & $\log_{10}(E_{\rm cut})$ \\ \hline
    \end{tabular}
    \label{tab:model_a1_pvals}
    \caption{ Posterior predictive p-values for the correlation coefficients between observables for  Model A1 which has $L = \eta E_{\rm cut}^{a_{\gamma}} B^{b_{\gamma}} \dot{E}^{d_{\gamma}}$. }
    \end{center}
\end{table}

\begin{table}
\centering
\label{tab:model_a6_pvals}
\begin{tabular}{|c||c|c|c|c|c|ll} 
\cline{1-2}
 $\log_{10}(\omega)$       & $0.426$                   & \multicolumn{1}{l}{} & \multicolumn{1}{l}{} & \multicolumn{1}{l}{} & \multicolumn{1}{l}{}  &                               &                                                 \\ 
\cline{1-3}
 $\log_{10}(F)$            & $0.615$                   & $0.158$              & \multicolumn{1}{l}{} & \multicolumn{1}{l}{} & \multicolumn{1}{l}{}  &                               &                                                 \\ 
\cline{1-4}
 $\log_{10}(P)$            & $0.138$                   & $0.941$              & $0.743$              & \multicolumn{1}{l}{} & \multicolumn{1}{l}{}  &                               &                                                 \\ 
\cline{1-5}
 $\log_{10}(\dot{P})$      & $0.891$                   & $0.265$              & $0.431$              & $0.145$              & \multicolumn{1}{l}{}  &                               &                                                 \\ 
\cline{1-6}
 $\mu$                     & $0.430$                   & $0.395$              & $0.282$              & $0.932$              & $0.366$               &                               &                                                 \\ 
\cline{1-7}
 $\log_{10}(E_{\rm cut})$  & $\textcolor{red}{0.992}$                   & $\textcolor{red}{0.007}$              & $0.697$              & $0.609$              & $0.216$               & \multicolumn{1}{c|}{$0.175$ } &                                                 \\ 
\hline
 $\Gamma$                  & $0.852$                   & $0.069$              & $0.879$              & $0.667$              & $0.223$               & \multicolumn{1}{c|}{$0.515$ } & \multicolumn{1}{c|}{$0.625$ }                   \\ 
\hhline{|=::=======|}
                           & $\log_{10}(\textrm{DM})$  & $\log_{10}(\omega)$  & $\log_{10}(F)$       & $\log_{10}(P)$       & $\log_{10}(\dot{P})$  & \multicolumn{1}{c|}{$\mu$ }   & \multicolumn{1}{c|}{$\log_{10}(E_{\rm cut})$ }  \\
\hline
\end{tabular}
\caption{ Posterior predictive p-values for the correlation coefficients between observables for  Model A6 which has $L = \eta P^{\alpha_{\gamma}} \dot{P}^{\beta_{\gamma}}$. Values  greater than 0.99 and less than 0.01 are marked in red.  }
\end{table}

\begin{table}
\centering
\label{tab:model_a7_pvals}
\begin{tabular}{|c||c|c|c|c|c|ll}
\cline{1-2}
 $\log_{10}(\omega)$       & $0.384$                   & \multicolumn{1}{l}{} & \multicolumn{1}{l}{} & \multicolumn{1}{l}{}          & \multicolumn{1}{l}{}  &                               &                                                 \\ 
\cline{1-3}
 $\log_{10}(F)$            & $0.719$                   & $0.120$              & \multicolumn{1}{l}{} & \multicolumn{1}{l}{}          & \multicolumn{1}{l}{}  &                               &                                                 \\ 
\cline{1-4}
 $\log_{10}(P)$            & $0.057$                   & $0.965$              & $0.646$              & \multicolumn{1}{l}{}          & \multicolumn{1}{l}{}  &                               &                                                 \\ 
\cline{1-5}
 $\log_{10}(\dot{P})$      & $0.938$                   & $0.225$              & $0.502$              & $0.157$                       & \multicolumn{1}{l}{}  &                               &                                                 \\ 
\cline{1-6}
 $\mu$                     & $0.412$                   & $0.415$              & $0.244$              & $0.963$                       & $0.316$               &                               &                                                 \\ 
\cline{1-7}
 $\log_{10}(E_{\rm cut})$  & $0.942$                   & $0.017$              & $0.583$              & $\textcolor{red}{0.004}$                       & $0.819$               & \multicolumn{1}{c|}{$0.339$ } &                                                 \\ 
\hline
 $\Gamma$                  & $0.889$                   & $0.059$              & $0.906$              & $\bf \textcolor{red}{0.000}$  & $0.894$               & \multicolumn{1}{c|}{$0.509$ } & \multicolumn{1}{c|}{$0.661$ }                   \\ 
\hhline{|=::=======|}
                           & $\log_{10}(\textrm{DM})$  & $\log_{10}(\omega)$  & $\log_{10}(F)$       & $\log_{10}(P)$                & $\log_{10}(\dot{P})$  & \multicolumn{1}{c|}{$\mu$ }   & \multicolumn{1}{c|}{$\log_{10}(E_{\rm cut})$ }  \\
\hline
\end{tabular}
\caption{
Same as Table~\ref{tab:model_a6_pvals} except for
  Model A7 which has $L = \eta E_{\rm cut}^{a_{\gamma}} B^{b_{\gamma}} \dot{E}^{d_{\gamma}}$, $a_{E_{\rm cut}} = a_{\Gamma} = 0$.}
\end{table}

\begin{table}
\centering
\label{tab:model_a9_pvals}
\begin{tabular}{|c||c|c|c|c|c|ll}
\cline{1-2}
 $\log_{10}(\omega)$       & $0.422$                         & \multicolumn{1}{l}{} & \multicolumn{1}{l}{} & \multicolumn{1}{l}{} & \multicolumn{1}{l}{}  &                               &                                                 \\ 
\cline{1-3}
 $\log_{10}(F)$            & $0.670$                         & $0.141$              & \multicolumn{1}{l}{} & \multicolumn{1}{l}{} & \multicolumn{1}{l}{}  &                               &                                                 \\ 
\cline{1-4}
 $\log_{10}(P)$            & $\bf \textcolor{red}{0.000}$    & $\textcolor{red}{0.998}$              & $0.289$              & \multicolumn{1}{l}{} & \multicolumn{1}{l}{}  &                               &                                                 \\ 
\cline{1-5}
 $\log_{10}(\dot{P})$      & $\textcolor{red}{0.997}$                         & $0.104$              & $0.697$              & $0.140$              & \multicolumn{1}{l}{}  &                               &                                                 \\ 
\cline{1-6}
 $\mu$                     & $0.424$                         & $0.396$              & $0.275$              & $\textcolor{red}{0.998}$              & $0.155$               &                               &                                                 \\ 
\cline{1-7}
 $\log_{10}(E_{\rm cut})$  & ${\bf \textcolor{red}{1.000}}$  & $\textcolor{red}{0.002}$              & $0.869$              & $0.541$              & $0.316$               & \multicolumn{1}{c|}{$0.053$ } &                                                 \\ 
\hline
 $\Gamma$                  & $\textcolor{red}{0.998}$                         & $0.012$              & $0.978$              & $0.621$              & $0.354$               & \multicolumn{1}{c|}{$0.192$ } & \multicolumn{1}{c|}{$0.631$ }                   \\ 
\hhline{|=::=======|}
                           & $\log_{10}(\textrm{DM})$        & $\log_{10}(\omega)$  & $\log_{10}(F)$       & $\log_{10}(P)$       & $\log_{10}(\dot{P})$  & \multicolumn{1}{c|}{$\mu$ }   & \multicolumn{1}{c|}{$\log_{10}(E_{\rm cut})$ }  \\
\hline
\end{tabular}
\caption{Same as Table~\ref{tab:model_a6_pvals} except for  Model A9 which has $L=\eta$.}
\end{table}

Some of our luminosity functions are nested. We can go from Model A1 with 
$L = \eta E_{\rm cut}^{a_{\gamma}} B^{b_{\gamma}} \dot{E}^{d_{\gamma}}$ to Model A9 with $L=\eta$ by setting $a_{\gamma}=b_{\gamma}=d_{\gamma}=0$. As can be seen from the Model A1 fit in Fig.~\ref{fig:E_cut_B_E_dot_MCMC_Params_1} and Table \ref{tab:mcmc_results_parameters}, the data prefer the $a_\gamma=1.2\pm 0.3$ and $d_\gamma=0.5\pm0.1$ parameters to be larger than zero at high significance.  This is consistent with the results of Table~\ref{tab:WAIC} where Model~A9 has a $\Delta{\rm WAIC}=42.2$ relative to
Model~A1. 
 Comparing the posterior predictive corner plots 
 for Model~A1 in 
 Fig.~\ref{fig:E_cut_B_E_dot_MSP_Params} and Model~A9 in Fig.~\ref{fig:efficiency_only_MSP_Params} we can see that Model~A9  does not capture some of the correlations in the data, 
 particularly those between distance (shown indirectly in the form of dispersion measure) and the spectral parameter
 $E_{\rm cut}$ 
 as well as between distance and period. A comparison of correlation coefficients between the real data and simulated data is shown in Fig.~\ref{fig:E_cut_B_E_dot_vs_efficiency_only_log_DM_log_P_correlation_coefficients}. 
 More detail can be seen by comparing Tables~\ref{tab:model_a1_pvals} and \ref{tab:model_a9_pvals} which list the posterior predictive p-values for the correlations between observables for Models A1 and A9 respectively.
Following  the recommendation given in Section 6.3 of ref.~\cite{Gelman2013}, we consider posterior predictive p-values below 1\% and above 99\% to be of concern.
 As can be seen, the posterior predictive p-values for Model A1 are all comfortably within the 1\% to 99\% interval. In contrast to this, seven correlation coefficients for Model A9 are outside this interval.
 The correlation coefficient 
  between $\log_{10}(P)$ and $\log_{10}({\rm DM})$ and also the one between $\log_{10}(E_{\rm cut})$ and $\log_{10}({\rm DM})$ are both particularly discrepant with the data.
These logarithmic correlations are  associated with the flux threshold which implies that, the 
more distant the MSP, the larger the intrinsic luminosity it is required to have in order to have a significant probability of being resolved. 
The point distinguishing Model A1 from Model A9 in  regards to the relationship between distance and $E_{\rm cut}$ arises because of Model A1 having a significant positive $a_{\gamma}=1.2\pm 0.3$ in Eq.~\ref{eq:Model1}, while in Model A9 we have set $a_{\gamma}=0$.
{\textcolor{black}{Regarding the logarithmic correlation between distance  and $P$, it follows from
the presence of $P$ in the denominator of Eq.~\ref{eq:Edot} that 
 the significantly positive $d_{\gamma}=0.5\pm0.1$ in Eq.~\ref{eq:Model1} leads to the negative correlation with the distance measures in Model A1, while, again, in Model A9 we have $d_\gamma=0$ by construction.}}

Model A6 is related to model A1 as follows from Eqs.~\ref{eq:Edot} and \ref{eq:magnetic_field_strength} which
demonstrate
that we can go from the luminosity function in Model~A1 to the luminosity function in Model~A6 by setting $b_{\gamma }=\frac{\alpha _{\gamma }}{2}+\frac{3 \beta _{\gamma }}{2}$ and $d_{\gamma }=\frac{\beta _{\gamma }}{4}-\frac{\alpha _{\gamma }}{4}$. However, Model~A6 does not have a dependence on $E_{\rm cut}$ and therefore assumes $a_\gamma=0$. As can be see from Table~\ref{tab:mcmc_results_parameters}, $a_\gamma=1.2\pm0.3$ is significantly positive for Model~A1 and so  Model~A1 is preferred over Model~A6.
This is consistent with Model~A6 having a $\Delta{\rm WAIC=}11.0$ relative to Model A1. As can be seen from Table~\ref{tab:model_a6_pvals} the problem with Model A6 is that it does not predict the observed logarithmic correlation between distance and $E_{\rm cut}$.

 Model A7 is the same as Model A1 except that we set $a_{E_{\rm cut}} = a_{\Gamma} = 0$. 
 We see from Table~\ref{tab:mcmc_results_parameters} that this choice for Model A7 is disfavoured as  we found that Model A1 had a significantly positive  $a_\Gamma=0.41\pm0.08$
 for Eq.~\ref{eq:Gamma}. This preference for Model A1 over A7 is confirmed in Table~\ref{tab:WAIC} where it can be seen that Model A7 has a $\Delta{\rm WAIC}=19.4$ relative to Model A1.
We can see how Model A7  produced a worse fit to the data by comparing Figs.~\ref{fig:E_cut_B_E_dot_MSP_Params} and \ref{fig:E_cut_B_E_dot_NoSpectrumDependenceOnEdot_MSP_Params}: in the latter case where $a_{E_{\rm cut}} = a_{\Gamma} = 0$, the relationship between spectral index and period 
has clearly disappeared. 
This is confirmed in Table~\ref{tab:model_a7_pvals} where two of the correlation coefficients are of concern and, in particular, the correlation coefficient between $\Gamma$ and $\log_{\rm 10}(P)$ has a posterior predictive p-value of 0.000. 

Our best fitting model of the Galactic MSP population was Model A1 for which $L = \eta E_{\rm cut}^{a_{\gamma}} B^{b_{\gamma}} \dot{E}^{d_{\gamma}}$. From inspection of Figs.~\ref{fig:E_cut_B_E_dot_posterior_predictive_plots} and \ref{fig:E_cut_B_E_dot_MSP_Params} it is evident that this model generally provided a good fit to the resolved MSP data. This is also confirmed in Table~\ref{tab:model_a1_pvals} where all the posterior predictive p--values are within the 1\% to 99\% range.
We find $a_{\gamma} = 1.2\substack{+0.3 \\ -0.3}$, $b_{\gamma} = 0.1\substack{+0.4 \\ -0.4}$ and $d_{\gamma} = 0.5\substack{+0.1 \\ -0.1}$. These results are consistent with those obtained by 
Kalapotharakos et al.~\cite{Kalapotharakos_2019}. These authors performed a least squares fit to both the MSPs and young pulsars in the Second Fermi Pulsar Catalog \citep{TheFermi-LAT:2013ssa} and find $a_{\gamma} = 1.12 \pm 0.24$, $b_{\gamma} = 0.17 \pm 0.05$ and $d_{\gamma} = 0.41 \pm 0.08$. Kalapotharakos et al.~\cite{Kalapotharakos_2019} point out that their results are consistent with predicted values of $a_{\gamma} = 4/3$, $b_{\gamma} = 1/6$ and $d_{\gamma} = 5/12$ in the case that curvature radiation is the source of gamma-ray emission.
This stands in contrast to the case of synchrotron radiation for which $a_{\gamma} = 1$, $b_{\gamma} = 0$ and $d_{\gamma} = 1$ is expected. Our posterior distributions for $a_{\gamma}$ and $b_{\gamma}$ are consistent with both cases to within $2 \sigma$, but our $d_{\gamma}$ is inconsistent with synchrotron radiation. One major difference between the work of Kalapotharakos et al.~\cite{Kalapotharakos_2019} and ours is that we use many more MSPs, but no young pulsars. Overall we have a similar total number of pulsars. However, because of the Shklovskii effect, 
the intrinsic period derivative $\dot{P}_{\rm int}$ is poorly determined relative to the case of young pulsars.
Also, $B$ and $\dot{E}$ depend on $\dot{P}_{\rm int}$ through Eqs.~\ref{eq:magnetic_field_strength} and \ref{eq:Edot} respectively. It follows that some MSPs with smaller values of $\dot{P}$
may have poorly-determined period derivatives relative to the case presented by young pulsars.

\begin{figure}
    \centering
    \subfigure{\centering\includegraphics[width=0.49\linewidth]{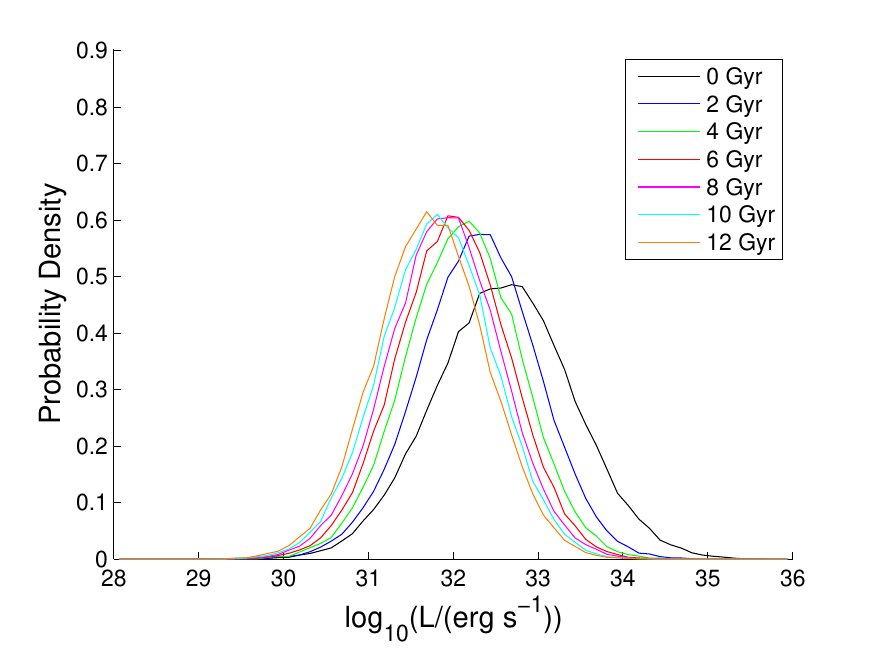}}
    \subfigure{\centering\includegraphics[width=0.49\linewidth]{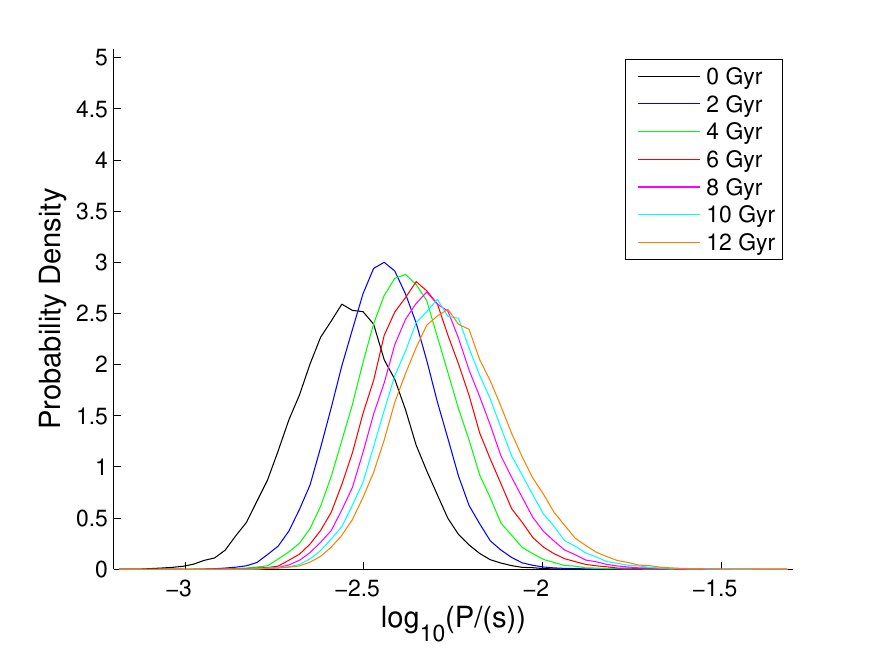}}
    \caption{Evolution of the luminosity and period distributions with time for Model A1 which has $L = \eta E_{\rm cut}^{a_{\gamma}} B^{b_{\gamma}} \dot{E}^{d_{\gamma}}$. The legend gives the age of the MSPs in the corresponding distribution. These figures were made by selecting randomly from the highest likelihood parameter sets in our eight Markov chains for this model.
    \label{fig:E_cut_B_E_dot_luminosity_period_evolution}}
\end{figure}

In Model A1, which has
$L = \eta E_{\rm cut}^{a_{\gamma}} B^{b_{\gamma}} \dot{E}^{d_{\gamma}}$,
relatively young members of the underlying population of MSPs are more likely to be resolved as they are brighter. This follows as $\dot{E}$, and therefore luminosity, decreases with age. As the magnetic field strength is assumed constant for each MSP over time, it can be seen from Eqs.~\ref{eq:magnetic_field_strength} and \ref{eq:Edot} that $\dot{E} \propto P^{-4}$. We show in Fig.~\ref{fig:E_cut_B_E_dot_luminosity_period_evolution} an example of how the $\log_{10}(L)$ and $\log_{10}(P)$ probability density functions evolve with age. 

As can be seen from Fig.~\ref{fig:E_cut_B_E_dot_MCMC_DTD_bin_Params},  %
the DTD peaks in the central $5.5$ to $8.3$ Gyr bin producing an age distribution that tends to plateau starting around $5$ Gyr ago as shown in Fig.~\ref{fig:E_cut_B_E_dot_age_distribution}.
As can be seen in Table \ref{tab:mcmc_results_parameters} the posterior distributions of the parameters are similar for the GCE and No GCE cases.
It can be seen in Table \ref{tab:WAIC} that
there is no significant difference in the WAIC for the DTD versus the uniform age distribution case.
As can be seen in Fig.~\ref{fig:E_cut_B_E_dot_age_distribution}, for the case with the GCE, the fitted DTD produces an age distribution that is similar to the uniform case. For the uniform age distribution, the probability density is $0.1$ for ages less than $10$ Gyr, and $0$ elsewhere. 
The models that were significantly worse when the GCE was included remained significantly worse when it was not. The model for which the luminosity obeys $L = \eta E_{\rm cut}^{a_{\gamma}} B^{b_{\gamma}} \dot{E}^{d_{\gamma}}$ remained the best model whether the GCE was included or not, i.e.,\ models A1 and A2 were the best models when the GCE was included and Models B1 and B2 were the best models when the GCE was not included. 

As can be seen from Fig.~\ref{fig:E_cut_B_E_dot_dtd_vs_uniform_boxy_bulge_spectra}, there is a small difference in the GCE spectrum for the Model A1 and A2. This is due to the spectral dependence on $\dot{E}$ in Eqs.~\ref{eq:Ecut} and \ref{eq:Gamma} and also the different star formation rates for the bulge and disk which are illustrated in Fig.~\ref{fig:sfr}. However, as can be seen in Fig.~\ref{fig:E_cut_B_E_dot_dtd_vs_uniform_boxy_bulge_spectra}, the differences in the predicted spectrum are negligible in comparison to the model prediction uncertainties. 
As can be seen in the top right panel of Fig.~\ref{fig:E_cut_B_E_dot_luminosity_distribution} the boxy bulge MSPs do have a median luminosity function that is less bright than the disk MSPs. However, the 95\% interval band encompasses zero which is the no difference case.
This shows that with the current levels of uncertainty the differences in the properties of the bulge and disk MSPs is not significant.

\begin{figure}
    \centering
    \includegraphics[width=0.75\linewidth]{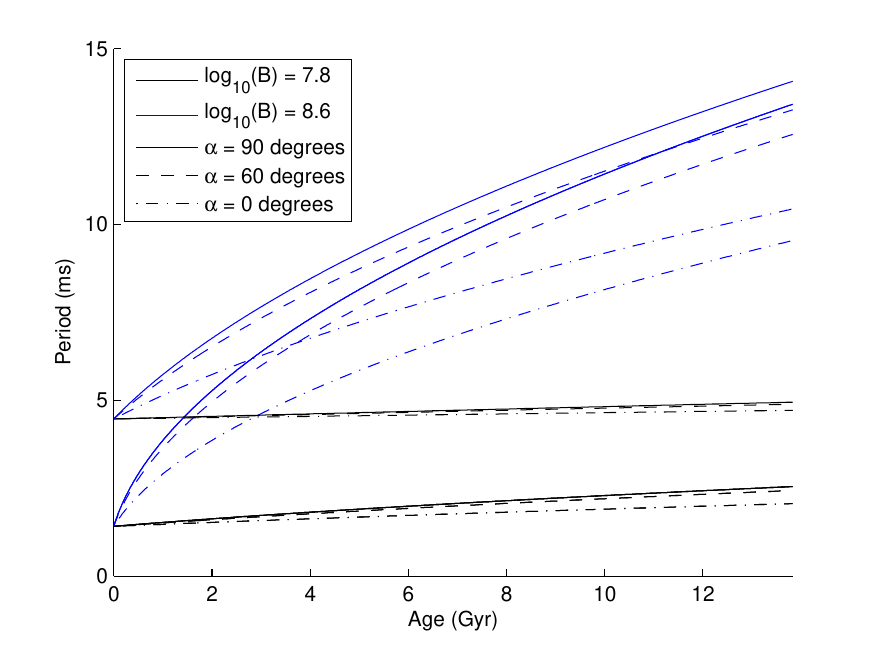}
    \caption{Evolution of period with time for various magnetic field strengths $B$, magnetic field axis angles $\alpha$ and initial periods $P_I$. The two $P_I$ values used were $10^{-2.85}$ s and $10^{-2.35}$ s.}
    \label{fig:period_evolution}
\end{figure}

In Fig.~\ref{fig:period_evolution} we show the evolution of period with time (Eq.\ \ref{eq:current_period}) for MSPs with different properties. This indicates where the constraints on our models of the initial period and age distribution come from. For a resolved MSP with a weak magnetic field, the current period will be near the initial period even if the MSP is old, so these MSPs should be approximately distributed like the initial period distribution. This is a consequence of the fact that an MSP cannot be older than the universe. On the other hand, MSPs with strong magnetic fields will quickly, within a couple of Gyr, move out of the initial period distribution. This (relatively) rapid evolution of period means the distributions of $P_I$ and $\alpha$ produce (for a given $P$ and $B$) a range of possible ages for each MSP.

In Fig.~\ref{fig:E_cut_B_E_dot_N_MSPs} it is shown that, at 68\% confidence level, for
Model A1
between around 23000 and 54000 MSPs
are needed in the bulge to produce the observed GCE. 
This is consistent with Gonthier et al.~\cite{Gonthier2018} in which they find, though with a different bulge density model and GCE spectrum, 34,200 MSPs are needed with 11,500 in the region of interest 
associated with
the Gordon et al.~\cite{Gordon:2013vta} GCE spectrum. In Ploeg et al.~\cite{Ploeg:2017vai} it was found that around $(4.0\pm 0.9)\times10^4$ MSPs with $L > 10^{32} \textrm{ erg s}^{-1}$ were needed to produce the GCE; here, as shown in Fig.~\ref{fig:E_cut_B_E_dot_N_MSPs_greater_than_log_L_32}, the required number is between about 13000 and 26000 at 95\% confidence interval. 
Note the difference between these two estimates is not statistically significant; the slight discrepancy is likely due to the more accurate bulge geometric model, luminosity function, and GCE spectrum used in the current study.

In order to check the sensitivity of the results to the particular GCE spectrum used we also fitted Model A1 to alternative spectra. Using Bartels et al.~\cite{Bartels2017}, we found a small number of changes where the median of a parameter was outside the $68\%$ ranges shown in Table \ref{tab:mcmc_results_parameters}. The ratio of nuclear bulge and boxy bulge MSPs is lower with $\log_{10}(N_{\rm nb} / N_{\rm bb}) = -0.90\substack{+0.06 \\ -0.06}$. The relationship between $\Gamma$ and $\dot{E}$ is steeper with $a_{\Gamma} = 0.52\substack{+0.07 \\ -0.08}$. The fifth DTD bin is constrained strongly to be near $0$ with $\textrm{DTD } p(11.1 \textrm{ - } 13.8 \textrm{ Gyr}) = 0.01\substack{+0.03 \\ -0.01}$. These changes are likely to be caused by the steep slope on both the low and high energy ends of the boxy bulge spectrum. For the spectrum in Calore et al.~\cite{Calore:2014xka}, we had no nuclear bulge, but otherwise the results were consistent with those using the spectra of Macias et al.~\cite{Macias19}.

It can be seen in Fig.~\ref{fig:E_cut_B_E_dot_N_observed_bulge_MSPs} that we find a probability of $0.16$ that no bulge MSPs would have been resolved at present, with a median of $2$ resolved. With a doubling and quadrupling of detection sensitivity, respectively, a median of $10$ and $35$ would be expected to be resolved. In Table \ref{tab:possible_resolved_bulge_msps} we provide a list of resolved MSPs with a significant probability (greater than $5\%$) of being bulge MSPs according to Model A1. These probabilities are worked out in Eq.~\ref{eq:bulgelikelihood} by evaluating the contribution of the bulge to the corresponding MSP's likelihood. 

The probability of resolved MSP $i$ being a bulge MSP can be thought of as a draw from a single trial binomial distribution with probability $p(\textrm{Bulge MSP}_i)$ and therefore also having an expectation value of $p(\textrm{Bulge MSP}_i)$. Using the linearity of expectation, this implies that the expected number of resolved bulge MSPs in the current data is $\sum_{i} p(\textrm{Bulge MSP}_i)=1.1$ where the sum is over all resolved MSPs including those listed in Table~\ref{tab:possible_resolved_bulge_msps}. This is at the lower end of, but consistent with, the range seen in Fig.~\ref{fig:E_cut_B_E_dot_N_observed_bulge_MSPs} produced based on the fitted model parameters.

Note that there is also some systematic uncertainty in the distance to the MSPs which is hard to quantify. For example, 
  in Table 4 of ref.~\cite{Camilo15} they have a distance of 3.4 kpc for J1747--4036 while we have a distance of $7.3\substack{+0.7 \\ -0.7}$~kpc. Also in Table 2.2 of ref.~\cite{Sanpaarsa2016} they have a distance of 3.1 kpc for J1855--1436 while we have a distance of $5.3\substack{+0.6 \\ -0.6}$~kpc. These difference may be due to a change in model of the Galactic free electron density. According to the ATNF online database\footnote{\url{https://www.atnf.csiro.au/research/pulsar/psrcat/}} J1747--4036 and J1855--1436  have distances of 7.15 kpc and 5.13 kpc respectively
  which are compatible with our values. Also, in Table 2 of ref.~\cite{Ng2020} they have a dispersion measure distance of 1.8 kpc for J1811--2405 while we have a parallax derived distance of $5\substack{+11 \\ -2}$~kpc. 

\begin{figure}
    \centering
    \subfigure{\centering\includegraphics[width=0.49\linewidth]{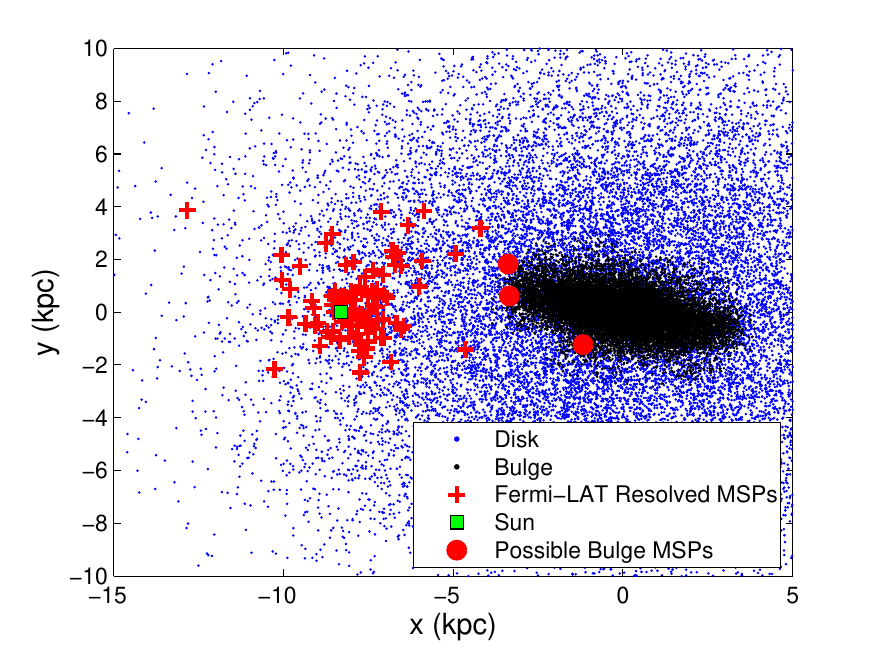}}
    \subfigure{\centering\includegraphics[width=0.49\linewidth]{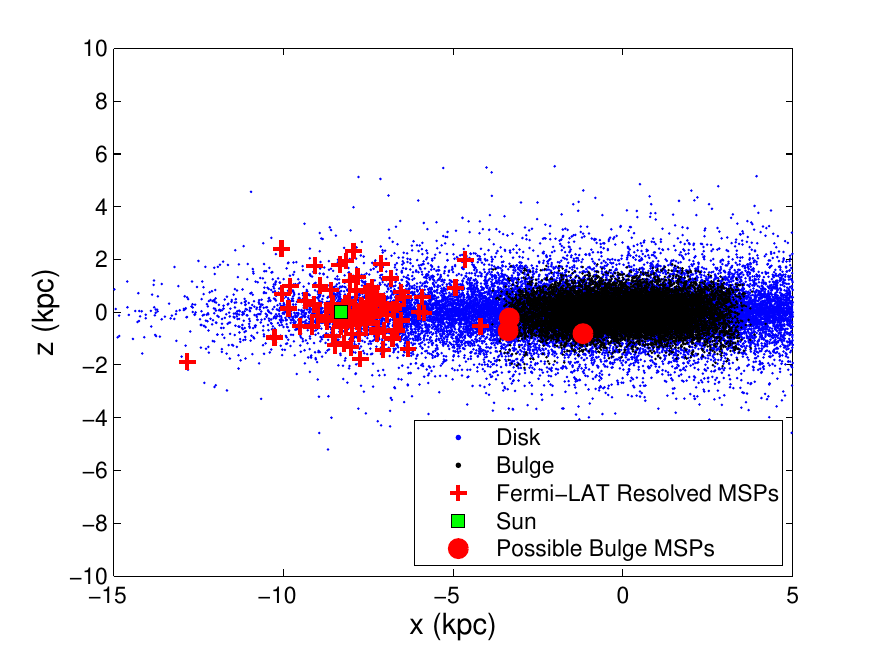}}
    \caption{Simulated distribution of disk and bulge MSPs for Model A1 ($L = \eta E_{\rm cut}^{a_{\gamma}} B^{b_{\gamma}} \dot{E}^{d_{\gamma}}$). Also shown are those resolved MSPs with available distance estimates. These figures were made by selecting randomly from the highest likelihood parameter sets in our eight Markov chains for this model. Red circles are the MSPs in Table \ref{tab:possible_resolved_bulge_msps}. }
    \label{fig:sample_msp_pop}
\end{figure}

In Fig.~\ref{fig:sample_msp_pop} the locations of resolved MSPs are shown along with a simulated distribution of disk and bulge MSPs. The elongated nature of the bulge geometry does not play a big role in the probability of having a resolved bulge MSP. This can be seen by changing the boxy bulge geometry in Eq.~\ref{eq:rho_main_bulge} to a spherically symmetric geometry with $\rho_{\rm boxy~bulge}\propto r^{-2.4}$ up to $r=3.1$~kpc and $\rho_{\rm boxy~bulge}=0$ for larger radii \cite{Ploeg:2017vai}.
We then find that $2\pm 2$ MSPs are expected to be resolved for Model A1. Also, for this spherically symmetric bulge case, the MSPs in Table~\ref{tab:possible_resolved_bulge_msps} have probabilities of 0.5, 0.2, and $4\times 10^{-4}$ respectively. So only PSR J1855-1436, with its high $l=20.4^\circ$, is significantly affected by the bulge geometry.

\begin{table}
\begin{center}
    \begin{tabular}{c|c|c|c|c|c|c|c}
         Name & Bulge Probability & $l$ (deg) & $b$ (deg) & $d$ (kpc) & $L$ ($\times 10^{34}$ erg s$^{-1}$) & $d$ percentile & $L$ percentile  \\ \hline \hline
         
         PSR J1747-4036 & 0.4 & $-9.8$ & $-6.4$ & $7.3\substack{+0.7 \\ -0.7}$ & $7.7\substack{+1.8 \\ -1.5}$ & $100$ & $99$ \\ \hline
         PSR J1811-2405 & 0.5 & $7.1$ & $-2.5$ & $5\substack{+11 \\ -2}$ & $5\substack{+47 \\ -4}$ & $96$ & $97$ \\ \hline
         PSR J1855-1436 & 0.1 & $20.4$ & $-7.6$ & $5.3\substack{+0.6 \\ -0.6}$ & $1.8\substack{+0.5 \\ -0.4}$ & $98$ & $91$ \\ \hline
    \end{tabular}
    \label{tab:possible_resolved_bulge_msps}
    \caption{ Details of MSPs with significant probability (greater than $5\%$) of coming from bulge population for Model A1. The names are from the Public List of LAT-Detected Gamma-Ray Pulsars\cref{footnote:pulsars}.
Distances are medians and $68\%$ intervals found by sampling the parallax uncertainty distribution if available or otherwise using the dispersion measure and randomly sampling the parameters of the YMW16 electron density model. Luminosities are medians and $68\%$ intervals found by additionally sampling from the flux uncertainty distribution. Percentiles are the fraction of resolved MSPs with smaller or equal luminosity/distance. This excludes MSPs without available distance estimates. }
    \end{center}
\end{table}

As shown in Fig.~\ref{fig:E_cut_B_E_dot_N_MSPs_per_solar_mass}, the disk, nuclear bulge  and boxy bulge have a consistent MSP to stellar mass ratio. This is a good confirmation of our assumption that %
{\textcolor{black}{the population of individual, resolved MSPs belonging mostly to the disk population, on the one hand, and the apparently diffuse $\gamma$-ray emission from the GCE, on the other, can be self-consistently explained as arising from}}
MSPs drawn from the same underlying luminosity function given by Eq.~\ref{eq:Model1}. %
\section{Conclusion}
\label{sec:conclusion}
We  compared a wide variety of luminosity function models for the Fermi-LAT gamma-ray MSP data. We found a convincing preference for Model A1 for which
$L = \eta E_{\rm cut}^{a_{\gamma}} B^{b_{\gamma}} \dot{E}^{d_{\gamma}}$
with a significantly positive $a_\gamma=1.2\pm0.3$ and $d_\gamma=0.5\pm 0.1$. Thus we confirm the result obtained by
Kalapotharakos et al.~\cite{Kalapotharakos_2019} that MSP gamma-ray  emission is consistent with curvature radiation and inconsistent with synchrotron radiation.
By comparing with other models, we showed that the main source of the positive 
$a_\gamma$ result was the need to account for a significant logarithmic correlation in the data between the $E_{\rm cut}$ and distance in the form of the dispersion measure. We also showed that the main source of the positive $d_\gamma$
was due to the significant logarithmic correlation in the data between the period and the distance in the form of the dispersion measure.

Additionally, we found that it was %
warranted
to include a relationship between the spectral parameters and $\dot{E}$. In particular we found that a linear relationship between the mean of the spectral index $\mu_\Gamma$ and the $\log_{10}(\dot{E})$, as specified in Eq.~\ref{eq:Gamma}, had a significantly positive slope of $a_\Gamma=0.4\pm0.08$. We identified the source of this positive slope to be the need to explain the significant correlation between $\Gamma$ and $\log_{\rm 10}(P)$ seen in the data.

We non-parametrically estimated the delay time distribution of the MSPs but found the current data does not strongly constrain it.
Our results obtained using a DTD prescription are also not significantly different from those obtained assuming a uniform age distribution for the MSPs.

Our results demonstrate that the population of MSPs that can explain the gamma-ray signal from the resolved MSPs in the Galactic disk and the unresolved MSPs in the boxy bulge and nuclear bulge
{\textcolor{black}{can consistently be described as arising from a common evolutionary trajectory for some subset of astrophysical sources common to all these different environments. 
We do not require that there is anything unusual about inner Galaxy MSPs to explain the GCE. We also found that the current data is not accurate enough to be sensitive to the small differences between the bulge and disk MSPs. 
}}

We estimated that there are 
between about 20 and 50 thousand
 MSPs in the boxy bulge at 68\% confidence interval.
We identified three candidate resolved MSPs (J1747-4036, J1811-2405, J1855-1436) that have significant probabilities (0.4, 0.5 and 0.1 respectively) of being members of the boxy bulge population. We estimate that this number would increase to 9 and 31 resolved boxy bulge MSPs were the 
{\textcolor{black}{point source}}
sensitivity of the gamma-ray observations doubled or quadrupled, respectively. 

\bibliographystyle{JHEP}
\bibliography{references}

\section*{Acknowledgments}
HP is supported by a University of Canterbury Doctoral Scholarship. 
RMC acknowledges an Australian Research Council Discovery award (DP190101258).
O.M. acknowledges support by JSPS KAKENHI Grant Numbers JP17H04836, JP18H04340, JP18H04578, and JP20K14463.
This work was supported by World Premier International Research Center Initiative (WPI Initiative), MEXT, Japan. 
\appendix
\section{Likelihood probability density function of resolved MSPs}
\label{app:probability}

In this appendix we implicitly assume that all
probabilities are conditioned on the parameters
($\thetab$). 
Also, as all of our expression here are also for an individual MSP we leave the subscript on each observational quantity as implicit.
As an example of these conventions, $p({\rm obs} \vert l, b, F)$ is equivalent to $p({\rm obs} \vert l_i, b_i, F_i,\thetab)$.

We assume the likelihood of observed MSPs with/without a parallax distance measurement depends only on distance $d$:
\begin{equation}
    \begin{multlined}
        p({\rm obs}, l, b, d, P, \dot{P}, \mu_l, \mu_b, F, E_{\rm cut}, \Gamma, \textrm{parallax/not parallax}) = p({\rm obs}, l, b, d, P, \dot{P}, \mu_l, \mu_b, F, E_{\rm cut}, \Gamma) \\ \times p(\textrm{parallax/not parallax} \vert d)
    \end{multlined}
\end{equation}
where the probability of a parallax measurement given distance $d$ is given by Eq.~\ref{eq:parallax_model}.
In terms of the various components of the MSP model, the probability density function of resolved MSPs at $l$, $b$, $d$, $P$, $\dot{P}$, $\mu_l$, $\mu_b$, $F$, $E_{\rm cut}$ and $\Gamma$ is: 
\begin{equation}
\label{eq:prob_density_int_init_period_alpha}
    p({\rm obs}, l, b, d, P, \dot{P}, \mu_l, \mu_b, F, E_{\rm cut}, \Gamma) =  \int \int   p({\rm obs}, l, b, d, P, \dot{P}, \mu_l, \mu_b, F, E_{\rm cut}, \Gamma, \alpha, P_I)\,\dd \alpha\, \dd P_I
\end{equation}
\noindent where we have integrated over the unknown magnetic axis angle and initial period. Then:
\begin{equation}
    \begin{multlined}
        p({\rm obs}, l, b, d, P, \dot{P}, \mu_l, \mu_b, F, E_{\rm cut}, \Gamma, \alpha, P_I) = p({\rm obs} \vert l, b, F) p(F \vert l, b, d, P, \dot{P}, \mu_l, \mu_b, E_{\rm cut}, \alpha) \\ \times p(E_{\rm cut}, \Gamma \vert l, b, d, P, \dot{P}, \mu_l, \mu_b) p(P, \dot{P} \vert l, b, d, \mu_l, \mu_b, \alpha, P_I) p(\mu_l, \mu_b \vert l, b, d) \\ \times p(l, b, d) p(\alpha) p(P_I)
    \end{multlined}
\end{equation}
\noindent The probability of observing an MSP with flux $F$ at $l$ and $b$ is given by Eq.\ \ref{eq:detection_probability}, so:
\begin{equation}
    p({\rm obs} \vert l, b, F) = p(F_{\rm th} \leq F \vert l, b)
\end{equation}

 Some of the data used is given in the form of $\pmb{y}_j$ where $\pmb{y}$ is not one of the directly modeled MSP parameters. In this case we perform a transformation to the observed parameter:
\begin{equation}
    \label{eq:uncertainty_integral_transformed}
    p(..., \pmb{y}_j, ...) = \int \dd \pmb{y} p(\pmb{y}_j \vert \pmb{y}) \lvert \pmb{J} \rvert p(..., \pmb{x}(\pmb{y}), ...)
\end{equation}
\noindent where $\lvert \pmb{J} \rvert$ is the Jacobian determinant of the transformation from $\pmb{y}$ to $\pmb{x}$, so $\pmb{J}_{i,j} = \partial x_i (\pmb{y}) / \partial y_j$. This transformation is needed for integrating over uncertainty in distance where we have either parallax ($\omega$), where $\omega = 1/d$, or dispersion measure (DM), where $\textrm{DM} = \int_0^d n_e(s) \dd s$ with $n_e(s)$ being the free electron density at distance $s$. 

The probability density function of an MSP having flux $F$ conditional upon it's other parameters is, given $\log(F) = \log(\eta) + f(...)$ where $f(...)$ is some function which doesn't depend on $\eta$:
\begin{equation}
\begin{split}
    p(F \vert l, b, d, P, \dot{P}, \mu_l, \mu_b, E_{\rm cut}, \alpha) &= p(\eta \vert l, b, d, P, \dot{P}, \mu_l, \mu_b, E_{\rm cut}, \alpha) \frac{\partial \eta}{\partial F} \\ &= p(\eta \vert l, b, d, P, \dot{P}, \mu_l, \mu_b, E_{\rm cut}, \alpha) \frac{\eta}{F}
\end{split}
\end{equation}
\noindent 
After a change of variables from period $P$ and observed period derivative $\dot{P}$ to magnetic field strength $B$ and age $t$, we get:
\begin{equation}
\label{eq:PPPdot}
    p(P, \dot{P} \vert l, b, d, \mu_l, \mu_b, \alpha, P_I) = p(B, t \vert l, b, d, \alpha, P_I) 
    \left\lvert \frac{\partial B}{\partial P} \frac{\partial t}{\partial \dot{P}} - \frac{\partial B}{\partial \dot{P}} \frac{\partial t}{\partial P} \right\rvert
\end{equation}
\noindent
where to evaluate the Jacobian in the above equation we
rewrite Eq.~\ref{eq:Shklovskii} as
\begin{equation}
\label{eq:Shklovskii1}
    \dot{P}_{\rm Shklovskii} = C_1 P
\end{equation}
where $C_1$ is a term that is independent of $P$ and $\dot{P}$. We also rewrite Eq.~\ref{eq:period_derivative_Galactic_correction} as
\begin{equation}
\label{eq:period_derivative_Galactic_correction1}
    \dot{P}_{\rm Galactic} = C_2 P
\end{equation}
where $C_2$ is also independent of $P$ and $\dot{P}$. 
We then obtain an equation for $B$ in terms of $P$ and $\dot{P}$ using
the above two equations with Eqs.~\ref{eq:period_derivative_observed_corrections} and  \ref{eq:magnetic_field_strength} to get 
\begin{equation}
\label{eq:B}
B^2=\frac{c^3 I P \left(- P({C}_1+{C}_2)+\dot{P}\right)}{\pi ^2 R^6 \left(\sin ^2(\alpha )+1\right)}
\end{equation}
Next we obtain an equation for $t$ in terms of $P$ and $\dot{P}$ by substituting the above equation into Eq.~\ref{eq:current_period} and solving $t$ to get
\begin{equation}
    t= \frac{P_I^2-P^2}{2 P \left(\left(C_1+{C}_2\right) P-\dot{P}\right)}
\end{equation}
Using the above two equations we can then solve for the Jacobian term in Eq.~\ref{eq:PPPdot} to get 
\begin{equation}
  \left\lvert \frac{\partial B}{\partial P} \frac{\partial t}{\partial \dot{P}} - \frac{\partial B}{\partial \dot{P}} \frac{\partial t}{\partial P} \right\rvert = \frac{c^6 I^2 P^2}{2 \pi^4 R^{12} (1 + \sin^2(\alpha))^2 B^3}
\end{equation}
where we have used Eq.~\ref{eq:B} to eliminate $C_1$ and $C_2$.
\noindent Similarly, using Eq.\ \ref{eq:proper_motion_to_linear_velocity}, $\partial v / \partial \mu \propto d$, so we find for proper motion:
\begin{equation}
   p(\mu_l, \mu_b \vert l, b, d) \propto p(v_l, v_b \vert l, b, d)  d^2
\end{equation}
\noindent where the proportionality constant is independent of our parameters and data and so does not affect our results. For position:
\begin{equation}
  p(l, b, d) \propto p(x, y, z) d^2 \cos(b)
\end{equation}
\noindent where $d^2 \cos(b)$ is proportional to the Jacobian of the change of variables from $x$, $y$ and $z$ to $l$, $b$ and $d$:
\begin{equation}
    \begin{split}
        x &= -R_0 + d \cos(l) \cos(b) \\
        y &= d \sin(l) \cos(b) \\
        z &= d \sin(b) \\
    \end{split}
\end{equation}
The density of resolved MSPs is the sum of the density in the disk, boxy bulge and nuclear bulge populations: 
\begin{equation}
\label{eq:density_sum_over_model_components}
    \rho(\textrm{obs}, ...) = N_{\rm disk} p_{\rm disk}(\textrm{obs}, ...) + N_{\rm bb} p_{\rm bb}(\textrm{obs}, ...) + N_{\rm nb} p_{\rm nb}(\textrm{obs}, ...)
\end{equation}
\noindent where we can calculate the number in each population using the parameters $\lambda_{\rm res}$, $\log_{10}(N_{\rm disk} / N_{\rm bulge})$ and $\log_{10}(N_{\rm nb} / N_{\rm bb})$ and solving with:
\begin{equation}
    \lambda_{\rm res} = N_{\rm disk} p_{\rm disk}(\textrm{obs}) + N_{\rm bb} p_{\rm bb}(\textrm{obs}) + N_{\rm nb} p_{\rm nb}(\textrm{obs})
\end{equation}
\noindent where each $p({\rm obs})$ is evaluated using Eq.~\ref{eq:pobs} for the corresponding spatial distribution.

In evaluating the likelihood for a given set of parameters, we used importance sampling to estimate integrals. 
As an example, in the case of the integrals over $P_I$ and $\alpha$ in Eq.~\ref{eq:prob_density_int_init_period_alpha} this integral becomes:
\begin{equation}
    \begin{split}
    p({\rm obs}, l, b, d, P, \dot{P}, \mu_l, \mu_b, F, E_{\rm cut}, \Gamma) &=  \int \int p({\rm obs}, l, b, d, P, \dot{P}, \mu_l, \mu_b, F, E_{\rm cut}, \Gamma, \alpha, P_I)\,\dd \alpha\, \dd P_I \\
    &=  \int \int p({\rm obs}, l, b, d, P, \dot{P}, \mu_l, \mu_b, F, E_{\rm cut}, \Gamma \vert \alpha, P_I)  p(\alpha) p(P_I) \,\dd \alpha\, \dd P_I \\
    &\approx \frac{1}{N} \sum_{i=1}^N p({\rm obs}, l, b, d, P, \dot{P}, \mu_l, \mu_b, F, E_{\rm cut}, \Gamma \vert \alpha_i, P_{Ii})
    \end{split}
\end{equation}
\noindent where we sum over $N$ samples $\alpha_i$ and $P_{Ii}$ from the probability distributions $p(\alpha)$ and $p(P_I)$.

\section{Measurement Uncertainties}
\label{appendix:uncertainties}
For the measurements of $F$, $E_{\rm cut}$ and $\Gamma$ we use the 4FGL covariance matrices\footnote{Kindly provided to us by Dr Jean Ballet of the Fermi-LAT collaboration.}
for the uncertainty in $N_0$, $\gamma_1$ and $a$, with the spectrum of the form\footnote{"PLSuperExpCutoff2" at https://fermi.gsfc.nasa.gov/ssc/data/analysis/scitools/source\_models.html}:
\begin{equation}
\label{eq:4fgl_xml_file_spectrum}
    \frac{\dd N}{\dd E} = N_0 \left(\frac{E}{E_0} \right)^{\gamma_1} \exp(-a E^{\gamma_2})
\end{equation}
\noindent where $E_0$ and $\gamma_2=2/3$ are fixed for each MSP.

For the small number of MSPs in 4FGL-DR2 that were not fitted with spectra in the form of Eq.\ \ref{eq:4fgl_xml_file_spectrum}, we simply use the reported flux and its error, treating $E_{\rm cut}$ and $\Gamma$ as missing. Since the catalog required $\Gamma > 0$, we also treat $E_{\rm cut}$ and $\Gamma$ as missing in the case where some MSPs had $\Gamma$ fixed near $0$. If, based on central estimates of $\dot{P}$, $\mu$ and $d$, $\dot{P}_{\rm int}$ appears to be negative for a particular MSP (i.e., the MSP is apparently spinning up) we treat $\dot{P}$ as missing. 
An apparently negative intrinsic period derivatives can result 
in the case that the true proper motion and/or distance is lower than the value we use, therefore causing the Shklovskii effect to be overestimated. An example of an updated proper motion measurement fixing this issue is given for PSR J1231-1411 in Abdo et al.~\cite{TheFermi-LAT:2013ssa}. A second potential cause of negative period derivative is radial acceleration in the direction of the sun significantly in excess of that accounted for by the $\dot{P}_{\rm Galactic}$ term in equations \ref{eq:period_derivative_observed_corrections} and \ref{eq:period_derivative_Galactic_correction}.

In the case of distance uncertainties, for the dispersion measure we use a method similar to that of Bartels et al.~\cite{Bartels2018}: The relative uncertainty in the dispersion measure is typically very small, so we assume that it is measured exactly. The main source of distance uncertainty is, therefore, associated with the model of the free electron density along the line of sight for each MSP. We use the YMW16 model of Yao et al.~\cite{Yao2017} and integrate over the uncertainty in the model parameters $\theta_{\rm YMW16}$ so that:
\begin{equation}
\begin{split}
    p(..., \textrm{DM}_j,...) &= \int \dd \theta_{\rm YMW16} p(\theta_{\rm YMW16}) p(..., \textrm{DM}_j, ... \vert \theta_{\rm YMW16}) \\
    &=  \!\begin{multlined}[t]
    \int \dd \theta_{\rm YMW16} p(\theta_{\rm YMW16}) \\ \times \left\lvert \frac{\partial}{\partial \textrm{DM}} d(\textrm{DM}_j, \theta_{\rm YMW16}) \right\rvert p(..., d(\textrm{DM}_j, \theta_{\rm YMW16}), ...) 
    \end{multlined} \\
    &= \int \dd \theta_{\rm YMW16} \frac{p(\theta_{\rm YMW16})}{n_e(d(\textrm{DM}_j, \theta_{\rm YMW16}), \theta_{\rm YMW16})} p(..., d(\textrm{DM}_j, \theta_{\rm YMW16}), ...)
\end{split}
\end{equation}


From Eqs.~\ref{eq:rhoD}, \ref{eq:likelihood_integral_over_true_vals} and \ref{eq:density_sum_over_model_components} we can work out the fraction of the likelihood contributed by the boxy bulge and nuclear bulge components for resolved  MSP $i$ given the model parameters and data uncertainties:
\begin{equation}
    p(\textrm{Bulge MSP}_i) = \frac{\rho_{\rm bb}(\D{i}) + \rho_{\rm nb}(\D{i})}{\rho(\D{i})}
    \label{eq:bulgelikelihood}
\end{equation}
\noindent where $\rho_{\rm bb}(\D{i}) = N_{\rm bb} p_{\rm bb}(\textrm{obs}, \D{i})$ and $\rho_{\rm nb}(\D{i}) = N_{\rm nb} p_{\rm nb}(\textrm{obs}, \D{i})$.

\section{Sampling methods}
\label{appendix:sampling}
The disk model, given in Eq.~\ref{eq:rho_disk}, can be sampled from by sampling in cylindrical coordinates a random $R$, $z$ and $\phi$. The radial coordinate $R$ is drawn from:
\begin{equation}
    p(R) = \frac{1}{\sigma_r^2} \exp(-R^2 / 2 \sigma_r^2) R
\end{equation}
\noindent for which, using the inverse of the cumulative distribution function of $R$, if $u$ is a uniformly random draw from $\left[0, 1 \right]$:
\begin{equation}
    R = \sigma_r \sqrt{-2 \log(1 - u)}
\end{equation}
\noindent The height $z$ is drawn from:
\begin{equation}
    p(z) = \frac{1}{2 z_0} \exp(-\abs{z} / z_0)
\end{equation}
\noindent which can be done by drawing $\abs{z}$ from an exponential distribution and choosing either a positive or negative sign each with $0.5$ probability. Finally, $\phi$ is drawn from a uniform distribution on $\left[0, 2 \pi \right]$.
For the boxy bulge and nuclear bulge distributions, we sampled the density using MCMC.

To sample from the age distribution, we used MCMC. We used standard library functions to sample from the various Gaussian distributions. To sample $\alpha$, inverse transform sampling was utilised.

\section{Watanabe-Akaike Information Criterion (WAIC)}
\label{appendix:WAIC}
The WAIC is defined in terms of the log pointwise predictive density (lppd) as \citep{Gelman2013}:
\begin{equation}
    \label{eq:WAIC_definition}
    {\rm WAIC} = -2 ({\rm lppd} - p_\text{WAIC})
\end{equation}
\noindent where for $N$ MSPs' data values ${\bf y}_i$ and $S$ parameter sets $\thetab^s$ in our Markov chain:
\begin{equation}
    \label{eq:lppd_definition}
    {\rm lppd} = \sum_{i=1}^N \log(\frac{1}{S} \sum_{s=1}^S p\left({\bf y}_i \;\middle\vert\; \thetab^s\right))
\end{equation}
\noindent and using the WAIC1 option  from  Gelman et al.~\cite{Gelman2013} 
\begin{equation}
    \label{eq:pwaic_definition}
    p_\text{WAIC} = 2 \sum_{i=1}^N \left( \log(\frac{1}{S} \sum_{s=1}^S p\left({\bf y}_i \;\middle\vert\; \thetab^s\right)) - \frac{1}{S} \sum_{s=1}^S \log(p\left({\bf y}_i \;\middle\vert\; \thetab^{s}\right)) \right)
\end{equation}

For the GCE contribution to the WAIC, for each set of parameters $\thetab^s$ we use the Gaussian likelihood for each bin as in Eq.\ \ref{eq:likelihood_GCE}. If we use, for the resolved MSPs, a Poisson distribution for bins in the several dimensions in which we have data, then define $\lambda_{i,s}$ as the expectation value for bin $i$ for parameter set $s$ and $n_i$ the number of observations in bin $i$, then:
\begin{equation}
    \label{eq:waic_poisson}
    \begin{split}
        p\left(n_i \;\middle\vert\; \lambda_{i,s}\right) &= \frac{\exp(-\lambda_{i,s}) \lambda_i^{n_i}}{n_i !} \\
        &= \frac{\exp(-\delta \rho_{i,s}) (\delta \rho_{i,s})^{n_i}}{n_i !}
    \end{split}
\end{equation}
\noindent where $\lambda_{i,s} = \delta \rho_{i,s}$ with $\delta$ the bin volume and $\rho_{i,s}$ the average density within bin $i$. Then if we choose $\delta$ small enough such that $\delta \rho_{i,s} \ll 1$ and $n_i$ is either $0$ or $1$, we can derive the contribution to the ${\rm lppd}$ from the resolved MSP data:
\begin{equation}
    \label{eq:lppd_res_msps1}
    \begin{split}
        {\rm lppd} &= \sum_{i=1}^N \log(\frac{1}{S} \sum_{s=1}^S p\left(n_i \;\middle\vert\; \lambda_{i,s}\right)), \mbox{ using Eq.~\ref{eq:lppd_definition}}  \\
        &= -N \log(S) + \sum_{i=1}^N \log(\sum_{s=1}^S \exp(-\delta \rho_{i,s}) (\delta \rho_{i,s})^{n_i}),\mbox{ using Eq.~\ref{eq:waic_poisson} and that  $n_i$ is either $0$ or $1$ } \\
        &\approx -N \log(S) + \sum_{i=1}^N \log(\sum_{s=1}^S (1 - \delta \rho_{i,s}) (\delta \rho_{i,s})^{n_i}), \mbox{ using $\delta \rho_{i,s} \ll 1$} \\
        &= -N \log(S) + \sum_{i=1}^N
       \begin{cases}
              \log(\sum_{s=1}^S (1 - \delta \rho_{i,s})) & n_i = 0 \\
              \log(\sum_{s=1}^S (1 - \delta \rho_{i,s}) (\delta \rho_{i,s})) & n_i = 1 
       \end{cases} \\
       &\approx -N \log(S) + \sum_{i=1}^N
       \begin{cases}
              \log(S (1 - \frac{1}{S} \sum_{s=1}^S \delta \rho_{i,s})) & n_i = 0 \\
              \log(\sum_{s=1}^S \delta \rho_{i,s}) & n_i = 1,  \mbox{ using $\delta \rho_{i,s} \ll 1$} 
       \end{cases} \\
       &= -N \log(S) + \sum_{i=1}^N
       \begin{cases}
              \left(\log(S) + \log(1 - \frac{1}{S} \sum_{s=1}^S \delta \rho_{i,s})\right) & n_i = 0 \\
              \left(\log(\delta) + \log(\sum_{s=1}^S \rho_{i,s})\right) & n_i = 1 
       \end{cases} \\
       &= -N \log(S) + N_{n_i=0} \log(S) + N_{n_i=1} \log(\delta) + \sum_{i=1}^N
       \begin{cases}
              \log(1 - \frac{1}{S} \sum_{s=1}^S \delta \rho_{i,s}) & n_i = 0 \\
              \log(\sum_{s=1}^S \rho_{i,s}) & n_i = 1 
       \end{cases}
       \end{split}
       \end{equation}
       where $N_{n_i=0}$ is the number of bin with zero counts and $N_{n_i=1}$ for the number of bins with one count.
       
       Using $-N  + N_{n_i=0}=-N_{n_i=1}$ in Eq.~\ref{eq:lppd_res_msps1} gives
       \begin{equation}
       \begin{split}
       {\rm lppd} &= -N_{n_i=1} \log(S) + N_{n_i=1} \log(\delta) + \sum_{i=1}^N
       \begin{cases}
              \log(1 - \frac{1}{S} \sum_{s=1}^S \delta \rho_{i,s}) & n_i = 0 \\
              \log(S \frac{1}{S} \sum_{s=1}^S \rho_{i,s}) & n_i = 1 
       \end{cases} \\
       &\approx -N_{n_i=1} \log(S) + N_{n_i=1} \log(\delta) + \sum_{i=1}^N
       \begin{cases}
              -\frac{1}{S} \sum_{s=1}^S \delta \rho_{i,s} & n_i = 0, \mbox{ using $\delta \rho_{i,s} \ll 1$}  \\
              \left(\log(S) + \log(\frac{1}{S} \sum_{s=1}^S \rho_{i,s})\right) & n_i = 1 
       \end{cases} \\
       &= N_{n_i=1} \log(\delta) + \sum_{i=1}^N
       \begin{cases}
              -\frac{1}{S} \sum_{s=1}^S \delta \rho_{i,s} & n_i = 0 \\
              \log(\frac{1}{S} \sum_{s=1}^S \rho_{i,s}) & n_i = 1 
       \end{cases} \\
       \end{split}
       \end{equation}
As most voxels will have $n_i=0$ we can approximate $\lambda_s=\sum_{n_i=0,n_i=1} \delta \rho_{i,s}\approx\sum_{n_i=0} \delta \rho_{i,s}$
     where $\lambda_s$ is simply the total expected number of resolved MSPs parameter.
     Therefore,
     \begin{equation}
     \begin{split}
     \label{eq:lppd_res_msps2}
    {\rm lppd}   &\approx N_{n_i=1} \log(\delta) - \frac{1}{S} \sum_{s=1}^S \lambda_s + \sum_{i=1}^N
       \begin{cases}
              0 & n_i = 0 \\
              \log(\frac{1}{S} \sum_{s=1}^S \rho_{i,s}) & n_i = 1 
       \end{cases} \\
       &= N_{n_i=1} \log(\delta) - \frac{1}{S} \sum_{s=1}^S \lambda_s + \sum_{j=1}^{N_{\rm res}} \log(\frac{1}{S} \sum_{s=1}^S N_{{\rm tot},s} p\left({\rm obs},\D{j} \;\middle\vert\; \thetab^s \right)) \\
    \end{split}
\end{equation}
\noindent where in the last line we have used Eq.~\ref{eq:rhoD}.
The first term in the above equation can be ignored as long as we are comparing models fitted using the same data. 

The other term we need to evaluate is given by Eq.~\ref{eq:pwaic_definition}
which we write in our notation as
\begin{equation}
      p_\text{WAIC} = 2 \sum_{i=1}^N \left( \log(\frac{1}{S} \sum_{s=1}^S p\left(n_i \;\middle\vert\; \lambda_{i,s}\right)) - \frac{1}{S} \sum_{s=1}^S \log(p\left(n_i \;\middle\vert\; \lambda_{i,s}\right)) \right)\,.
\end{equation}
We substitute Eq.~\ref{eq:waic_poisson}  to get
\begin{equation}
   p_\text{WAIC} = -2 N \log(S) + 2 \sum_{i=1}^N \left( \log(\sum_{s=1}^S \exp(-\delta \rho_{i,s}) (\delta \rho_{i,s})^{n_i}) - \frac{1}{S} \sum_{s=1}^S \log(\exp(-\delta \rho_{i,s}) (\delta \rho_{i,s})^{n_i}) \right) 
\end{equation}
where we have used the fact that $n_i=0$ or $n_i=1$ so $n_i!=1$ in either case. Next we separate out the two possible values for $n_i$ and make use of $\delta\rho_{i,s}\ll1$ as follows:
\begin{equation}
    \begin{split}
      p_\text{WAIC} &= -2 N \log(S) + 2 \sum_{i=1}^N
       \begin{cases}
              \left( \log(\sum_{s=1}^S \exp(-\delta \rho_{i,s})) - \frac{1}{S} \sum_{s=1}^S \log(\exp(-\delta \rho_{i,s})) \right) & n_i = 0 \\
              \left( \log(\sum_{s=1}^S \exp(-\delta \rho_{i,s}) (\delta \rho_{i,s})) - \frac{1}{S} \sum_{s=1}^S \log(\exp(-\delta \rho_{i,s}) (\delta \rho_{i,s})) \right) & n_i = 1 
       \end{cases} \\
       &\approx -2 N \log(S) + 2 \sum_{i=1}^N
       \begin{cases}
              \left( \log(\sum_{s=1}^S (1 - \delta \rho_{i,s})) - \frac{1}{S} \sum_{s=1}^S (-\delta \rho_{i,s}) \right) & n_i = 0 \\
              \left( \log(\sum_{s=1}^S (1 - \delta \rho_{i,s}) (\delta \rho_{i,s})) - \frac{1}{S} \sum_{s=1}^S \log((1 - \delta \rho_{i,s}) (\delta \rho_{i,s})) \right) & n_i = 1 
       \end{cases} \\
       &\approx -2 N \log(S) + 2 \sum_{i=1}^N
       \begin{cases}
              \left( \log(S (1 - \frac{1}{S} \sum_{s=1}^S \delta \rho_{i,s})) - \frac{1}{S} \sum_{s=1}^S (-\delta \rho_{i,s}) \right) & n_i = 0 \\
              \left( \log(S \frac{1}{S} \sum_{s=1}^S \delta \rho_{i,s}) - \frac{1}{S} \sum_{s=1}^S \log(\delta \rho_{i,s}) \right) & n_i = 1 
       \end{cases} \\
       &= -2 N \log(S) + 2 \sum_{i=1}^N
       \begin{cases}
              \left( \log(S) + \log(1 - \frac{1}{S} \sum_{s=1}^S \delta \rho_{i,s}) + \frac{1}{S} \sum_{s=1}^S \delta \rho_{i,s} \right) & n_i = 0 \\
              \left( \log(S) + \log(\frac{1}{S} \sum_{s=1}^S \delta \rho_{i,s}) - \frac{1}{S} \sum_{s=1}^S \log(\delta \rho_{i,s}) \right) & n_i = 1 
       \end{cases} \\
       &\approx 2 \sum_{i=1}^N
       \begin{cases}
              \left(-\frac{1}{S} \sum_{s=1}^S \delta \rho_{i,s} + \frac{1}{S} \sum_{s=1}^S \delta \rho_{i,s} \right) & n_i = 0 \\
              \left(\log(\frac{1}{S} \sum_{s=1}^S \delta \rho_{i,s}) - \frac{1}{S} \sum_{s=1}^S \log(\delta \rho_{i,s}) \right) & n_i = 1 
       \end{cases} \\
       &= 2 \sum_{i=1}^N
       \begin{cases}
              0 & n_i = 0 \\
              \left(\log(\frac{1}{S} \sum_{s=1}^S \rho_{i,s}) - \frac{1}{S} \sum_{s=1}^S \log(\rho_{i,s}) \right) & n_i = 1\,. 
       \end{cases} 
    \end{split}
\end{equation}
Substituting Eq.~\ref{eq:rhoD} into the above equation gives
\begin{equation}
\label{eq:p_waic1_res_msps}
     p_\text{WAIC}=  
       2 \sum_{j=1}^{N_{\rm res}} \left(\log(\frac{1}{S} \sum_{s=1}^S N_{{\rm tot},s} p\left({\rm obs}, \D{j} \;\middle\vert\; \thetab^s \right))  - \frac{1}{S} \sum_{s=1}^S \log(N_{{\rm tot},s} p\left({\rm obs},  \D{j} \;\middle\vert\; \thetab^s \right)) \right)\, .
\end{equation}
Substituting Eqs.~\ref{eq:p_waic1_res_msps} and \ref{eq:lppd_res_msps2} into Eq.~\ref{eq:WAIC_definition} gives
\begin{equation}
\label{eq:WAIC}
\begin{split}
    {\rm WAIC}=&2\left[
     \frac{1}{S} \sum_{s=1}^S \lambda_s - \sum_{j=1}^{N_{\rm res}} \log(\frac{1}{S} \sum_{s=1}^S N_{{\rm tot},s} p\left({\rm obs},\D{j} \;\middle\vert\; \thetab^s \right))\right]\\
    &
    +4\left\{ \sum_{j=1}^{N_{\rm res}} \left[\log(\frac{1}{S} \sum_{s=1}^S N_{{\rm tot},s} p\left({\rm obs}, \D{j} \;\middle\vert\; \thetab^s \right))  - \frac{1}{S} \sum_{s=1}^S \log(N_{{\rm tot},s} p\left({\rm obs},  \D{j} \;\middle\vert\; \thetab^s \right)) \right]
    \right\}\,.
    \end{split}
\end{equation}

\newpage

\label{lastpage}

\end{document}